\documentclass[10pt]{article}

\usepackage[dvips]{epsfig}
\usepackage{amssymb,amsmath,amsthm,mathrsfs}
\usepackage{graphicx}
\usepackage{lineno}

\usepackage[authoryear,sort&compress]{natbib}
\usepackage{url}
\usepackage{enumerate}
\usepackage{colordvi,color,pspicture}
\usepackage{psfrag}

\newcommand{\X}{\mathcal{X}}
\newcommand{\U}{\mathcal{U}}
\newcommand{\I}{\mathbf{I}}

\newcommand{\E}{\mathbf{E}}
\newcommand{\Var}{\mathbf{Var}}
\newcommand{\Cov}{\mathbf{Cov}}

\newcommand{\al}{\alpha}

\newcommand{\bh}{\widehat{\beta}}

\theoremstyle{plain}
\newtheorem{theorem}{Theorem}[section]

\theoremstyle{definition}

\theoremstyle{remark}

\newtheorem{remark}[theorem]{Remark}

\begin{document}


\title{Technical Report \# KU-EC-08-6:\\
New Tests of Spatial Segregation Based on Nearest Neighbor Contingency Tables}
\author{
Elvan Ceyhan
\thanks{Department of Mathematics, Ko\c{c} University, 34450 Sar{\i}yer, Istanbul, Turkey.
{\it e-mail:} elceyhan@ku.edu.tr}
}
\date{\today}
\maketitle


\begin{abstract}
\noindent
The spatial clustering of points from two or
more classes (or species) has important implications in many fields
and may cause the spatial patterns of segregation and association,
which are two major types of spatial interaction between the classes.
The null patterns we consider are random labeling (RL)
and complete spatial randomness (CSR) of points from two or more classes,
which is called \emph{CSR independence}, henceforth.
The segregation and association patterns can be studied
using a nearest neighbor contingency table (NNCT)
which is constructed using the frequencies of nearest neighbor (NN) types
in a contingency table.
Among NNCT-tests (i.e., tests based on NNCTs),
Pielou's test is equivalent to the usual (Pearson's) test
of independence for contingency tables,
but is liberal under CSR independence or RL patterns.
On the other hand, Dixon's test of segregation
has the desired significance level under the RL pattern.
We propose three new multivariate clustering tests based on NNCTs using the
appropriate sampling distribution of the cell counts in a NNCT
and suggest a simple correction for Pielou's test for data with rectangular support.
We compare the finite sample performance of these new tests with Pielou's and Dixon's tests
and Cuzick \& Edward's $k$-NN tests
in terms of empirical size under the null cases and
empirical power under various segregation and association alternatives
and provide guidelines for using the tests in practice.
We demonstrate that the newly proposed
NNCT-tests perform relatively well compared to their competitors
and illustrate the tests using three example data sets.
Furthermore, we compare the NNCT-tests with the second-order methods such as Ripley's
$L$-function and pair correlation function using these examples.
\end{abstract}

\noindent
{\small {\it Keywords:} Association; spatial clustering; complete spatial randomness;
independence; nearest neighbor methods; random labeling; second-order analysis; spatial pattern

\vspace{.25 in}





\section{Introduction}
\label{sec:intro}
Spatial point patterns have important implications in various fields
such as epidemiology, population biology, and ecology.
It is of practical interest to investigate the pattern of one class
not only with respect to the ground but also with respect to
the other classes (\cite{pielou:1961}, \cite{whipple:1980},
and \cite{dixon:1994,dixon:NNCTEco2002}).
For convenience and generality,
we refer to the different types of points as ``classes",
but the class can stand for any characteristic of an observation
or a point at a particular location.
For example, the spatial segregation pattern
has been investigated for \emph{plant species} (\cite{diggle:2003}),
\emph{age classes} of plants (\cite{hamill:1986}),
\emph{fish species} (\cite{herler:2005}),
and \emph{sexes} of dioecious plants (\cite{nanami:1999}).
Many of the epidemiological applications are for a two-class system of
case and control labels (\cite{waller:2004}).

Many tests of spatial segregation have been developed in literature (\cite{orton:1982}).
These include comparison of Ripley's $K$- or $L$-functions (\cite{ripley:2004}),
comparison of nearest neighbor (NN) distances (\cite{diggle:2003} and \cite{cuzick:1990}),
and analysis of nearest neighbor contingency tables (NNCTs)(\cite{pielou:1961} and \cite{meagher:1980}).
NNCTs are constructed using the NN frequencies of classes.
\cite{kulldorff:2006} provides an extensive review of tests of spatial randomness that
adjust for an inhomogeneity of the densities of the underlying populations.
In the presence of numerous tests available,
one advantage of NNCT-tests (i.e., tests based on NNCTs) is a theoretical one:
their asymptotic distributions are available.
However, given the available computational tools,
it might be a marginal advantage in practice.
Most of the tests in literature use Monte Carlo simulation
or randomization tests (\cite{kulldorff:2006}).
In fact, our NNCT-tests could also employ these simulation methods.
A major consideration is the power of the newly proposed NNCT-tests
in comparison with the currently available tests.
For better understanding of the power performance,
it is desirable to know the finite sample moments and
empirical size performance of the tests under the null cases.
Usually, the tests of spatial randomness involve a parameter,
which forces the user to resort to and then adjust for multiple testing.
The NNCT-tests combine four tests in $2 \times 2$ case,
and $q^2$ tests in $q \times q$ case,
so by construction avoid the problem of multiple testing.
Furthermore, they are potentially more powerful compared to other NN tests
which use less of the information provided.
The effects of homogeneity or lack of it and
how non-stationarity affects the results of spatial pattern tests
have recently been discussed in some detail in the ecological context
(\cite{perry:2006}) and for other methods (e.g., Ripley's $K$-function; \cite{baddeley:2000b}).
Some of the tests for spatial randomness are designed only for homogeneous populations,
while the tests we consider adjust for any inhomogeneity of the data,
in the sense that, these tests are appropriate for both homogeneous and
inhomogeneous populations.
For the $q$-class case with $q>2$ classes,
the overall segregation tests provide information on the (small-scale)
multivariate spatial interaction in one compound summary measure;
while the Ripley's $L$-function or pair correlation function requires performing all bivariate
spatial interaction analysis.
When the overall test is significant,
one can also use the cell-specific NNCT-tests for pairwise post hoc analysis (\cite{ceyhan:cell}).

In this article, for simplicity, we describe the spatial point patterns for two
classes only; the extension to multi-class case is straightforward.
We consider two major types of spatial clustering patterns, namely
\emph{association} and \emph{segregation}.
The null pattern is usually one of the two (random) pattern
types: \emph{complete spatial randomness} (CSR) or \emph{random labeling} (RL).
The NNCT-tests do not suffer from the problem of incorrect
specification of the null hypothesis (CSR independence or RL),
since they are for testing a more refined null hypothesis:
``the randomness in the NN structure".
In this article, we introduce correction (for dependence in cell counts)
strategies for Pielou's test.
The first type of correction methods
is derived analytically based on the correct
distribution of the cell counts under CSR independence or RL,
while the second type is based on Monte Carlo simulations.
We only consider \emph{completely mapped data},
i.e., the locations of all events in a defined space are observed.
There have been some reservations on the appropriateness of Pielou's
tests for completely mapped data (see \cite{meagher:1980} and \cite{dixon:1994}). 
Pielou's test assumes independence of cell counts in the NNCT,
but these cell counts are dependent since it is more likely for a point
to be a NN of its NN (reflexivity in NN structure).

\section{Null and Alternative Patterns}
\label{sec:null-and-alt}
The appropriate null pattern for NNCT-tests is $H_o:$ \emph{randomness in the NN structure}.
This null hypothesis usually results from one of the two (random) pattern
types: \emph{random labeling} (\emph{RL})
of a set of fixed points with two classes or
\emph{complete spatial randomness} (\emph{CSR}) of points from two classes,
which is called ``CSR independence", henceforth.
That is, when the points from each
class are assumed to be uniformly distributed over the region of
interest, then randomness in the NN structure is implied by the CSR independence pattern.
This CSR independence pattern is also referred to as ``population independence"
in literature (\cite{goreaud:2003}).
Note that CSR independence is equivalent
to the case that RL procedure is applied to a
given set of points from a CSR pattern,
in the sense that after points are generated uniformly
in the region, the class labels are assigned randomly.
When only the labeling of a set of fixed points
(the allocation of the points could be regular, aggregated, or clustered, or of lattice type)
is random, the null hypothesis is implied by RL pattern.
The distinction between CSR independence and RL is very important
when defining the appropriate null model for each empirical case;
i.e., the null model depends on the particular ecological context.
\cite{goreaud:2003} discuss the differences between these two null hypotheses
and demonstrate that the misinterpretation is very common.
They also propose some guidelines to help decide
which null hypothesis is appropriate and when.
They assert that under CSR (independence) the (locations of the points from)
two classes are \emph{a priori}
the result of different processes
(e.g., individuals of different species or age cohorts),
whereas under RL, some processes affect \emph{a posteriori}
the individuals of a single population
(e.g., diseased versus non-diseased individuals of a single species).
Notice also that although CSR independence and RL are not same,
they lead to the same null model (i.e., randomness in NN structure) for tests using NNCT,
which does not require spatially-explicit information.
In general CSR refers to a univariate pattern and implies that the spatial distribution of
points is random over the study area,
but makes no assumption about the distribution of the
labels within the set (e.g., points could conform to CSR and the marks
or labels could also be segregated).
To emphasize the distinction between univariate and multivariate CSR patterns,
the univariate pattern is called just ``CSR",
while the multivariate CSR pattern is called ``CSR independence".
Hence in this article, CSR independence pattern refers to the spatial distribution of points
from each of the classes is random and uniform over the region of interest.
RL suggests that different labels are assigned to the points at random,
but makes no assumption about the spatial arrangement of points
(e.g., points could be spatially clumped, segregated or associated).

We consider two major types of (bivariate) spatial clustering patterns of
\emph{association} and \emph{segregation} as alternative patterns.
{\em Association} occurs if the NN of an individual is more
likely to be from another class.
For example, in plant biology, the two classes of points
might represent the coordinates of mutualistic plant species,
so the species depend on each other to survive.
As another example, one class of points
might be the geometric coordinates of
parasitic plants exploiting  the other plant
whose coordinates are of the other class.
In epidemiology, one class of points might
be the geographical coordinates of contaminant sources,
such as a nuclear reactor or a factory emitting toxic waste,
and the other class of points might be the coordinates of the residences of cases
(i.e., incidences) of certain diseases, e.g., some type of cancer caused by the contaminant.
{\em Segregation} occurs if the
NN of an individual is more likely to be of the same class as the individual;
i.e., the members of the same class tend to be clumped or clustered
(see, e.g., \cite{pielou:1961}).
For instance, one type of plant might not grow well
around another type of plant and vice versa.
In plant biology, one class of points might represent
the coordinates of trees from a species with large canopy,
so that other plants (whose coordinates are the points from the other class)
that need light cannot grow around these trees.
See, for instance, (\cite{dixon:1994}, \cite{coomes:1999}) for more detail.
The segregation and association patterns are not symmetric in the sense that,
when two classes are segregated, they do not necessarily
exhibit the same degree of segregation
or when two classes are associated,
one class could be more associated with the other.
Many different forms of segregation (and
association) are possible.
Although it is not possible to list all
types of segregation,
its existence can be tested by an analysis of the
NN relationships between the classes (\cite{pielou:1961}).
Both patterns might result from differences
between first-order or second-order stationarity
of the two classes or species.
In general departures from first-order homogeneity are likely to be important drivers of segregation;
and association would usually result from the second order effects.
However, when discussing the various examples (in Section \ref{sec:examples})
we refrain from mentioning any causality,
which is not necessarily implied by these tests.

We describe the construction of NNCTs in Section \ref{sec:NNCT},
provide NNCT-tests in Section \ref{sec:NNCT-tests},
Pielou's and Dixon's tests in Sections
\ref{sec:pielou-tests} and \ref{sec:dixon-tests},
respectively, and
the new versions of segregation tests in Section \ref{sec:new-seg-tests},
other tests of spatial clustering in Section \ref{sec:other-spat-tests},
empirical significance levels of the tests in Section \ref{sec:emp-sign-level},
empirical power analysis in Section \ref{sec:emp-power},
three illustrative examples in Section \ref{sec:examples},
and discussion and conclusions with guidelines in using the tests in
Section \ref{sec:disc-conc}.

\section{Nearest Neighbor Contingency Tables and Related Tests}
\subsection{Construction of the Nearest Neighbor Contingency Tables}
\label{sec:NNCT}
NNCTs are constructed using the NN frequencies of classes.
We describe the construction of NNCTs for two classes; extension to multi-class
case is straightforward.
Consider two classes with labels $\{1,2\}$.
Let $N_i$ be the number of points from class $i$ for $i \in \{1,2\}$ and
$n$ be the total sample size, so $n=N_1+N_2$.
If we record the class
of each point and the class of its NN, the NN
relationships fall into four distinct categories:
$(1,1),\,(1,2);\,(2,1),\,(2,2)$ where in cell $(i,j)$, class $i$ is
the \emph{base class}, while class $j$ is the class of its
\emph{NN}.
That is, the $n$ points constitute $n$
(base,NN) pairs.
Then each pair can be categorized with respect to
the base label (row categories) and NN label (column categories).
Denoting $N_{ij}$ as the frequency of cell $(i,j)$ for $i,j \in
\{1,2\}$, we obtain the NNCT in Table \ref{tab:NNCT-2x2} where $C_j$
is the sum of column $j$; i.e., number of times class $j$ points
serve as NNs for $j \in \{1,2\}$.
Furthermore, $N_{ij}$ is the cell count for
cell $(i,j)$ that is the count of all (base,NN) pairs each of which
has label $(i,j)$.
Note also that
$n=\sum_{i,j}N_{ij}$; $n_i=\sum_{j=1}^2\, N_{ij}$; and
$C_j=\sum_{i=1}^2\, N_{ij}$.
By construction, if $N_{ij}$ is larger (smaller) than expected,
then class $j$ serves as NN more (less) to class $i$ than expected,
which implies (lack of) segregation if $i=j$ and (lack of) association of class $j$
with class $i$ if $i\not=j$.
Furthermore, we adopt the convention that variables denoted by upper (lower) case letters are random (fixed) quantities
throughout the article.
Hence, column sums and cell counts are random, while row sums and the overall sum are fixed
quantities in a NNCT.

\begin{table}[ht]
\centering
\begin{tabular}{cc|cc|c}
\multicolumn{2}{c}{}& \multicolumn{2}{c}{NN class}& \\
\multicolumn{2}{c|}{}& class 1 &  class 2 & sum  \\
\hline
&class 1 &    $N_{11}$  &   $N_{12}$  &   $n_1$  \\
\raisebox{1.5ex}[0pt]{base class}
&class 2 &    $N_{21}$ &  $N_{22}$    &   $n_2$  \\
\hline
& sum    &    $C_1$   & $C_2$         &   $n$  \\
\end{tabular}
\caption{
\label{tab:NNCT-2x2}
NNCT for two classes.}
\end{table}

Observe that, under segregation, the diagonal entries, i.e.,
$N_{ii}$ for $i=1,2$, tend to be larger than expected; under
association, the off-diagonals tend to be larger than expected.
The general alternative is that some cell counts are different
than expected under CSR independence or RL.

\subsection{A Review of NNCT-Tests in Literature}
\label{sec:NNCT-tests}
\cite{pielou:1961} proposed tests (for segregation, symmetry, niche specificity,
etc.) and Dixon introduced an
overall test of segregation, cell-, and class-specific tests based on NNCTs
for the two-class case (\cite{dixon:1994}) and extended his tests to
multi-class (or multi-species) case (\cite{dixon:NNCTEco2002}).
\cite{pielou:1961} used the usual Pearson's $\chi^2$-test of
independence for detecting the segregation of the two classes.
Due to the ease in computation and interpretation, Pielou's test of
segregation is frequently used for
both completely mapped and sparsely sampled data (\cite{meagher:1980});
indeed it is more frequently used than Dixon's test.
For example, Pielou's test is used for the segregation
of males and females in dioecious species (e.g., \cite{herrera:1988}
and \cite{armstrong:1989}), and of different species (\cite{good:1982}).
\cite{dixon:1994} points out two problems with Pielou's test:
(i) it fails to identify certain types of segregation (e.g., mother-daughter processes)
and
(ii) the sampling distribution of cell counts is not appropriate.
The assumption for the use
of chi-square test for NNCTs is the independence between cell-counts
(hence rows and columns), which is violated for NNCTs
based on the two classes from CSR independence or RL patterns.
\cite{dixon:1994} derived the appropriate (asymptotic) sampling
distribution of cell counts using Moran join count statistics (\cite{moran:1948}) and
hence the appropriate test which also has a $\chi^2$-distribution (\cite{dixon:1994}).
Problem (ii) was first
noted by \cite{meagher:1980} who identify the main cause
of it to be reflexivity of (base, NN) pairs.
A pair of points is \emph{reflexive} if each point in the pair
is the NN of the other point in the pair,
regardless of the class of the two individuals.
As an alternative, they
suggest using Monte Carlo simulations for Pielou's test.
\cite{dixon:1994} also argues that Pielou's test is not appropriate for completely mapped
data, but is appropriate for sparsely sampled data.

For the two-class case, \cite{ceyhan:overall} compared these tests,
and in addition to the previously mentioned reservations on the use of Pielou's tests
for completely mapped data, demonstrated that
Pielou's tests are liberal under CSR independence and RL and are
only appropriate for a random sample of (base,NN) pairs.


\subsection{Pielou's Test of Segregation}
\label{sec:pielou-tests}
In the two-class case, Pielou used Pearson's $\chi^2$-test
of independence to detect any deviation from CSR independence or RL (\cite{pielou:1961}).
The test statistic is
\begin{equation}
\label{eqn:piel-seg-2x2}
\X_P^2=\sum_{i=1}^2\sum_{j=1}^2\frac{\left( N_{ij}-\mathbf {E_P}[N_{ij}] \right)^2}{\mathbf {E_P}[N_{ij}]}.
\end{equation}
When NNCT is based on a random sample of (base,NN) pairs,
in Equation \eqref{eqn:piel-seg-2x2},
$\mathbf {E_P}[N_{ij}]=n_i\,c_j/n$ and $c_j$ is the sum for column $j$;
and $\X_P^2$ is approximately distributed as $\chi^2_1$ (i.e.,
$\chi^2$ distribution with 1 degrees of freedom) for large $n_i$.
Rejecting $H_o:$ \emph{independence of cell counts} for large values
of $\X_P^2$ (for $\X_P^2>\chi^2_1(1-\al)$, the $(1-\al)$ quantile of
$\chi^2_1$ distribution) yields a consistent test.
But, under CSR independence or RL, this test is liberal;
i.e., it has larger size than the desired level (\cite{ceyhan:overall}).

\subsection{Dixon's NNCT-Tests}
\label{sec:dixon-tests}
Dixon proposed a series of tests
for segregation based on NNCTs (\cite{dixon:1994}).
For Dixon's tests, the probability of an individual from
class $j$ serving as a NN of an individual from class $i$
depends only on the class sizes (i.e., row sums),
but not the total number of times class $j$ serves as NNs (i.e., column sums).

\subsubsection{Dixon's Cell-Specific Tests}
\label{sec:Dixon-cell-spec}
The level of segregation is estimated by
comparing the observed cell counts to the expected cell counts
under RL of points that are fixed. 
Dixon demonstrates that under RL,
one can write down the cell frequencies as Moran join count
statistics (\cite{moran:1948}).
He then derives the means, variances, and covariances of
the cell counts (frequencies) in a NNCT (\cite{dixon:1994,dixon:NNCTEco2002}).

The null hypothesis of RL implies
\begin{equation}
\label{eqn:Exp[Nij]}
\mathbf {E}[N_{ij}]=
\begin{cases}
n_i(n_i-1)/(n-1) & \text{if $i=j$,}\\
n_i\,n_j/(n-1)     & \text{if $i \not= j$.}
\end{cases}
\end{equation}
Observe that the expected cell counts depend only on the size of
each class (i.e., row sums), but not on column sums.

The cell-specific test statistics suggested by Dixon are given by
\begin{equation}
\label{eqn:dixon-Zij}
Z^D_{ij}=\frac{N_{ij}-\mathbf {E}[N_{ij}]}{\sqrt{\Var[N_{ij}]}},
\end{equation}
where
\begin{equation}
\label{eqn:VarNij}
\Var[N_{ij}]=
\begin{cases}
(n+R)\,p_{ii}+(2\,n-2\,R+Q)\,p_{iii}+(n^2-3\,n-Q+R)\,p_{iiii}-(n\,p_{ii})^2 & \text{if $i=j$,}\\
n\,p_{ij}+Q\,p_{iij}+(n^2-3\,n-Q+R)\,p_{iijj} -(n\,p_{ij})^2         & \text{if $i \not= j$,}
\end{cases}
\end{equation}
with $p_{xx}$, $p_{xxx}$, and $p_{xxxx}$
are the probabilities that a randomly picked pair,
triplet, or quartet of points, respectively, are the indicated classes and
are given by
\begin{align}
\label{eqn:probs}
p_{ii}&=\frac{n_i\,(n_i-1)}{n\,(n-1)},  & p_{ij}&=\frac{n_i\,n_j}{n\,(n-1)},\nonumber\\
p_{iii}&=\frac{n_i\,(n_i-1)\,(n_i-2)}{n\,(n-1)\,(n-2)}, &
p_{iij}&=\frac{n_i\,(n_i-1)\,n_j}{n\,(n-1)\,(n-2)},\\
 p_{iijj}&=\frac{n_i\,(n_i-1)\,n_j\,(n_j-1)}{n\,(n-1)\,(n-2)\,(n-3)},&
p_{iiii}&=\frac{n_i\,(n_i-1)\,(n_i-2)\,(n_i-3)}{n\,(n-1)\,(n-2)\,(n-3)}.\nonumber
\end{align}
Furthermore, $Q$ is the number of points with shared  NNs,
which occur when two or more
points share a NN and
$R$ is twice the number of reflexive pairs.
Then $Q=2\,(Q_2+3\,Q_3+6\,Q_4+10\,Q_5+15\,Q_6)$
where $Q_k$ is the number of points that serve
as a NN to other points $k$ times.
One-sided and two-sided tests are possible for each cell $(i,j)$
using the asymptotic normal approximation of $Z^D_{ij}$ given in
Equation (\ref{eqn:dixon-Zij}) (\cite{dixon:1994}).
The test in Equation \eqref{eqn:dixon-Zij} is the same as
Dixon's $Z_{AA}$ when $i=j=1$;
same as $Z_{BB}$ when $i=j=2$ (\cite{dixon:1994}).
Note also that in Equation \eqref{eqn:dixon-Zij} four different tests are defined
as there are four cells and each is testing deviation from the null case
for the respective cell.
These four tests are combined and
used in defining an overall test of segregation in Section \ref{sec:dix-overall}.

Under CSR independence, the null hypothesis, the test statistics, and
the variances are as in the RL case for the cell-specific tests,
except that the variances are conditional on $Q$ and $R$.

\begin{remark}
\label{rem:QandR}
\textbf{The Status of $Q$ and $R$ under CSR Independence and RL:}
Note the difference  in status of the variables $Q$ and $R$ under CSR independence and RL models.
Under RL, $Q$ and $R$ are fixed quantities;
while under CSR independence, they are random.
The quantities given in Equations \eqref{eqn:Exp[Nij]}, \eqref{eqn:VarNij},
and all the quantities depending on these expectations also depend on $Q$ and $R$.
Hence these expressions are appropriate under the RL pattern.
Under the CSR independence pattern,
they are conditional variances and
covariances obtained by conditioning on $Q$ and $R$.
The unconditional variances and covariances can be obtained
by replacing $Q$ and $R$ with their expectations.

Unfortunately, given the difficulty of calculating the
expectations of $Q$ and $R$ under CSR independence,
it is reasonable and convenient to use test statistics employing the
conditional variances and covariances even when assessing their
behavior under CSR independence.
Alternatively, one can estimate the expected values of $Q$ and $R$ empirically
and substitute these estimates in the expressions.
For example, for the homogeneous planar Poisson process (conditional on the sample size),
we have $\E[Q/n] \approx .632786$ and $\E[R/n] \approx 0.621120$.
(estimated empirically based on 1000000 Monte Carlo simulations
for various values of $n$ on unit square).
When $Q$ and $R$ are replaced by $0.63 \, n$ and $0.62 \,n$, respectively,
we obtain the so-called \emph{QR-adjusted} tests.
However, as shown in \cite{ceyhanarXivQRAdjust:2008},
QR-adjustment does not improve on the unadjusted NNCT-tests.
$\square$
\end{remark}

\subsubsection{Dixon's Overall Test of Segregation}
\label{sec:dix-overall}
Dixon's overall test of segregation tests the hypothesis that expected
cell counts in the NNCT are as in Equation \eqref{eqn:Exp[Nij]}.
In the two-class case, he calculates
$\displaystyle Z_{ii}=(N_{ii}-\E[N_{ii}])\big/\sqrt{\Var[N_{ii}]}$ for both $i \in
\{1,2\}$ and combines these test statistics into a statistic
that is asymptotically distributed as $\chi^2_2$ under RL (\cite{dixon:1994}).
Under RL, the suggested test statistic is given by
\begin{equation}
\label{eqn:dix-chisq-2x2}
C_D=\mathbf{Y}'\Sigma^{-1}\mathbf{Y}=
\left[
\begin{array}{c}
N_{11}-\E[N_{11}] \\
N_{22}-\E[N_{22}]
\end{array}
\right]'
\left[
\begin{array}{cc}
\Var[N_{11}] & \Cov[N_{11},N_{22}] \\
\Cov[N_{11},N_{22}] & \Var[N_{22}] \\
\end{array}
\right]^{-1}
\left[
\begin{array}{c}
N_{11}-\E[N_{11}] \\
N_{22}-\E[N_{22}]
\end{array}
\right],
\end{equation}
where $\E[N_{ii}]$ are as in Equation (\ref{eqn:Exp[Nij]}),
$\Var[N_{ii}]$ are as in Equation (\ref{eqn:VarNij}),
and
\begin{equation}
\label{eqn:cov-N11-N22}
\Cov[N_{11},\,N_{22}]=(n^2-3\,n-Q+R)\,p_{1122}-n^2\,p_{11}\,p_{22}.
\end{equation}
Dixon's $C_D$ statistic given in Equation \eqref{eqn:dix-chisq-2x2} can also be written as
$$C_D=\frac{\left(Z^D_{11} \right)^2+\left(Z^D_{22} \right)^2-2rZ^D_{11}Z^D_{22}}{1-r^2},$$
where $r=\Cov[N_{11},N_{22}] \Big /\sqrt{\Var[N_{11}]\Var[N_{22}]}$ (\cite{dixon:1994}).

Under CSR independence, the expected values, variances and covariances are as in the RL case.
However, the variance and covariance terms include $Q$ and $R$ which are random
under CSR independence and fixed under RL.
Hence Dixon's test statistic $C_D$ asymptotically has a $\chi^2_1$-distribution under CSR independence
conditional on $Q$ and $R$.

\subsection{New Overall Segregation Tests Based on NNCTs}
\label{sec:new-seg-tests}
First, we propose tests based on the correct sampling distribution of
the cell counts in a NNCT under RL or CSR independence.
Then, we suggest a transformation based on Monte Carlo simulations
to correct for the effect of the dependence between the cell counts.
Thereby, we adjust Pielou's test for location and scale to render it
have the desired level under the null case.
In defining the new segregation or clustering tests, we follow
a track similar to that of Dixon's (\cite{dixon:1994}).
For each cell, we define a new type of cell-specific test statistic, and then
combine these four tests into one overall test.

\subsubsection{First Version of the New Segregation Tests}
\label{sec:seg-test-I}
Recall that in Equation \eqref{eqn:piel-seg-2x2},
$\mathbf {E_P}[N_{ij}]=n_i\,c_j/n$.
Asymptotically, $\X_P^2$ has
$\chi^2_1$-distribution only when the NNCT is based on a random sample of
(base,NN) pairs, which is not the case under CSR independence or RL (\cite{ceyhan:overall}).

Similar to Dixon's cell-specific tests in Section \ref{sec:Dixon-cell-spec},
we consider the following test statistics for cells in the NNCT
\begin{equation}
\label{eqn:seg-test-I}
T^I_{ij}=N_{ij}-\frac{n_i\,C_j}{n}.
\end{equation}
Under RL, row sums $N_i=n_i$ are fixed while
column sums $C_j$ are random quantities.
Hence, conditional on $C_j=c_j$, $T^I_{ij}=N_{ij}-n_i\,c_j/n$;
and let
\begin{equation}
\label{eqn:ana-correct-I}
N^I_{ij}=\frac{T^I_{ij}}{\sqrt{n_i\,c_j/n}}=
\frac{\left( N_{ij}-n_i\,c_j/n \right)}{ \sqrt{n_i\,c_j/n} },
\end{equation}
then $\X^2_P=\sum_{i=1}^2\sum_{j=1}^2 \left( N^I_{ij} \right)^2$.
Under RL, we find that
\begin{equation}
\label{eqn:Exp-T-Iij}
\E\left[ T^I_{ij} \right]=
\begin{cases}
\frac{n_i\,(n_i-n)}{n\,(n-1)} & \text{if $i=j$,} \vspace{.05 in} \\
\frac{n_i\,n_j}{n\,(n-1)}     & \text{if $i \not= j$.}
\end{cases}
\end{equation}
Notice that under RL, $\E\left[ T^I_{ij} \right] \not=0$
which implies that $\E\left[ N^I_{ij} \right] \not=0$.
Furthermore,
\begin{equation}
\lim_{n_i,n_j \rightarrow \infty} \E\left[ T^I_{ij} \right] =
\begin{cases}
\nu_i \, (1-\nu_i) & \text{if $i=j$,} \vspace{.05 in} \\
\nu_i \, \nu_j    & \text{if $i \not= j$,}
\end{cases}
\end{equation}
where $\nu_i$ is the probability that an individual is of class $i$
and $n_i,n_j \rightarrow \infty$ means that $\min(n_i,n_j) \rightarrow \infty$.
However, although, $\E\left[ N^I_{ij} \right]$ is not analytically tractable,
$\lim_{n_i,n_j \rightarrow \infty} \E\left[ N^I_{ij} \right]=0$ since
$\sqrt{n_i\,c_j/n^2} \rightarrow \sqrt{\nu_i^2}=\nu_i$
which implies $1\big/\sqrt{n_i\,c_j/n} \rightarrow 0$ as $n_i,n_j \rightarrow \infty$.

Let $\mathbf{N_I}$ be the vector of $N^I_{ij}$ values
concatenated row-wise and let $\Sigma_I$ be the
variance-covariance matrix of $\mathbf{N_I}$ based on the correct
sampling distribution of the cell counts.
That is,
$\Sigma_I=\left(
\Cov\left[ N^I_{ij},N^I_{kl} \right] \right)$ where
$\displaystyle \Cov\left[ N^I_{ij},N^I_{kl} \right]=
\frac{n}{\sqrt{n_i\,c_j\,n_k\,c_l}}\Cov\left[ N_{ij},N_{kl} \right]$ with
$\Cov\left[ N_{ij},N_{kl} \right]$ is as in Equation \eqref{eqn:VarNij} if
$(i,j)=(k,l)$ and as in Equation \eqref{eqn:cov-N11-N22} if
$(i,j)=(1,1)$ and $(k,l)=(2,2)$.
Since $\Sigma_I$ is not invertible,
we use its generalized inverse, $\Sigma_I^-$ (\cite{searle:2006}).
Then the proposed test statistic for overall segregation is the quadratic form
\begin{equation}
\label{eqn:X-square-I}
\X_I^2=\mathbf{N'_I}\Sigma_I^-\mathbf{N_I}
\end{equation}
which asymptotically has a $\chi^2_1$ distribution.

The test statistic $\X_I^2$ can be obtained by adding a correction term to $\X^2_P$.
Recall that
$$\X^2_P=\sum_{i=1}^2\sum_{j=1}^2\left( N^I_{ij} \right)^2=\mathbf{N'_I}\mathbf{N_I},$$
hence
$$\X_I^2=\mathbf{N'_I}\Sigma_I^-\mathbf{N_I}=
\mathbf{N'_I}(\Sigma_I^- -I_2 +I_2)\mathbf{N_I}=
\mathbf{N'_I}\mathbf{N_I}+\mathbf{N'_I}(\Sigma_I^- -I_2)\mathbf{N_I}=
\X^2_P+\Delta_c$$
where $\Delta_c=\mathbf{N'_I}(\Sigma_I^- -I_2)\mathbf{N_I}$ and $I_2$
is the $2 \times 2$ identity matrix.
Furthermore, $\Sigma_I$ can be obtained from $\Sigma$
in Equation \eqref{eqn:dix-chisq-2x2} by multiplying $\Sigma$ entry-wise with the matrix
$\displaystyle C^I_M=\left( \frac{n}{\sqrt{n_i\,c_j\,n_k\,c_l}} \right)$.
Since $T^I_{ij}$ is conditional on $C_j=c_j$,
segregation test with $\X_I^2$ is conditional on the column sums.

Under CSR independence, the expected values, variances, and covariances related to $\X_I^2$ are as in the RL case,
except they are not only conditional on column sums (i.e., on $C_j=c_j$),
but also conditional on $Q$ and $R$.
Hence $\X_I^2$ has asymptotically $\chi^2_1$ distribution
conditional on column sums, $Q$, and $R$ under CSR independence.

\subsubsection{Second Version of the New Segregation Tests}
\label{sec:seg-test-II}
For large $n$, we have $n_i\,c_j/n \approx n\,\nu_i\,\kappa_j$,
where $\kappa_j$ is the probability that a NN is of class $j$.
Under CSR independence or RL, $\nu_i=\kappa_i$ for all $i=1,2$,
then $n_i\,c_j/n \approx n\,\nu_i\,\nu_j$ for large $n$.
This suggests the following test statistics for the four cells,
\begin{equation}
\label{eqn:seg-test-II}
T^{II}_{ij}=N_{ij}-\frac{n_i\,n_j}{n}.
\end{equation}
Let
\begin{equation}
\label{eqn:ana-correct-II}
N^{II}_{ij}=\frac{T^{II}_{ij}}{\sqrt{n_i\,n_j/n}}=
\frac{\left( N_{ij}-n_i\,n_j/n \right)}{ \sqrt{n_i\,n_j/n}}.
\end{equation}
Under RL, we find that
\begin{equation}
\label{eqn:Exp-T-IIij}
\E\left[ T^{II}_{ij} \right]=
\begin{cases}
\frac{n_i\,(n_i-n)}{n\,(n-1)} & \text{if $i=j$,} \vspace{.05 in} \\
\frac{n_i\,n_j}{n\,(n-1)}     & \text{if $i \not= j$.}
\end{cases}
\end{equation}
Hence
\begin{equation}
\lim_{n_i,n_j \rightarrow \infty} \E\left[ T^{II}_{ij} \right]=
\begin{cases}
\nu_i\,(1-\nu_i) & \text{if $i=j$,} \vspace{.05 in} \\
\nu_i\,\nu_j     & \text{if $i \not= j$.}
\end{cases}
\end{equation}
Notice that under RL, $\E\left[ T^{II}_{ij} \right]=\E\left[ T^I_{ij} \right] \not=0$
which implies that $\E\left[ N^{II}_{ij} \right] \not=0$.
Furthermore,
\begin{equation}
\label{eqn:Exp-N-IIij}
\E\left[ N^{II}_{ij} \right]=
\begin{cases}
\frac{(n_i-n)}{\sqrt{n}\,(n-1)} & \text{if $i=j$,} \vspace{.05 in} \\
\frac{\sqrt{n_i\,n_j}}{\sqrt{n}\,(n-1)}     & \text{if $i \not= j$.}
\end{cases}
\end{equation}
Thus, $\lim_{n_i,n_j \rightarrow \infty} \E\left[ N^{II}_{ij} \right]=0$.

Note that the sum of the squares of $N^{II}_{ij}$ does not equal $\X^2_P$.
Let $\mathbf{N_{II}}$ be the vector of
$N^{II}_{ij}$ concatenated row-wise and let
$\Sigma_{II}$ be the variance-covariance matrix of
$\mathbf{N_{II}}$ based on the correct sampling distribution of the cell counts.
That is, $\Sigma_{II}=\left(
\Cov\left[ N^{II}_{ij},N^{II}_{kl} \right] \right)$ where
$\displaystyle \Cov\left[ N^{II}_{ij},N^{II}_{kl} \right]=
\frac{n}{\sqrt{n_i\,n_j\,n_k\,n_l}}\Cov\left[ N_{ij},N_{kl} \right]$.
Since $\Sigma_{II}$ is not invertible,
we use its generalized inverse $\Sigma_{II}^-$.
Then the proposed test statistic for overall segregation is the quadratic form
\begin{equation}
\label{eqn:X-square-II}
\X_{II}^2=\mathbf{N'_{II}}\Sigma_{II}^-\mathbf{N_{II}}
\end{equation}
which asymptotically has a $\chi^2_2$ distribution under RL.
Note that $\Sigma_{II}$ can be obtained from $\Sigma$
by multiplying $\Sigma$ entry-wise with the matrix
$\displaystyle C^{II}_M=\left( \frac{n}{\sqrt{n_i\,n_j\,n_k\,n_l}} \right)$.
This version of the segregation test is asymptotically equivalent
to Dixon's test of segregation.

Under CSR independence, the expectations, variances, and covariances related to $\X^2_{II}$ are as in the RL case,
except the variances and covariances are conditional on $Q$ and $R$.
Hence, the asymptotic $\chi^2_2$ distribution of $\X^2_{II}$ is also conditional on $Q$ and $R$.

\subsubsection{Third Version of the New Segregation Tests}
\label{sec:seg-test-III}
Among the first two versions we discussed so far,
version I of the new tests is a conditional test (conditional on column sums),
while version II of the new tests is asymptotically equivalent to Dixon's test,
although different from it for finite samples.
Furthermore, both Dixon's test and
version II of the new tests incorporate only row sums (class sizes) in the NNCTs.

Now, for the NNCT cells, we suggest the following test statistics which use both the row
and column sums (i.e., number of times a class serves as NN) and are not
conditional on the column sums:
\begin{equation}
\label{eqn:seg-test-III}
T^{III}_{ij}=
\begin{cases}
N_{ij}-\frac{(n_i-1)}{(n-1)}\,C_j & \text{if $i=j$,}\\
N_{ij}-\frac{n_i}{(n-1)}\,C_j     & \text{if $i \not= j$.}
\end{cases}
\end{equation}

Note that $\E\left[ T^{III}_{ij} \right]=0$ under RL,
but the sum of the squares of $T^{III}_{ij}$ does not equal $\X^2_P$.
Let $\mathbf{N_{III}}$ be the vector of $T^{III}_{ij}$ values
concatenated row-wise and let $\Sigma_{III}$ be the
variance-covariance matrix of $\mathbf{N_{III}}$ based on the correct
sampling distribution of the cell counts.
That is,
$\Sigma_{III}=\left( \Cov\left[ T^{III}_{ij},T^{III}_{kl} \right] \right)$
where $\Cov\left[ T^{III}_{ij},T^{III}_{kl} \right]$ has the following
forms based on the pairs $(i,j)$ and $(l,m)$.
\begin{itemize}
\item[Case (1)]
For $i=j,\,k=l$,
\begin{multline*}
\Cov\left[ T^{III}_{ii},T^{III}_{kk} \right]=
\Cov\left[ N_{ii}-\frac{(n_i-1)}{(n-1)}\,C_i,N_{kk}-\frac{(n_k-1)}{(n-1)}\,C_k \right]\\
=\Cov[N_{ii},N_{kk}]-\frac{(n_k-1)}{(n-1)}\Cov[N_{ii},C_k] -
\frac{(n_i-1)}{(n-1)}\Cov[N_{kk},C_i]+ \frac{(n_i-1)(n_k-1)}{(n-1)^2}\Cov[C_i,C_k].
\end{multline*}

\item[Case (2)]
For $i=j,\,k\not=l$,
\begin{multline*}
\Cov\left[ T^{III}_{ii},T^{III}_{kl} \right]=
\Cov\left[ N_{ii}-\frac{(n_i-1)}{(n-1)}\,C_i,N_{kl}-\frac{n_k}{(n-1)}\,C_l \right]\\
=\Cov[N_{ii},N_{kl}]-\frac{n_k}{(n-1)}\Cov[N_{ii},C_l] -
\frac{(n_i-1)}{(n-1)}\Cov[N_{kl},C_i]+ \frac{(n_i-1)n_k}{(n-1)^2}\Cov[C_i,C_l].
\end{multline*}

\item[Case (3)]
For $i\not=j,\,k=l$,
\begin{multline*}
\Cov\left[ T^{III}_{ij},T^{III}_{kk} \right]=
\Cov\left[ N_{ij}-\frac{n_i}{(n-1)}\,C_j,N_{kk}-\frac{(n_k-1)}{(n-1)}\,C_k \right]\\
=\Cov[N_{ij},N_{kk}]-\frac{(n_k-1)}{(n-1)}\Cov[N_{ij},C_k] -
\frac{n_i}{(n-1)}\Cov[N_{kk},C_j]+ \frac{n_i(n_k-1)}{(n-1)^2}\Cov[C_j,C_k].
\end{multline*}

\item[Case (4)]
For $i\not=j,\,k\not=l$,
\begin{multline*}
\Cov\left[ T^{III}_{ij},T^{III}_{kl} \right]=
\Cov\left[ N_{ij}-\frac{n_i}{(n-1)}\,C_j,N_{kl}-\frac{n_k}{(n-1)}\,C_l \right]\\
=\Cov[N_{ij},N_{kl}]-\frac{n_k}{(n-1)}\Cov[N_{ij},C_l] -
\frac{n_i}{(n-1)}\Cov[N_{kl},C_j]+ \frac{n_i\,n_k}{(n-1)^2}\Cov[C_j,C_l],
\end{multline*}
\end{itemize}
where
$$\Cov[N_{ij},C_k]=\Cov[N_{ij},N_{1k}+N_{2l}]= \Cov[N_{ij},N_{1k}]+\Cov[N_{ij},N_{2l}],$$
and
\begin{multline}
\Cov[C_j,C_l]=\Cov[N_{1j}+N_{2j},N_{1l}+N_{2m}]\\
=\Cov[N_{1j},N_{1l}]+\Cov[N_{1j},N_{2m}]+\Cov[N_{2j},N_{1l}]+\Cov[N_{2j},N_{2m}].
\end{multline}

Since $\Sigma_{III}$ is not invertible,
we use its generalized inverse $\Sigma_{III}^-$.
Then the proposed test statistic for overall segregation is the quadratic form
\begin{equation}
\label{eqn:X-square-III}
\X_{III}^2=\mathbf{N'_{III}}\Sigma_{III}^-\mathbf{N_{III}}
\end{equation}
which asymptotically has a $\chi^2_1$ distribution.


Under CSR independence, the discussion related to and derivation of $\X^2_{III}$ are as in the RL case,
however, the variance and covariance terms (hence the
asymptotic distribution) are conditional on $Q$ and $R$.

\subsubsection{Correcting Pielou's Test for CSR Independence Based on Monte Carlo Simulations}
\label{sec:MC-correction}
For the null case, we simulate the CSR independence case only with
classes 1 and 2 (i.e., $X$ and $Y$) of sizes $n_1$ and $n_2$, respectively.
At each of $N_{mc}=10000$ replicates, under $H_o$,
we generate data for the pairs of $(n_1,n_2) \in
\{(10,10),(10,30),(10,50),(30,30),(30,50),(50,50),(100,100),(200,200)\}$
points iid (independently and identically distributed)
from $\U((0,1)\times (0,1))$, the uniform distribution on the unit square.
These sample size combinations are chosen so that one can examine the
influence of small and large samples, and similar and very different sample
sizes on the tests. 
The corresponding test statistics are recorded at each Monte Carlo
replication for each sample size combination.
In the Appendix, in Figure \ref{fig:Piel-chi-scores},
the kernel density estimates for Pielou's test statistic
and the density plot of the $\chi^2_1$-distribution are provided in order to make
distributional comparisons.
The histograms (not presented),
follow the trend of a chi-square distribution, but need an adjustment for location and scale.
Based on the means and variances of Pielou's test
statistics for each sample size combination which are also provided in Table \ref{tab:means-variances}.
Using these statistics,
we transform the $\X^2_P$ scores
by adjusting on location and scaling as
\begin{equation}
\label{eqn:piel-MC-trans}
\X^2_{P,mc} := \frac{\X^2_P+0.013}{1.643}
\end{equation}
so that the transformed statistic will be approximately distributed as $\chi^2_1$.

By construction this Monte Carlo correction
is only appropriate when the null pattern is the CSR independence with rectangular study regions.
Furthermore, the location and scale adjustments are based on large
sample mean and variance estimates of Pielou's test of segregation
for similar sample sizes.
In fact we used large $n_1=n_2$ values
in the Monte Carlo simulations.
\cite{meagher:1980} propose and illustrate calculating
critical values of Pielou's test statistic under RL using simulation.
This is similar to our Monte Carlo test, but not identical:
\cite{meagher:1980} use a Monte-Carlo computation of the critical value
while we use a Monte Carlo based moment adjustment,
which is intended as a simple and quick fix for Pielou's test.
Furthermore, a Monte Carlo hypothesis testing may not be easily applicable
for the CSR independence pattern (e.g., when the region is very complicated),
but a randomization test can easily be conducted for the RL pattern.

\begin{remark}
\label{rem:MC-correction-Dixon}
For Dixon's test and the new versions of the NNCT-tests,
such a correction for means and variances is not necessary,
as we start with the correct sampling distribution of the cell counts.
However, for small sample size combinations, the estimated variances of
$C_D$ and $\X^2_{II}$ are smaller than 4 (not presented),
while the means are around 2,
which explains the slightly conservative nature of these tests for small samples.
On the other hand, $\X^2_{I}$ and $\X^2_{III}$ have estimated means around 1,
while their estimated variances are smaller than 2.
For small samples, one can transform these tests to
have the appropriate variance, while retaining their means.
But, this seems to be not worth the effort.
$\square$
\end{remark}

\begin{remark}
\label{rem:extension-to-multiple-classes}
\textbf{Extension of NNCT-Tests to Multi-Class Case:}
In Sections \ref{sec:dix-overall} and \ref{sec:new-seg-tests},
we describe the segregation tests for the two class case in which the corresponding NNCT is of dimension $2\times 2$.
For $q$ classes with $q>2$, the NNCT will be of dimension $q \times q$.
Pielou's test readily extends to $q \times q$ contingency tables,
but it will still be inappropriate for use in $q \times q$ NNCTs.
The Monte Carlo corrected version in Section \ref{sec:MC-correction} is
designed for the two-class case for rectangular regions.
For more classes, such a correction can be carried out in a similar fashion.
The cell counts for the diagonal cells have asymptotic normality.
For the off-diagonal cells, although the asymptotic normality is supported by
extensive Monte Carlo simulation results (\cite{dixon:NNCTEco2002}),
it is not rigorously proven yet.
Nevertheless, if the asymptotic normality held for all $q^2$ cell counts in the NNCT,
under RL, Dixon's test and version II of the new tests would have $\chi^2_{q(q-1)}$ distribution,
versions I and III of the new tests would have $\chi^2_{(q-1)^2}$ distribution asymptotically.
Under CSR independence, these tests will have the corresponding asymptotic distributions
conditional on $Q$ and $R$.
$\square$
\end{remark}

\section{Other Tests of Spatial Clustering}
\label{sec:other-spat-tests}
There are many tests for spatial clustering
of points from one class or multiple classes in the literature (\cite{diggle:2003} and \cite{kulldorff:2006}).
Among them are Ripley's $K$ or $L$-functions (\cite{ripley:2004}),
Diggle's $D$-function which is a modified version of Ripley's $K$-function (\cite{diggle:2003} p. 131),
pair correlation function (\cite{stoyan:1994}),
the univariate $J$-function (\cite{lieshout:1996})
and multivariate $J$-function (\cite{lieshout:1999}),
and many other first and second order tests
(see \cite{perry:2006} for a detailed review of spatial pattern
tests in plant ecology).
There are also spatial pattern tests
that adjust for an inhomogeneity and
are mostly used for clustering of cases in epidemiology
(\cite{kulldorff:2006}).
Among them are Cuzick and Edward's $k$-NN tests (\cite{cuzick:1990}),
spatial scan statistic of \cite{kulldorff:1997},
Whittemore's test, Tango's MEET, Besag-Newel's $R$, Moran's $I$ (\cite{song:2003}).
An extensive survey of such tests is provided by \cite{kulldorff:2006}.

Among the above clustering tests, univariate tests are not comparable with NNCT-tests,
Moran's $I$ and Whittemore's tests are shown to perform poorly
in detecting some kind of clustering (\cite{song:2003})
and most of the tests require Monte Carlo simulation or randomization
methods to attach significance to their results.
Hence we only consider Cuzick-Edward's $k$-NN tests and
their combined versions (\cite{cuzick:1990}),
and compare NNCT-tests with these tests in an extensive simulation study.
We also compare NNCT-tests with Ripley's $L$-function, Diggle's $D$-function,
and the pair correlation functions for the appropriate null hypotheses in the examples,
as they are perhaps the most commonly used
tests for spatial interaction at various scales,
although they are based on Monte Carlo simulation.

Cuzick-Edward's $k$-NN test is defined as
$T_k=\sum_{i=1}^n \delta_i d_i^k$,
where
\begin{equation}
\label{eqn:delta-i}
\delta_i=
\begin{cases}
1 & \text{if $z_i$ is a case,} \vspace{.05 in} \\
0 & \text{if $z_i$ is a control,}
\end{cases}
\end{equation}
with $z_i$ being the $i^{th}$ point and $d_i^k$ is the number of $k$ NNs which are cases.
Since in practice, the correct choice of $k$ is not known in advance,
\cite{cuzick:1990} also suggest combining information for various $T_k$ values.
Assuming multivariate normality of $T_k$ values and $T_k$ being a mixture
of shifts all in the same direction under an alternative,
the combined test statistic is given by
\begin{equation}
\label{eqn:Tcomb}
T^{comb}_S=\mathbf 1' \Sigma^{-1/2} \mathbf{T}
\end{equation}
where $S=\{k_1,k_2,\ldots,k_m\}$ and $\mathbf{T}=(T_{k_1},T_{k_2},\ldots,T_{k_m})'$
(i.e., $T^{comb}_S$ is the test obtained by combining $T_k$ whose indices are in $S$),
$\mathbf 1'=(1,1,\ldots,1)$, $\Sigma=\Cov[\mathbf{T}]$ is the
variance-covariance matrix of $\mathbf{T}$.
Under $H_o:$ \emph{RL of cases and controls to the given locations in the study region},
$T_k$ converges in law to $N(\E[T_k],\Var[T_k]/n_0)$;
similarly, $T^{comb}_S$ converges in law to $N(\E[T^{comb}_S],\Var[T^{comb}_S])$ when number of cases $n_0$
goes to infinity.
The expected values $\E[T_k]$ and $\E[T^{comb}_S]$ and variances
$\Var[T_k]$ and $\Var[T^{comb}_S]$ are provided in (\cite{cuzick:1990}).

The computational order of Cuzick-Edward's $k$-NN test is of $O(n^2)$,
while if $r$ distinct $T_k$ tests are combined it is $O\left( r^2n^2 \right)$.
Although theoretically, both versions are of the same order for fixed $r$,
in practice it might make a big difference in computation time.
In fact, the computation of $T_k$ for $k=1,2,\ldots,5$ for $n_1=n_2=50$,
10000 times took about a day,
while $T^{comb}_{1-5}$ took about 10 days in an Intel Pentium 4 2.4 GHz with 1 GB memory and 40 GB storage.

When the NNCT-tests and $k$-NN tests indicate
significant segregation or clustering,
one might also be interested in the (possible) causes of the segregation
and the type and level of interaction between the classes
at different scales (i.e., inter-point distances).
To answer such questions,
we calculate Ripley's (univariate) $L$-function which is the modified version of $K$-function,
which are denoted by $L_{ii}(t)$ and $K_{ii}(t)$ for class $i$, respectively.
The univariate $L$-function is estimated as
$\widehat{L}_{ii}(t)=\sqrt{\left( \widehat{K}_{ii}(t)/\pi \right)}$
where $t$ is the  is the distance from a randomly chosen event and
$\widehat{K}_{ii}(t)$ is an estimator of
\begin{equation}
\label{eqn:Kii}
K_{ii}(t)=\lambda^{-1}\E[\text{\# of extra events within distance $t$ of a randomly chosen event}]
\end{equation}
with $\lambda$ being the density (number per unit area) of events
and is calculated as
\begin{equation}
\label{eqn:Kiihat}
\widehat{K}_{ii}(t)=\widehat{\lambda}^{-1}\sum_i\sum_{j \not= i}w(i,d_{ij})\I(d_{ij}<t)/N
\end{equation}
where $\widehat{\lambda}=N/A$ is an estimate of density ($N$ is the observed number of points
and $A$ is the area of the study region),
$d_{ij}$ is the distance between points $i$ and $j$,
$\I(\cdot)$ is the indicator function,
$w(i,d_{ij})$ is the proportion of the circumference of the circle
centered at $l_i$ with radius $d_{ij}$ that falls in the study area,
which corrects for the boundary effects.
Under CSR independence, $L_{ii}(t)-t=0$ holds.
If the univariate pattern exhibits aggregation,
then $L_{ii}(t)-t$ tends to be positive;
if it exhibits regularity then $L_{ii}(t)-t$ tends to be negative.
See (\cite{diggle:2003}) for more detail.

We also provide Ripley's bivariate $L$-function, denoted by $L_{ij}(t)$ for classes $i$ and $j$
and estimated as $\widehat{L}_{ij}(t)=\sqrt{\left( \widehat{K}_{ij}(t)/\pi \right)}$
where $\widehat{K}_{ij}(t)$ is an estimator of
$$K_{ij}(t)=\lambda_j^{-1}\E[\text{\# of extra type $j$ events within distance $t$ of a randomly chosen type $i$ event}]$$
with $\lambda_j$ being the density of type $j$ events
and is calculated as
\begin{equation}
\label{eqn:Kijhat}
\widehat{K}_{ij}(t)=\left( \widehat{\lambda}_i\widehat{\lambda}_j A \right)^{-1}\sum_i\sum_{j}w(i_k,d_{i_k,j_l})\I(d_{i_k,j_l}<t)
\end{equation}
where $d_{i_k,j_l}$ is the distance between $k^{th}$ type $i$ and $l^{th}$ type $j$ points,
$w(i_k,d_{i_k,j_l})$ is the proportion of the circumference of the circle
centered at $k^{th}$ type $i$ point with radius $d_{i_k,j_l}$ that falls in the study area,
which is used for edge correction.
Under CSR independence, $L_{ij}(t)-t=0$ holds.
If the bivariate pattern is the segregation of the classes or species,
then $L_{ij}(t)$ tends to be negative,
if it is association of the classes or species then $L_{ij}(t)$ tends to be positive.
See (\cite{diggle:2003}) for more detail.

However, Ripley's $K$-function is cumulative,
so interpreting the spatial interaction at larger distances
is problematic (\cite{wiegand:2007} and ).
The pair correlation function $g(t)$
is better for this purpose (\cite{stoyan:1994}).
The pair correlation function of a (univariate)
stationary point process is defined as
$$g(t) = \frac{K'(t)}{2\,\pi\,t}$$
where $K'(t)$ is the derivative of $K(t)$.
For a univariate stationary Poisson process, $g(t)=1$;
values of $g(t) < 1$ suggest inhibition (or regularity) between points;
and values of $g(t) > 1$ suggest clustering (or aggregation).
The same definition of the pair correlation function
can be applied to Ripley's bivariate $K$ or $L$-functions as well.
The benchmark value of $K_{ij}(t) = \pi \, t^2$ corresponds to $g(t) = 1$;
$g(t) < 1$ suggests segregation of the classes;
and $g(t) > 1$ suggests association of the classes.
However the pair correlation function estimates might have critical behavior
for small $t$ if $g(t)>0$ since the estimator variance and hence
the bias are considerably large.
This problem gets worse especially in cluster processes (\cite{stoyan:1996}).
So pair correlation function analysis is more reliable for larger distances
and it might be safer to use $g(t)$ for distances larger than the average NN distance in the data set.

When the null case is the RL of
points from an inhomogeneous Poisson process,
Ripley's $K$- or $L$-functions in the general form are not appropriate
to test for the spatial clustering of the cases (\cite{kulldorff:2006}).
However, \cite{diggle:2003} suggests a version based on Ripley's univariate $K$-function
as $D(t)=K_{11}(t)-K_{22}(t)$. 
In this setup, ``no spatial clustering" is equivalent to RL of cases and controls
on the locations in the sample, which implies $D(t)=0$,
since $K_{22}(t)$ measures the degree of spatial aggregation of the controls
(i.e., the population at risk), while $K_{11}(t)$ measures
this same spatial aggregation plus any additional clustering due to the disease.
The test statistic $D(t)$ is estimated by
$\widehat D(t)=\widehat K_{11}(t)-\widehat K_{22}(t)$,
where $\widehat K_{ii}(t)$ is as in Equation \eqref{eqn:Kiihat}.

Among the tests we will consider, NNCT-tests summarize the spatial
interaction at the smaller scales (more specifically, for distances
about the average NN distance in the data set),
Cuzick-Edward's $k$-NN and combined tests
provide information on spatial interaction for distances about the
average $k$-NN distance between the points.
On the other hand, second order analysis by Ripley's $K$- or $L$-functions,
Diggle's $D$-function, and pair correlation function may provide
information on the univariate or bivariate patterns at all scales
(i.e., for all inter-point distances) of interest.

The NNCT-tests are designed for RL of classes to a set of given points.
For CSR independence of classes they are conditional on $Q$ and $R$ (essentially,
on the location of the points up to scale).
Cuzick-Edward's tests are designed to detect the clustering
of cases in the presence of inhomogeneity in the locations of both cases and controls.
Hence, they are appropriate for the RL of the points in a study area;
and similar to NNCT-tests, they are conditional on the locations of the points under CSR independence.
Both NNCT and Cuzick-Edward's tests appeal to asymptotic approximation of the test statistics,
although Monte Carlo simulation or randomization versions for them are readily available.
On the other hand, Ripley's $K$ or $L$-functions and pair correlation functions
are appropriate for the CSR independence null pattern,
while Diggle's $D$-function is appropriate for either CSR independence or RL patterns.
However, these tests are based on Monte Carlo simulation or randomization of points in the study area.

The order of classes in the construction of the NNCTs is irrelevant
for the values hence for the results of the NNCT-tests.
However, by construction, Cuzick-Edward's tests are more sensitive for clustering
of the cases in a case/control framework
(or the first class that is treated as cases in Equation \eqref{eqn:delta-i}
in the generalized two-class framework).
That is, they are not symmetric for the classes,
i.e., if one reverses the role of cases and controls in a data set,
the test statistics might give different results.
The order of the classes is inconsequential for Ripley's
bivariate $K$ or $L$-functions and pair correlation functions in theory,
as they are symmetric in the classes they pertain to.
But in practice edge corrections will render it slightly asymmetric,
i.e., $\widehat{L}_{ij}(t)\not=\widehat{L}_{ji}(t)$ for $i \not=j$.
Diggle's $D$-function is dependent on the order of the classes up to a sign difference,
in the sense that,
if one switches the roles of the two classes, the calculated test statistics
differ in sign only.

\section{Empirical Significance Levels under CSR Independence}
\label{sec:emp-sign-level}
We generate points from two classes under $H_o:$ \emph{CSR independence} as in Section \ref{sec:MC-correction}.
At each sample size combination,
we record how many times the $p$-value is at or below $\al=.05$ for each test
to estimate of the empirical size.
We present the empirical sizes for NNCT-tests in Table \ref{tab:MC-emp-sig-level-NNCT},
where $\widehat{\al}_P$ is the empirical significance level
for Pielou's test, $\widehat{\al}_D$ is for Dixon's test,
$\widehat{\al}_I,\,\widehat{\al}_{II}$ and $\widehat{\al}_{III}$
are for versions I, II, and III of the new tests, respectively,
and $\widehat{\al}_{P,mc}$ is for the Monte Carlo corrected version of Pielou's test
as in Equation \eqref{eqn:piel-MC-trans}.
The empirical size estimates are also plotted against the sample size
combinations in Figure \ref{fig:Emp-Size-NNCT-CSR} where the trend and performance of
the tests are easier to detect.

Observe that Pielou's test is extremely liberal in rejecting $H_o$
and its empirical size is severely affected by the difference
in the sample (or class) sizes.
That is, when the sample sizes are very different
(i.e.,$(n_1,n_2) \in \{(10,50),(50,10)\}$),
the empirical sizes are significantly smaller
than those for the other sample size combinations.
This seems to work in favor of Pielou's test when applied on NNCTs,
as it is extremely liberal,
and the difference in the sample sizes reduces its size significantly
toward the nominal level.
These results were also presented in more detail in (\cite{ceyhan:overall});
we include them here in order to compare Pielou's test with
the Monte Carlo corrected version $\X_{P,mc}^2$.
The NNCT-tests other than Pielou's test are about the desired level
(or size) when $n_1$ and $n_2$ are both $\ge 30$,
and mostly conservative otherwise.
However, in general if Pielou's test were at the desired level for
similar sample sizes, it would have been extremely conservative
for very different sample sizes.
The Monte Carlo corrected version of Pielou's test, $\X_{P,mc}^2$, has significantly
smaller empirical sizes than the uncorrected one, $\X_P^2$.
However, it is extremely conservative when at least one sample is too small (i.e., $<30$).
But Dixon's test and $\X_{II}^2$ are usually conservative when at least one sample is small
(i.e., $ \le 30$), liberal for one case ($(n_1,n_2)=(50,100)$),
and are about the appropriate nominal level $\al$ for the rest of the sample size combinations.
On the other hand,
$\X_I^2$ is also conservative when at least one sample is small (i.e., $ \le 30$),
liberal for small and equal sample sizes (i.e., $(n_1,n_2) \in \{(10,10),(30,30)\}$),
and about the nominal level for other cases.
Finally, $\X_{III}^2$ is usually conservative when at least one sample is small (i.e., $ \le 30$),
and is about the nominal level for other cases.
Pielou's test is extremely liberal, while Monte Carlo corrected version
is sporadic for smaller samples.
Version I of the new tests is appropriate for large samples,
but sporadic for small samples.
Dixon's test reveals less fluctuation, and is more appropriate for
larger samples.
Versions II and III of the new tests are conservative for small
sample sizes, and about the desired level for large samples.

\begin{table}[ht]
\centering
\begin{tabular}{|c||c|c||c|c|c|c|}
\hline
\multicolumn{7}{|c|}{Empirical significance levels of the NNCT-tests} \\
\hline
$(n_1,n_2)$ &  $\widehat{\al}_P$ & $\widehat{\al}_D$ & $\widehat{\al}_I$
& $\widehat{\al}_{II}$ & $\widehat{\al}_{III}$ & $\widehat{\al}_{P,mc}$  \\
\hline
 (10,10)   &  .1280$^{\ell}$ & .0432$^c$ & .0593$^{\ell}$ & .0461$^c$ & .0439$^c$ & .0608$^{\ell}$ \\
\hline
 (10,30)   &  .1429$^{\ell}$ & .0440$^c$ & .0451$^c$ & .0421$^c$ & .0410$^c$ & .0320$^c$ \\
\hline
 (10,50)   &  .0664$^{\ell}$ & .0482 & .0335$^c$ & .0423$^c$ & .0397$^c$ & .0292$^c$ \\
\hline
 (30,10)   &  .1383$^{\ell}$ &  .0390$^c$ &  .0411$^c$ &  .0383$^c$ &  .0391$^c$ &  .0282$^c$\\
\hline
 (30,30)   &  .1339$^{\ell}$ & .0464 & .0544$^{\ell}$ & .0476 & .0427$^c$ & .0552$^{\ell}$ \\
\hline
 (30,50)   &  .1319$^{\ell}$ & .0454$^c$ & .0507 & .0481 & .0504 & .0484 \\
\hline
 (50,10)   &  .0654$^{\ell}$ &  .0529 &  .0326$^c$ &  .0468 &  .0379$^c$ &  .0287$^c$\\
\hline
 (50,30)   &  .1275$^{\ell}$ &  .0429$^c$ &  .0494 &  .0468 &  .0469 &  .0477\\
\hline
 (50,50)   &  .1397$^{\ell}$ & .0508 & .0494 & .0497 & .0499 & .0494 \\
\hline
 (50,100)  &  .1223$^{\ell}$ &  .0560$^{\ell}$ &  .0501 &  .0564$^{\ell}$ &  .0516 &  .0499\\
\hline
 (100,50)  &  .1190$^{\ell}$ &  .0483 &  .0463$^c$ &  .0492 &  .0479 &  .0455$^c$\\
\hline
 (100,100) &  .1324$^{\ell}$ &  .0504 &  .0524 &  .0519 &  .0489 &  .0524\\
\hline

\end{tabular}
\caption{
\label{tab:MC-emp-sig-level-NNCT}
The empirical significance levels for Pielou's, Dixon's, and the new overall NNCT-tests
based on 10000 Monte Carlo simulations of the CSR independence pattern.
$\widehat{\al}_P$ stands for the empirical significance level
for Pielou's test, $\widehat{\al}_D$ for Dixon's test,
$\widehat{\al}_I,\,\widehat{\al}_{II}$ and $\widehat{\al}_{III}$
for versions I, II, and III of the new tests, respectively,
and $\widehat{\al}_{P,mc}$ for the Monte Carlo corrected version of Pielou's test
as in Equation \eqref{eqn:piel-MC-trans}.
($^c$: the empirical size is significantly smaller than .05; i.e., the test is conservative.
$^{\ell}$: the empirical size is significantly larger than .05; i.e., the test is liberal.)
}
\end{table}

\begin{figure}[ht]
\centering
\rotatebox{-90}{ \resizebox{2.1 in}{!}{\includegraphics{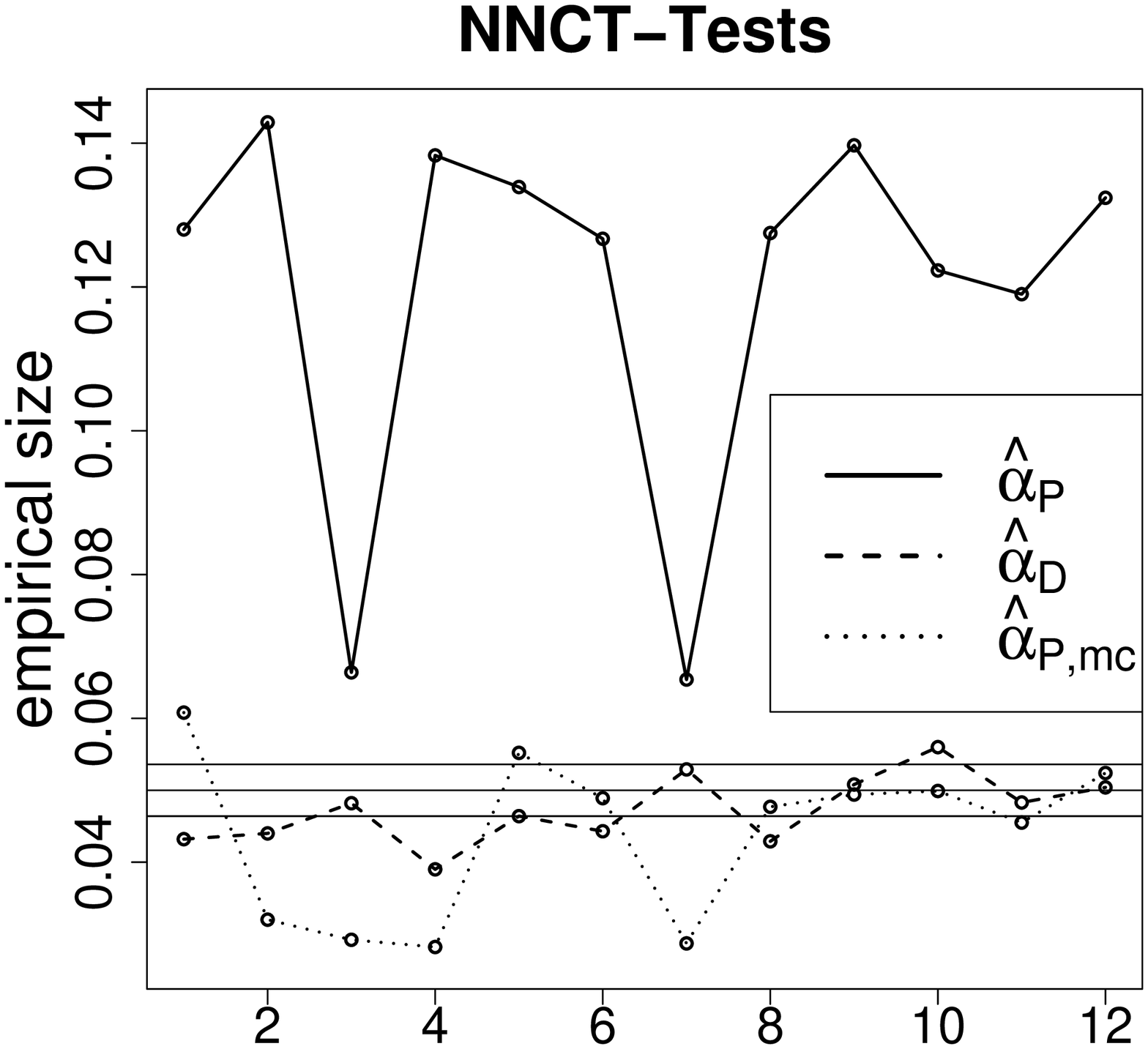} }}
\rotatebox{-90}{ \resizebox{2.1 in}{!}{\includegraphics{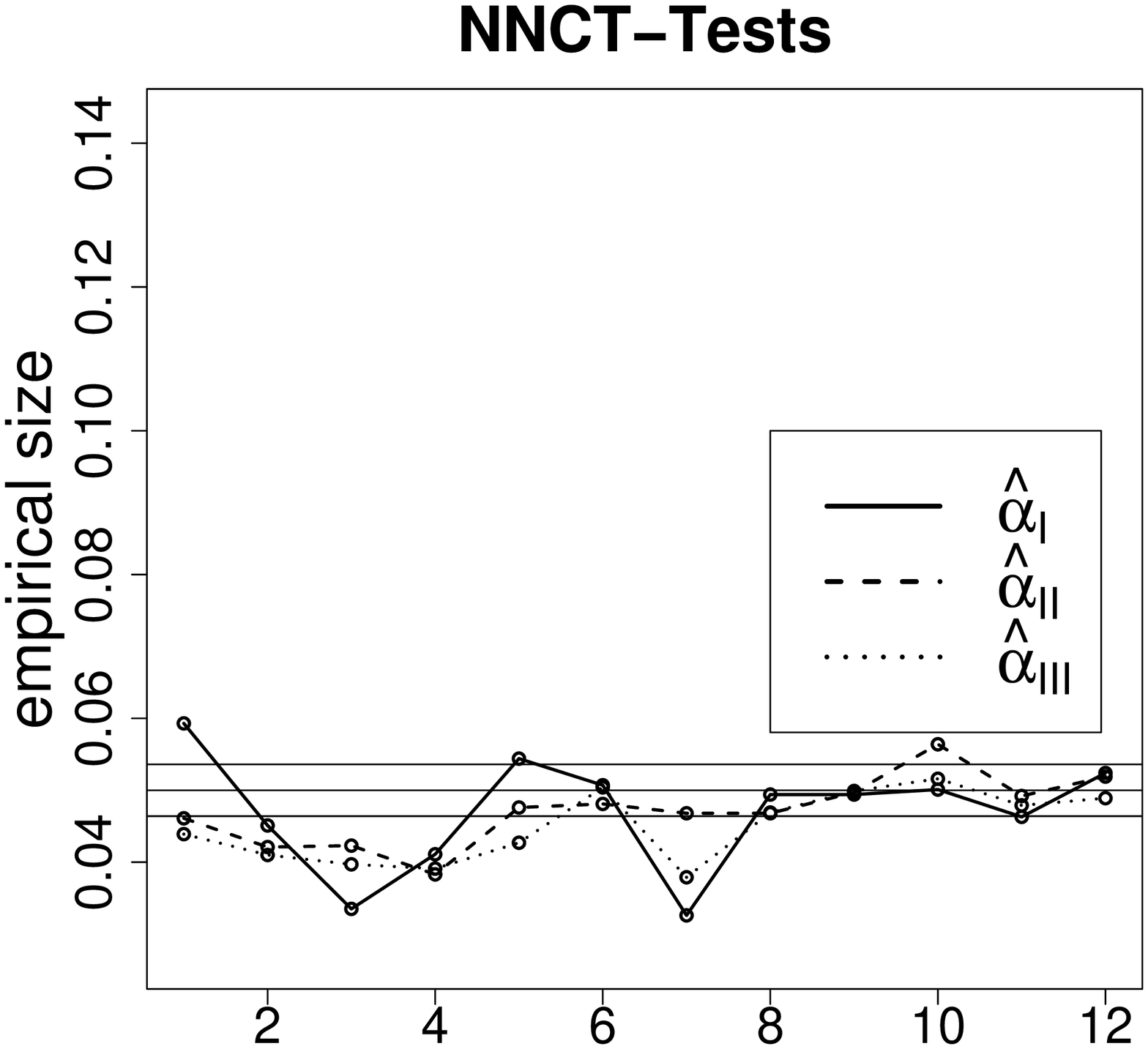} }}
 \caption{
\label{fig:Emp-Size-NNCT-CSR}
Empirical size estimates for the NNCT-tests based on
10000 Monte Carlo replicates under the CSR independence of the two classes, i.e.,
uniform data from two classes on the unit square.
The horizontal lines are located at .0464 (upper threshold for conservativeness),
.0500 (nominal level), and .0536 (lower threshold for liberalness).
The numbers in the horizontal axis labels represent sample (i.e., class) size combinations:
1=(10,10), 2=(10,30), 3=(10,50), 4=(30,10), 5=(30,30),
6=(30,50), 7=(50,10), 8=(50,30), 9=(50,50), 10=(50,100), 11=(100,50), 12=(100,100).
The empirical size labeling is as in Table \ref{tab:MC-emp-sig-level-NNCT}.
Notice that they are arranged in the increasing order for the first and then the second entries.
The size values for discrete sample size combinations are joined
by piecewise straight lines for better visualization.}
\end{figure}

\begin{table}[ht]
\centering
\begin{tabular}{|c||c|c|c|c|c||c|c|c|c|}
\hline
\multicolumn{10}{|c|}{Empirical significance levels of Cuzick-Edward's $k$-NN and combined tests} \\
\hline
$(n_1,n_2)$ &  $\widehat{\al}^{CE}_1$ & $\widehat{\al}^{CE}_2$ & $\widehat{\al}^{CE}_3$
& $\widehat{\al}^{CE}_4$ & $\widehat{\al}^{CE}_5$ & $\widehat{\al}^{comb}_{1-2}$
& $\widehat{\al}^{comb}_{1-3}$ & $\widehat{\al}^{comb}_{1-4}$ & $\widehat{\al}^{comb}_{1-5}$  \\
\hline
(10,10) & .0454$^c$ & .0398$^c$ & .0495 & .0474 & .0492  & .0478  & .0484  & .0477  & .0497\\
\hline
(10,30) & .0306$^c$ & .0495 & .0400$^c$ & .0458$^c$ & .0418$^c$  & .0334$^c$  & .0434$^c$  & .0434$^c$  & .0432$^c$\\
\hline
(10,50) & .0270$^c$ & .0541$^{\ell}$ & .0367$^c$ & .0490 & .0529  & .0438$^c$  & .0419$^c$  & .0413$^c$  & .0409$^c$\\
\hline
(30,10) & .0479 & .0493 & .0458$^c$ & .0497 & .0471  & .0462$^c$  & .0464  & .0479  & .0488\\
\hline
(30,30) & .0507 & .0529 & .0480 & .0475 & .0479  & .0467  & .0452$^c$  & .0455$^c$  & .0477\\
\hline
(30,50) & .0590$^{\ell}$ & .0416$^c$ & .0429$^c$ & .0485 & .0435$^c$  & .0471  & .0478  & .0458$^c$  & .0463$^c$\\
\hline
(50,10) & .0524 & .0492 & .0474 & .0509 & .0548$^{\ell}$  & .0507  & .0513  & .0511  & .0512\\
\hline
(50,30) & .0535 & .0483 & .0502 & .0485 & .0504  & .0489  & .0472  & .0477  & .0480\\
\hline
(50,50) & .0465 & .0490 & .0516 & .0545$^{\ell}$ & .0514  & .0522  & .0509  & .0511  & .0514\\
\hline
\end{tabular}
\caption{
\label{tab:MC-emp-sig-level-CE}
The empirical significance levels for Cuzick-Edward's $k$-NN tests $T_k$
for $k=1,2,\ldots,5$ and the combined tests $T^{comb}_S$ for $S=1-2, 1-3, 1-4$, and $1-5$.
$\widehat{\al}^{CE}_k$ stands for the empirical size for Cuzick-Edward's $k$-NN test for $k=1,2,\ldots,5$,
and $\widehat{\al}^{comb}_S$ for the combined test as in Equation \eqref{eqn:Tcomb}
for $S\in \{\{1,2\},\{1,2,3\},\{1,2,3,4\},\{1,2,3,4,5\}\}$.
Superscript labeling is as in Table \ref{tab:MC-emp-sig-level-NNCT}.}
\end{table}

\begin{figure}[ht]
\centering
\rotatebox{-90}{ \resizebox{2.1 in}{!}{\includegraphics{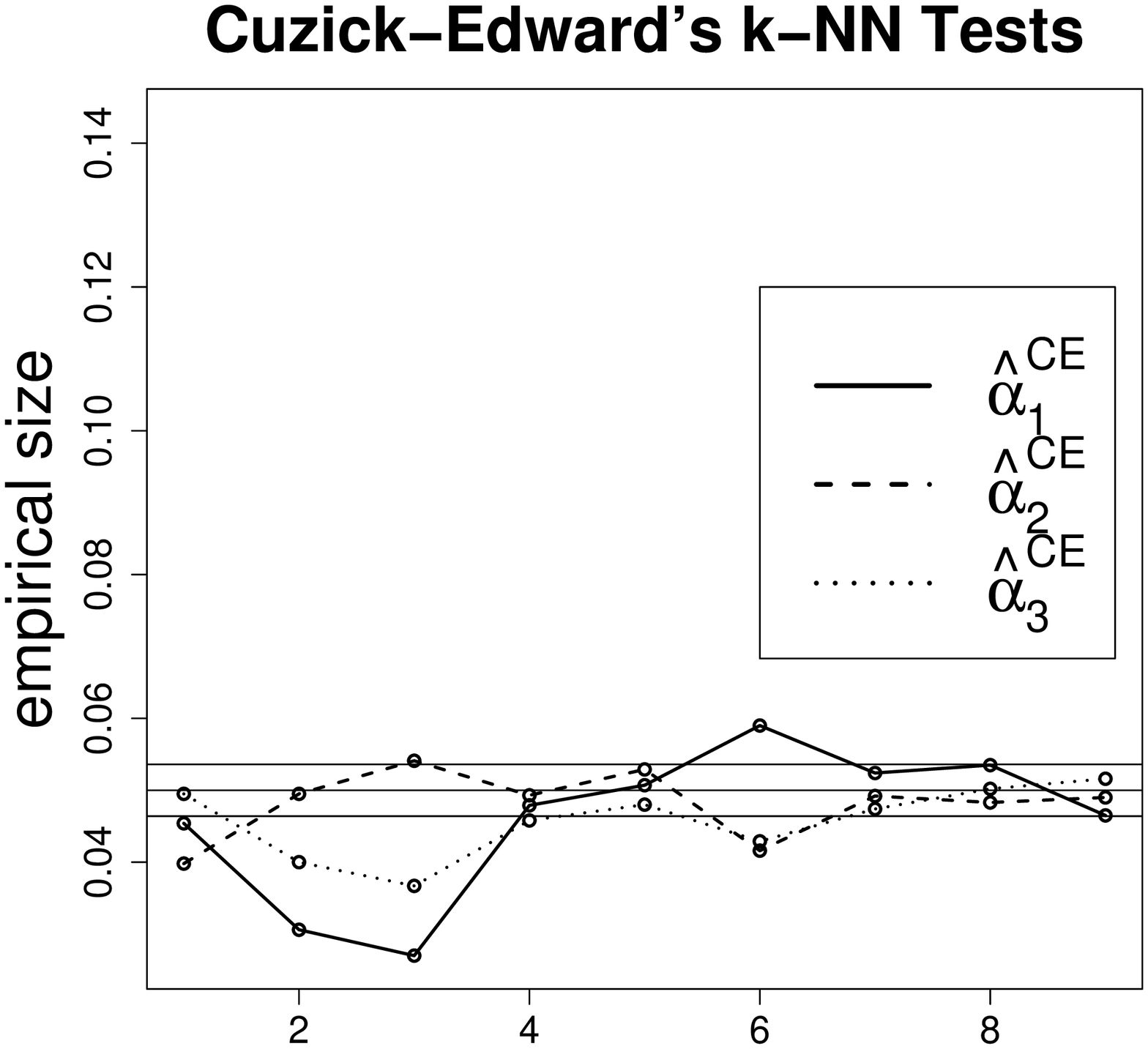} }}
\rotatebox{-90}{ \resizebox{2.1 in}{!}{\includegraphics{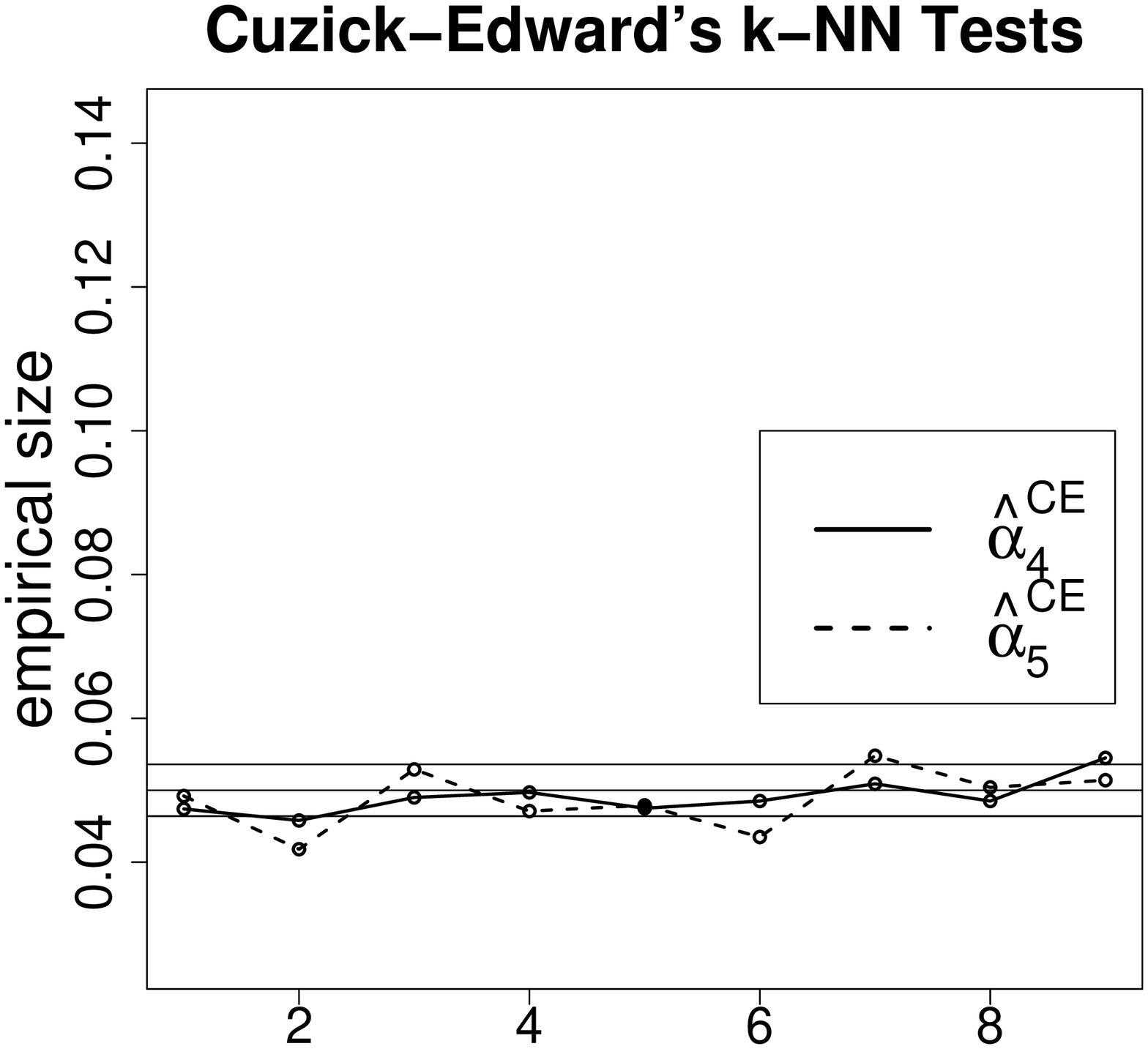} }}
\rotatebox{-90}{ \resizebox{2.1 in}{!}{\includegraphics{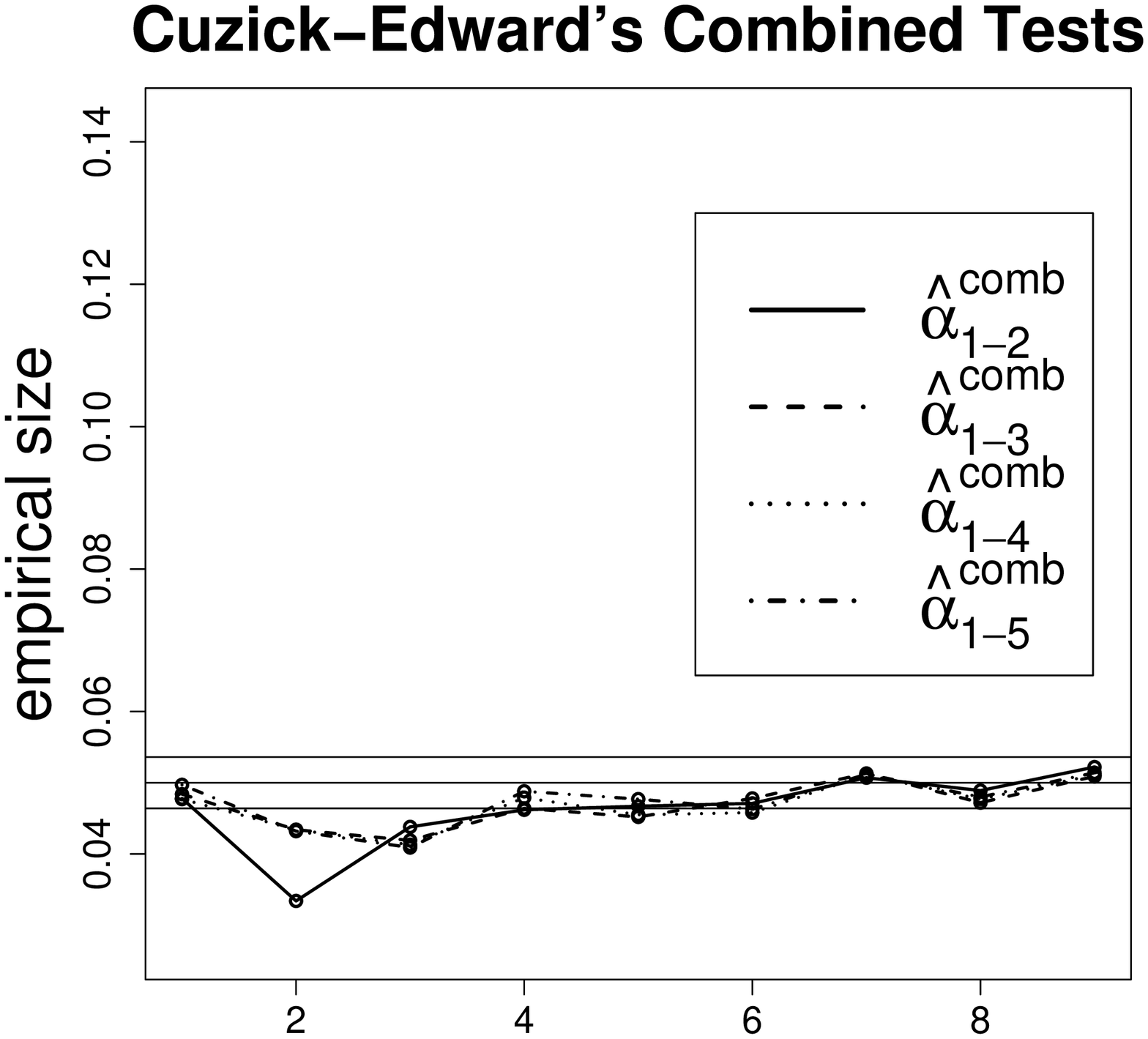} }}
 \caption{
\label{fig:Emp-Size-CE-CSR}
Empirical size estimates for Cuzick-Edward's $k$-NN tests and $T^{comb}_S$
(i.e., combined) tests based on 10000 Monte Carlo simulations
under the CSR independence pattern, i.e., uniform data in the unit square.
The horizontal lines are as in Figure \ref{fig:Emp-Size-NNCT-CSR}.
The numbers in the horizontal axis labels represent sample (i.e., class) size combinations:
1=(10,10), 2=(10,30), 3=(10,50), 4=(30,10), 5=(30,30),
6=(30,50), 7=(50,10), 8=(50,30), 9=(50,50).
The empirical size labeling is as in Table \ref{tab:MC-emp-sig-level-CE}.
}
\end{figure}

In the simulated patterns under CSR independence of classes,
class $X$ represents the cases and class $Y$ represents the controls
in the context of Cuzick-Edward's tests.
However, the simulated patterns are not realistic for the case/control framework,
since $X$ and $Y$ points are from a homogeneous Poisson process,
while case and control locations usually exhibit inhomogeneity in practice.
So Cuzick-Edward's tests are not used in the
conventional sense (as in case/control framework) here,
but instead are used to test deviations of the classes from CSR independence.
The empirical sizes for Cuzick-Edward's $k$-NN tests for $k \le 5$ and
$T^{comb}_S$ for 4 combinations of the
tests are presented in Figure \ref{fig:Emp-Size-CE-CSR},
where $\widehat{\al}^{CE}_k$ is the empirical size for Cuzick-Edward's $k$-NN test for $k=1,2,\ldots,5$,
and $\widehat{\al}^{comb}_S$ is the combined test as in \eqref{eqn:Tcomb}
for $S\in \{\{1,2\},\{1,2,3\},\{1,2,3,4\},\{1,2,3,4,5\}\}$.
For brevity in notation the index set $S$ is written as $1-j$ for $j=2,3,4,5$.
Due to the computational cost in time for Cuzick-Edward's test,
we only present 9 sample size combinations compared to 12 sample size combinations
for the NNCT-tests.

Observe that Cuzick-Edward's $k$-NN tests and $T^{comb}_S$ tests are usually
conservative when $n_1 < 30$ (recall that $n_1$ corresponds to the
number of cases in a case/control framework) and about the desired size
for other sample size combinations.
In particular $T_1$ and $T_2$ are more conservative than the other $T_k$ tests.
Cuzick-Edward's $k$-NN tests for $k > 2$ are about the desired size for most of the
sample size combinations.
The size performance of $T_k$ seems to get better as $k$ increases,
since as $k$ increases the size
gets closer and closer to the desired nominal level of .05.
However for most sample sizes $T_4$ seems to have the best empirical size performance.
Then comes $T_5$, and $T_3$, $T_2$, and $T_1$ in decreasing order of size performance.
On the other hand, the combined tests are conservative
when the number of cases is $<30$, and about the desired size for most of the
sample size combinations.
In particular $T^{comb}_{1-2}$ is most conservative among
the combined tests.
$T^{comb}_S$ tests usually have better size performance
compared to each $T_k$ for $k \in S$.
Furthermore, as the number of combined tests
(i.e., the size of the index set $S$) increases,
the size gets closer to the nominal level,
hence the $T^{comb}_{1-5}$ exhibits the best size performance.


\begin{remark}
\label{rem:MC-emp-size-CSR}
\textbf{Main Result of Monte Carlo Simulations under CSR Independence:}
Based on the simulation results under the CSR independence of the points,
we recommend the disuse of Pielou's test in practice, as it is extremely liberal,
hence might give false alarms when the pattern
is actually not significantly different from CSR independence.
None of the other NNCT-tests we consider
have the desired level when at least one sample size is small so that
the cell count(s) in the corresponding NNCT have a high probability of being $\le 5$.
This usually corresponds to the case that at least
one sample size is $\leq 10$ or the sample sizes (i.e., relative abundances)
are very different in the simulation study.
When sample sizes are small
(hence the corresponding cell counts are $\leq 5$),
the asymptotic approximation of the NNCT-tests is not appropriate.
However, when sample sizes are very different, cell counts are also more likely to be $\leq 5$,
compared to cell counts for similar sample sizes (roughly,
the sample sizes are similar when $\max_{i} (n_i)/ \min_{i} (n_i) \leq 2$.)
Dixon's test and versions II and III of the new tests tend to be conservative when the NNCT
contains cell(s) whose counts are $\leq 5$.
For larger samples  (i.e., the cell counts are larger than 5),
NNCT-tests yield empirical sizes that are about the desired nominal level.
Version I of the new tests and Monte Carlo corrected version
of Pielou's test are liberal when $n_1=n_2 \leq 30$,
and conservative for $n_1\not=n_2 \leq 30$, for other sample sizes they are about the desired level.
So \cite{dixon:1994} recommends Monte Carlo randomization
for his test when some cell count(s) are $\le 5$ in a NNCT.
We extend this recommendation for all the NNCT-tests (other than Pielou's test)
discussed in this article.
On the other hand, for large samples,
the asymptotic approximation or Monte Carlo randomization can be employed.

For Cuzick-Edward's tests, we recommend Monte Carlo randomization,
when $n_1<10$;
otherwise asymptotic approximation can also be employed.
Observe also that $k$-NN tests for $k>1$ and $T^{comb}_S$ tests
attain the normal approximation at smaller
sample size combinations
(i.e., they approach to normality faster)
compared to the NNCT-tests.
$\square$
\end{remark}

\subsection{Empirical Significance Levels under RL}
\label{sec:RL-emp-sign-2Cl}
Recall that the clustering tests we consider are conditional under the CSR independence pattern.
To better assess their empirical size performance,
we also perform Monte Carlo simulations under various RL patterns where
the tests are not conditional.
We consider the following three cases for the RL pattern.
In each RL case, we first determine the locations of points for which class labels
are to be assigned randomly.
Then we apply the RL procedure to these points for various sample size combinations.

\noindent
\textbf{RL Case (1):}
First, we generate $n=(n_1+n_2)$ points iid $\U((0,1) \times (0,1))$
for some combinations of $n_1,n_2 \in \{10,30,50,100\}$.
In each $(n_1,n_2)$ combination, the locations of these points
are taken to be the fixed locations
for which we assign the class labels randomly.
For each sample size combination $(n_1,n_2)$,
we randomly choose $n_1$ points (without replacement) and label them
as $X$ and the remaining $n_2$ points as $Y$ points.
We repeat the RL procedure $N_{mc}=10000$ times for each sample size combination.
At each Monte Carlo replication, we compute the NNCT-tests
and Cuzick-Edwards $k$-NN and combined tests.
Out of these 10000 samples the number of significant outcomes by each test is recorded.
The nominal significance level used in all these tests is $\alpha=.05$.
The empirical sizes are calculated as
the ratio of number of significant results to the number of Monte
Carlo replications, $N_{mc}$.

\noindent
\textbf{RL Case (2):}
We generate $n_1$ points iid $\U((0,2/3) \times (0,2/3))$ and
$n_2$ points iid $\U((1/3,1) \times (1/3,1))$
for some combinations of $n_1,n_2 \in \{10,30,50,100\}$.
The locations of these points are taken to be the fixed locations
for which we assign the class labels randomly.
The RL procedure is applied to these fixed points $N_{mc}=10000$
times for each sample size combination and
the empirical sizes for the tests are calculated similarly as in RL Case (1).

\noindent
\textbf{RL Case (3):}
We generate $n_1$ points iid $\U((0,1) \times (0,1))$
and $n_2$ points iid $\U((2,3) \times (0,1))$
for some combinations of $n_1,n_2 \in \{10,30,50,100\}$.
The RL procedure is applied and the empirical sizes for the tests
are calculated as in the previous RL Cases.

The locations for which the RL procedure is applied in RL Cases (1)-(3) are plotted
in Figure \ref{fig:RL-cases} for $n_1=n_2=100$.
Although there are many possibilities for the allocation of points to which RL procedure can
be applied, we only chose these three generic cases.
In RL Case (1), the allocation of the points are a realization of a homogeneous
Poisson process in the unit square;
in RL Case (2) the points are a realization of two overlapping clusters;
in RL Case (3) the points are a realization of two disjoint clusters.

\begin{figure}[ht]
\centering
\rotatebox{-90}{ \resizebox{2. in}{!}{\includegraphics{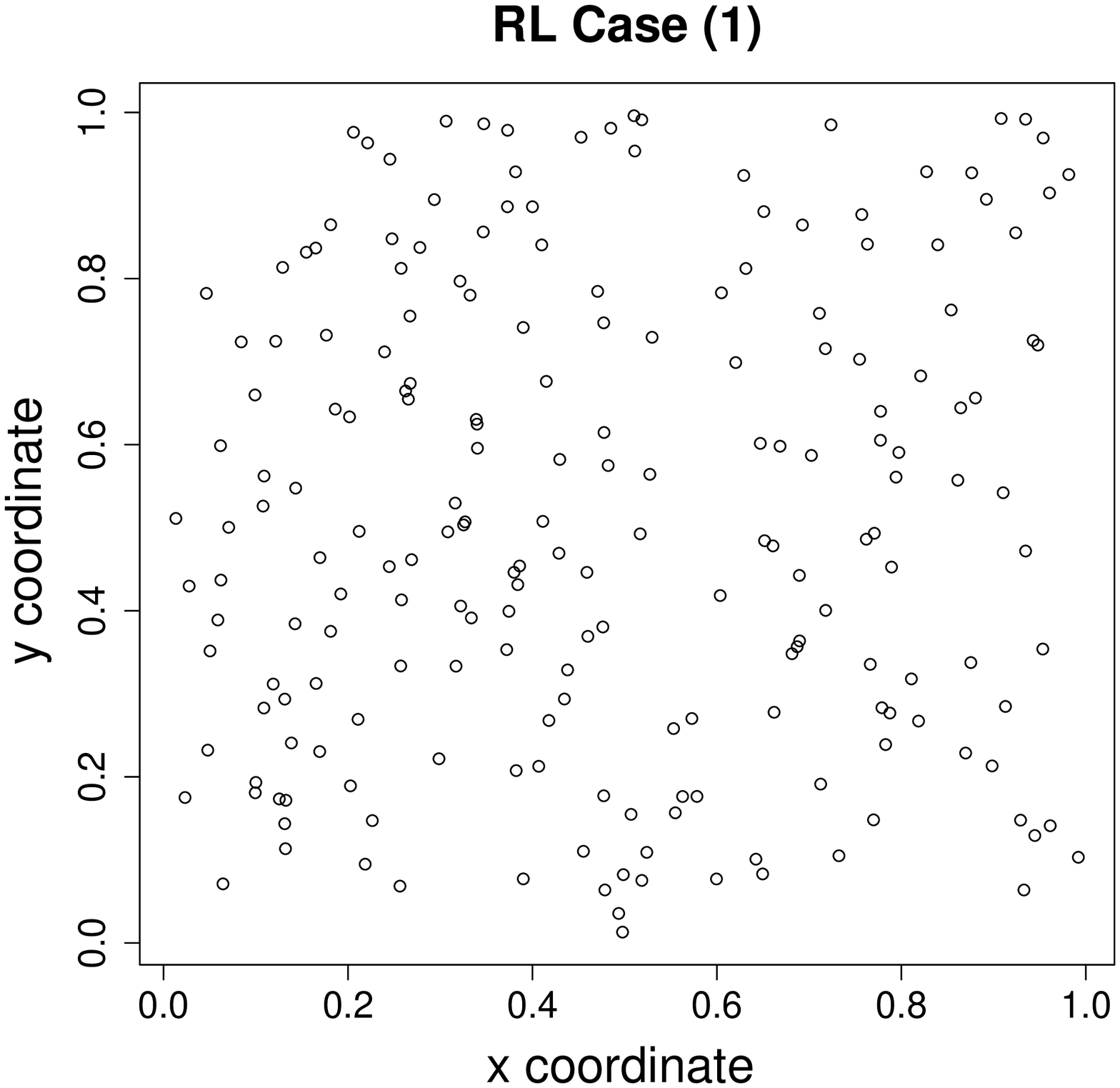} }}
\rotatebox{-90}{ \resizebox{2. in}{!}{\includegraphics{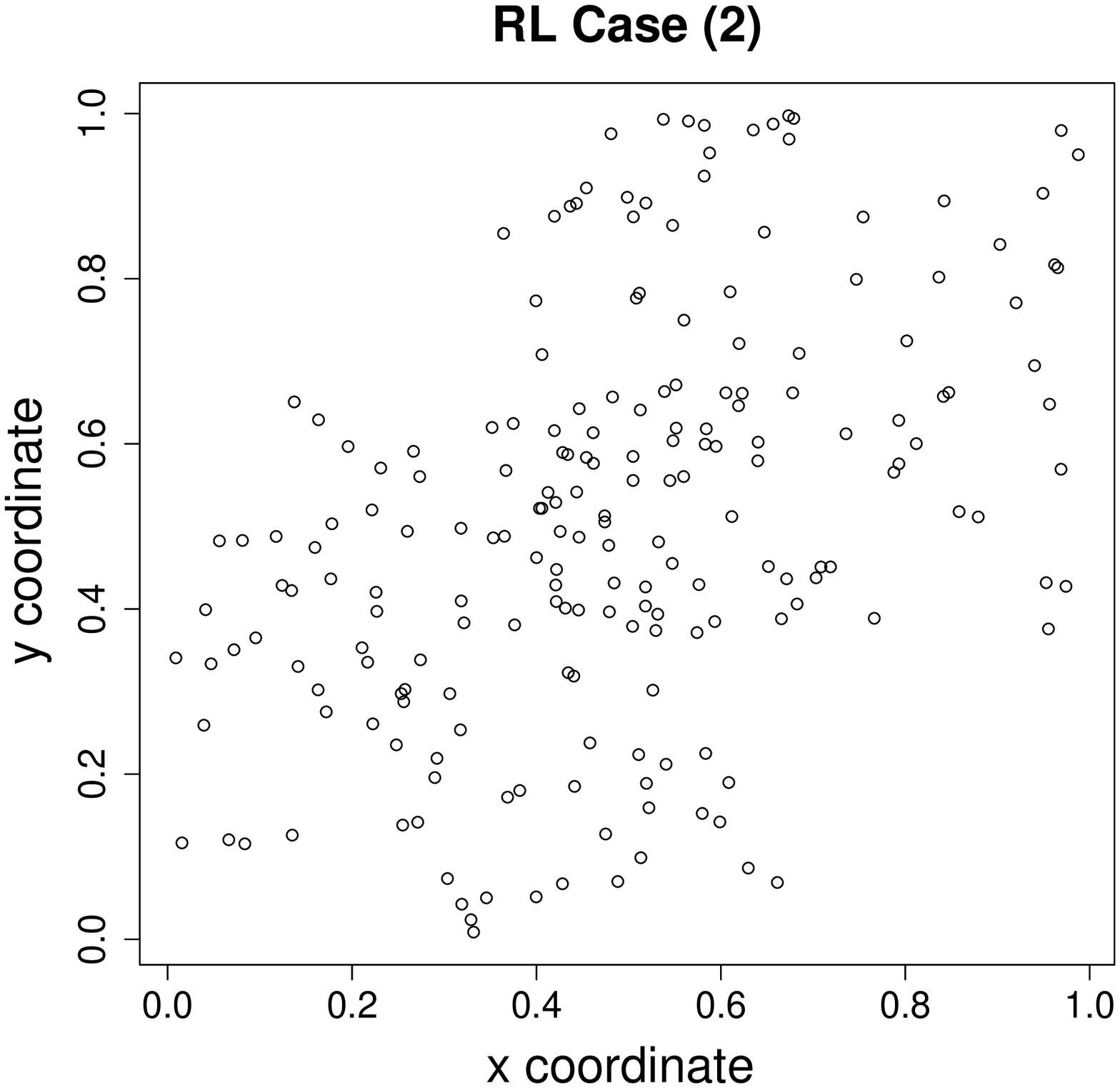} }}
\rotatebox{-90}{ \resizebox{2. in}{!}{\includegraphics{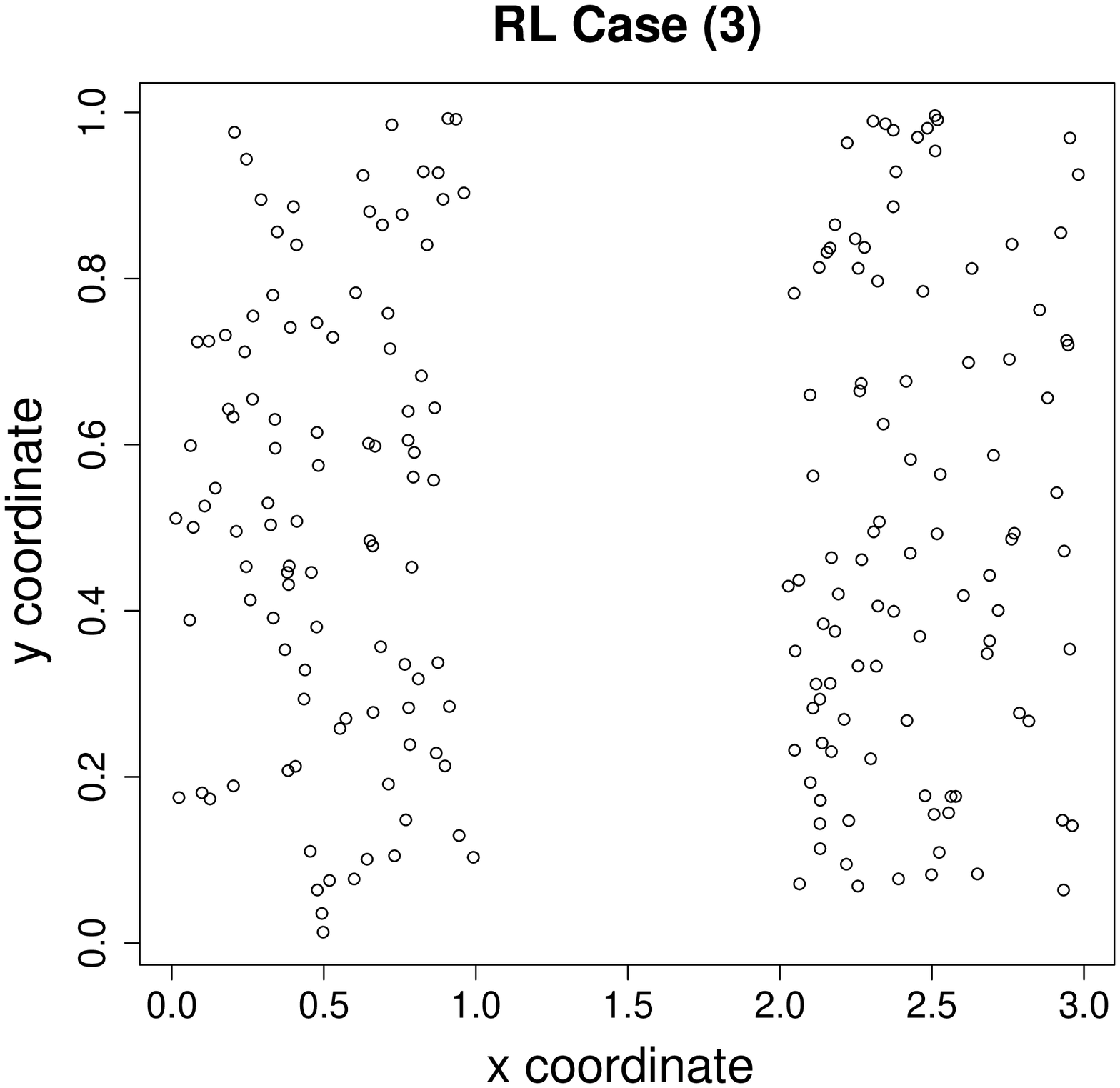} }}
 \caption{
\label{fig:RL-cases}
The fixed locations of points for which RL procedure is applied
for RL Cases (1)-(3) with $n_1=n_2=100$ in the two-class case.
Notice that $x$-axis for RL Case (3) is differently scaled than others.
}
\end{figure}

We present the empirical significance
levels for the NNCT-tests in Figure \ref{fig:Emp-Size-NNCT-RL},
where the empirical significance level labeling is as in Table \ref{tab:MC-emp-sig-level-NNCT}.
Observe that, as in the CSR independence case, Pielou's test is extremely liberal for each sample
size combination under each RL Case.
Monte Carlo corrected version of Pielou's test is conservative for most small sample size combinations,
and usually about the desired size for larger samples.
Dixon's test has the desired level for moderate to large sample sizes,
but conservative for small sample sizes.
The new versions seem to have the desired level for larger samples,
but they fluctuate between conservativeness and liberalness for smaller samples.
Dixon's test and version II of the new tests seem to
have the best empirical size performance
and for smaller samples we again recommend the Monte Carlo randomization version of the tests.

The RL Cases are more appropriate for the case/control framework of Cuzick-Edward's tests
compared to the CSR independence cases.
In the RL cases, the locations of the points could represent the locations
of the $n=(n_1+n_2)$ subjects so that $n_1$ of them are patients (i.e., cases)
while the rest are controls.
The empirical significance levels for Cuzick-Edwards $k$-NN and $T^{comb}_S$
tests are presented in Figure \ref{fig:Emp-Size-CE-RL},
where the empirical significance level labeling is as in Figure \ref{fig:Emp-Size-CE-CSR}.
Observe that, $T_1$ is at about the desired level for similar relative abundances,
but is conservative when $n_1>n_2$ and is liberal when $n_1<n_2$.
For $k>1$, the empirical size estimates of $T_k$ are about the nominal level.
In particular, as $k$ increases, the empirical size estimates of $T_k$
get to be closer to the nominal level.
$T^{comb}_S$ are conservative when $n_1 \le 10$ and they are about the desired level otherwise.
Furthermore, the combined tests $T^{comb}_S$ have better size performance than $T_k$.
When all the NN tests are considered,
$T_k$ for $k \ge 3$ and $T^{comb}_S$ have better size performance than NNCT-tests.

\begin{figure}[ht]
\centering
\rotatebox{-90}{ \resizebox{2.1 in}{!}{\includegraphics{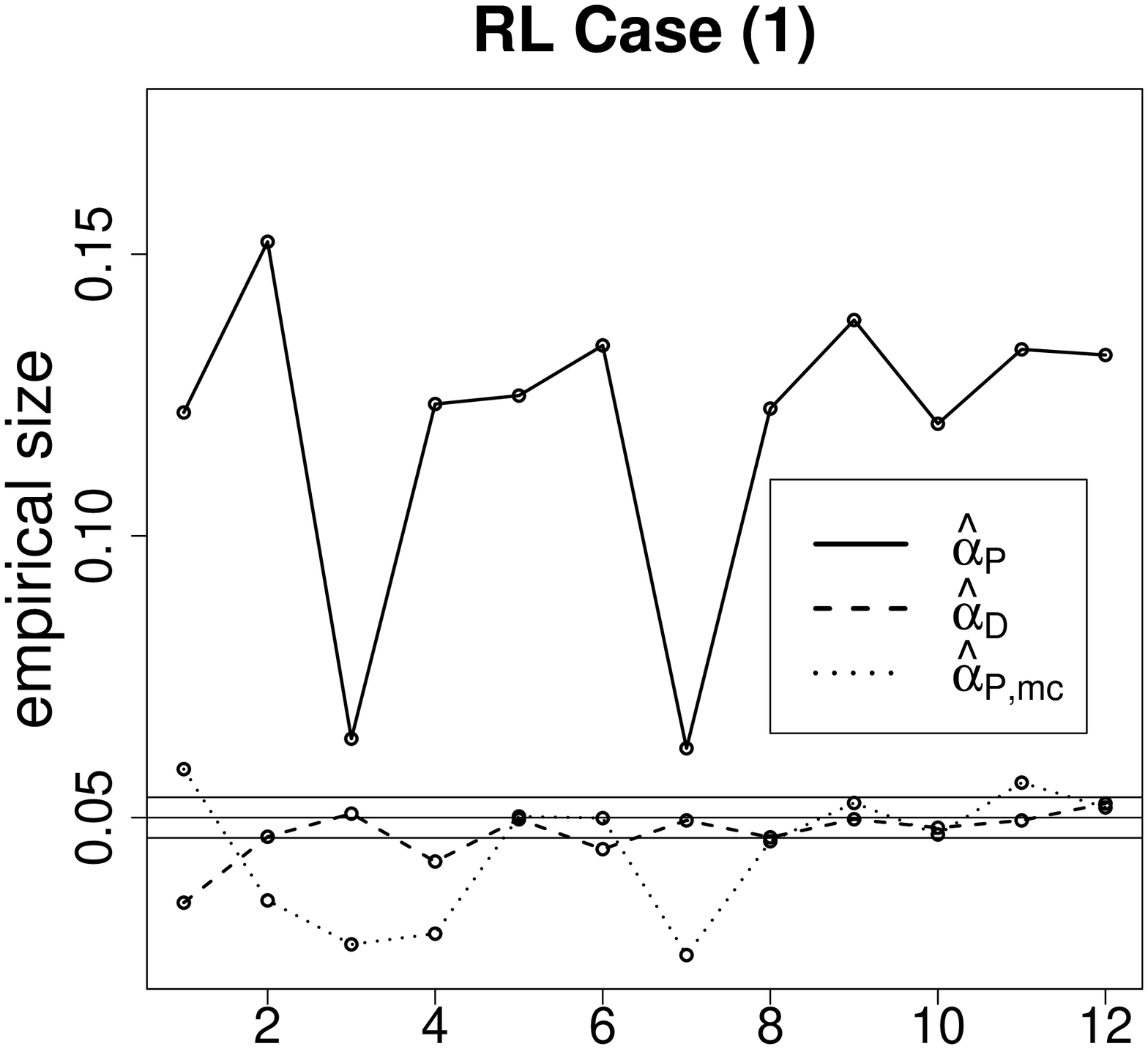} }}
\rotatebox{-90}{ \resizebox{2.1 in}{!}{\includegraphics{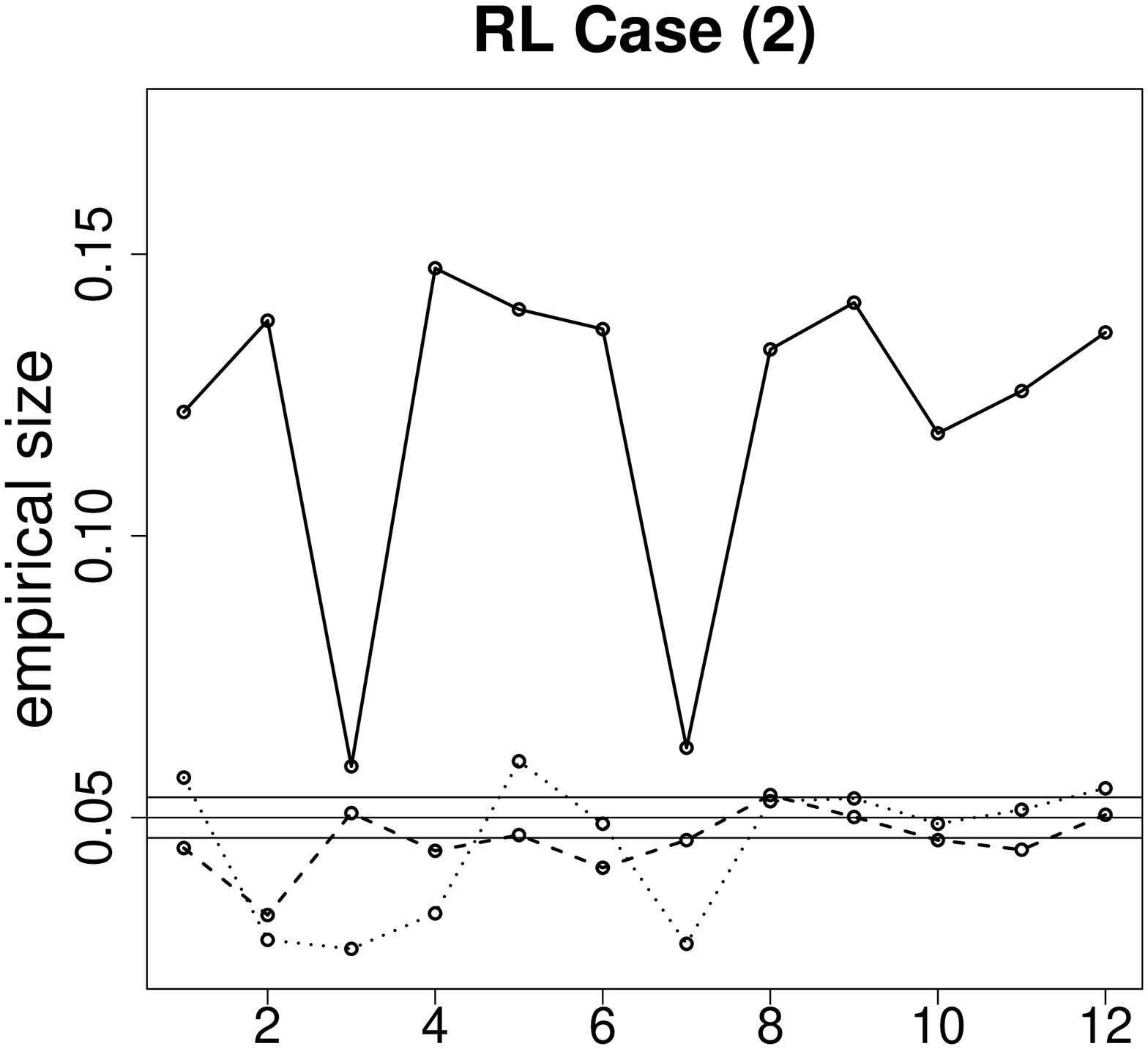} }}
\rotatebox{-90}{ \resizebox{2.1 in}{!}{\includegraphics{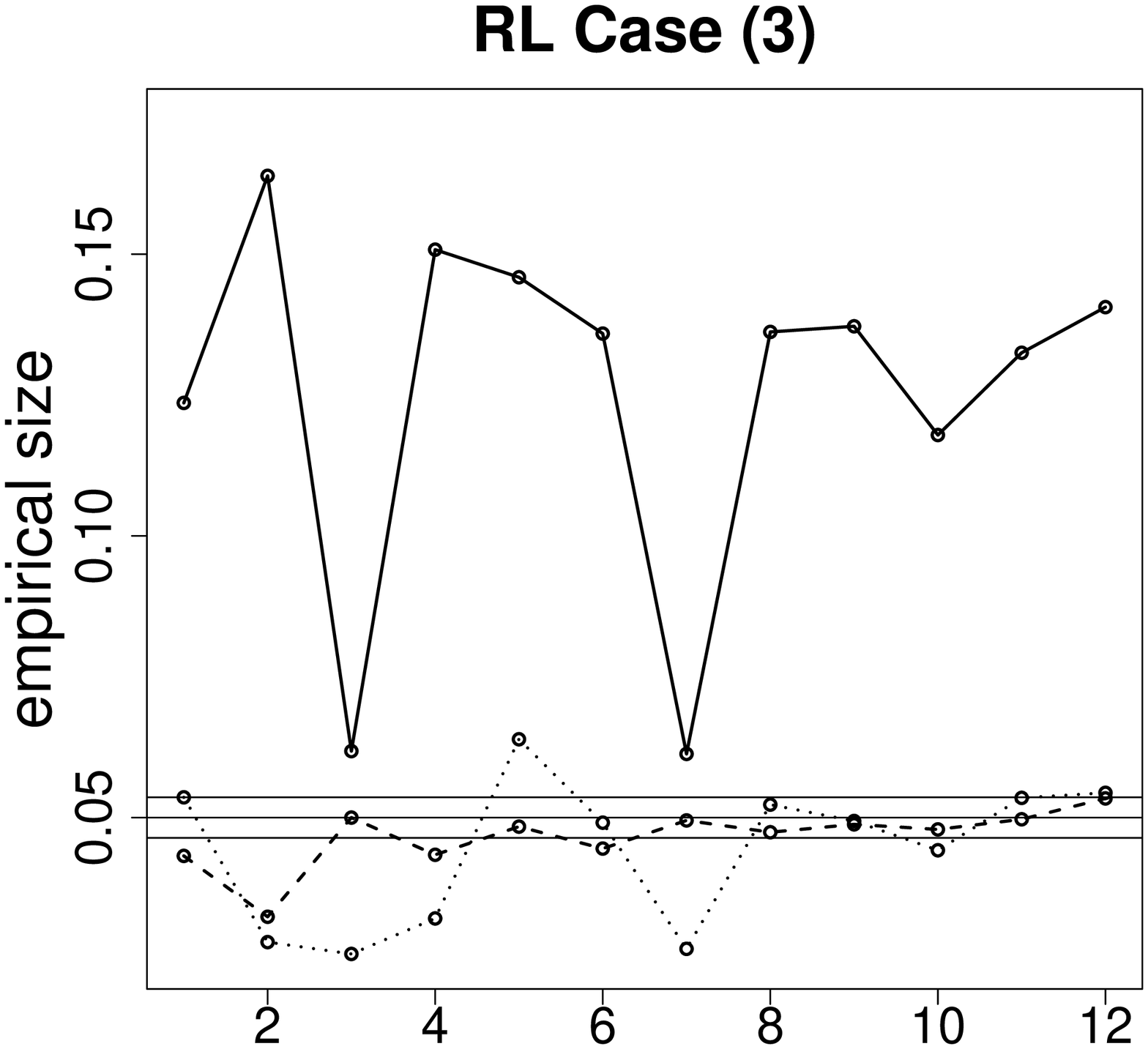} }}

\rotatebox{-90}{ \resizebox{2.1 in}{!}{\includegraphics{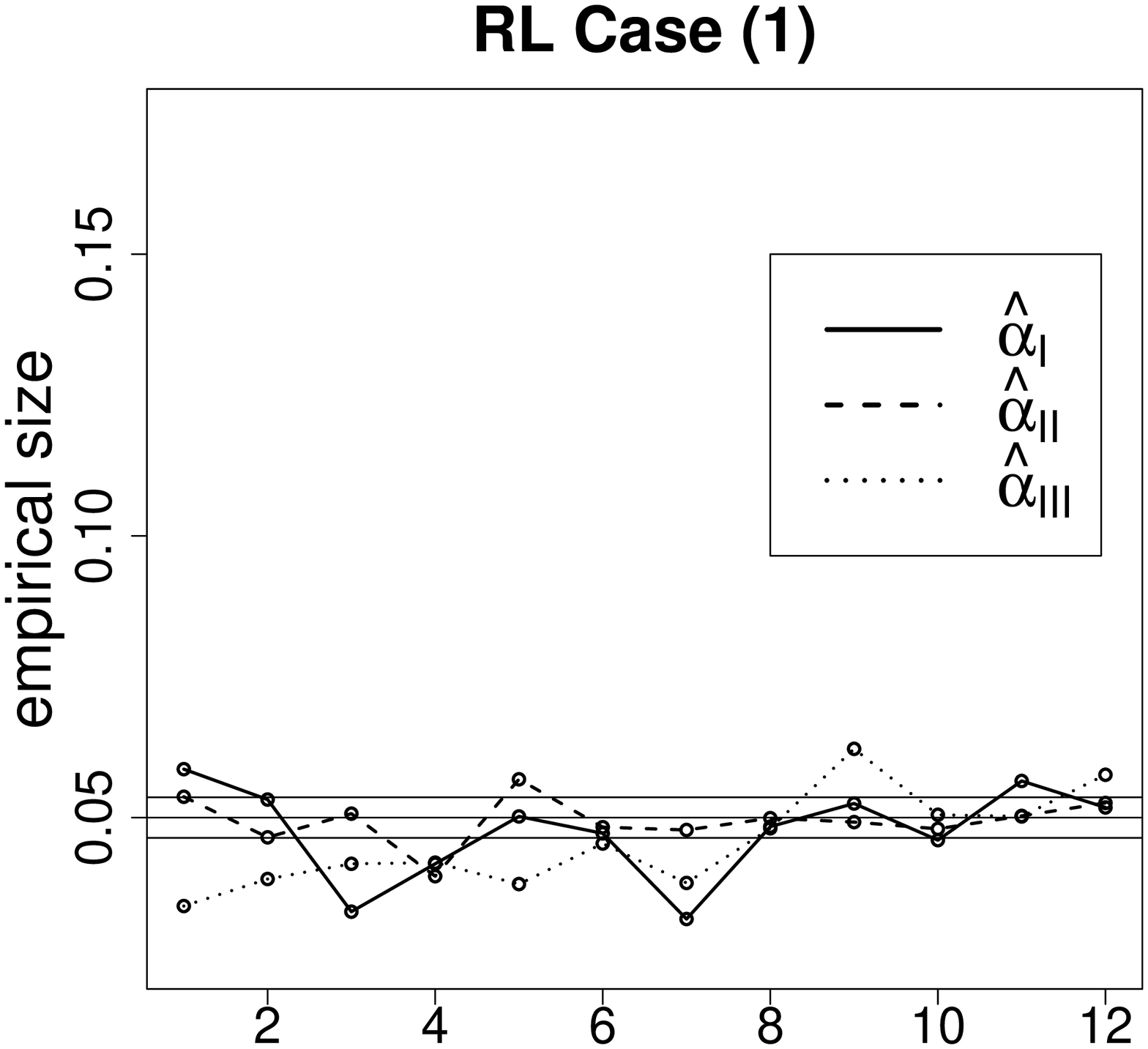} }}
\rotatebox{-90}{ \resizebox{2.1 in}{!}{\includegraphics{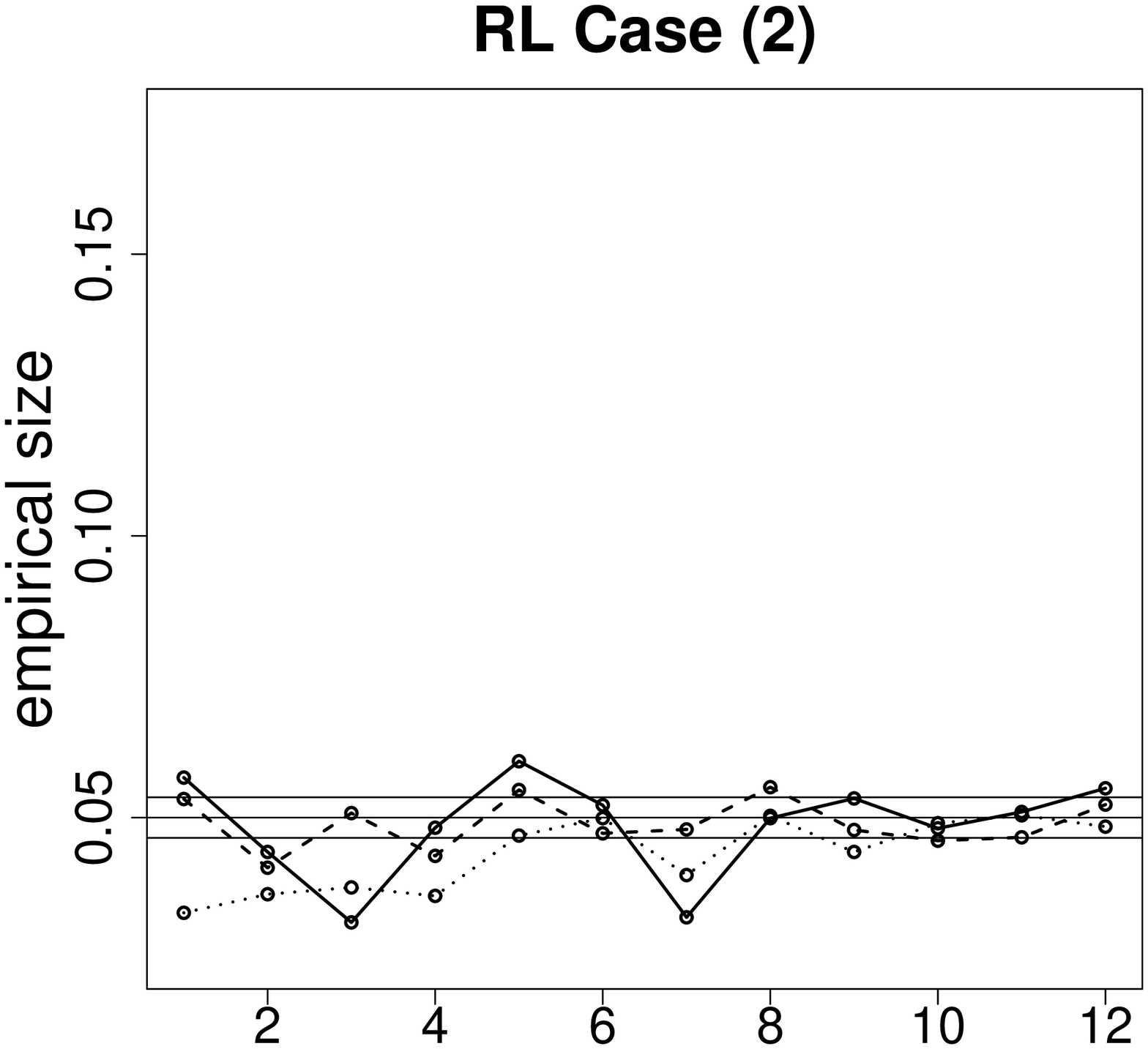} }}
\rotatebox{-90}{ \resizebox{2.1 in}{!}{\includegraphics{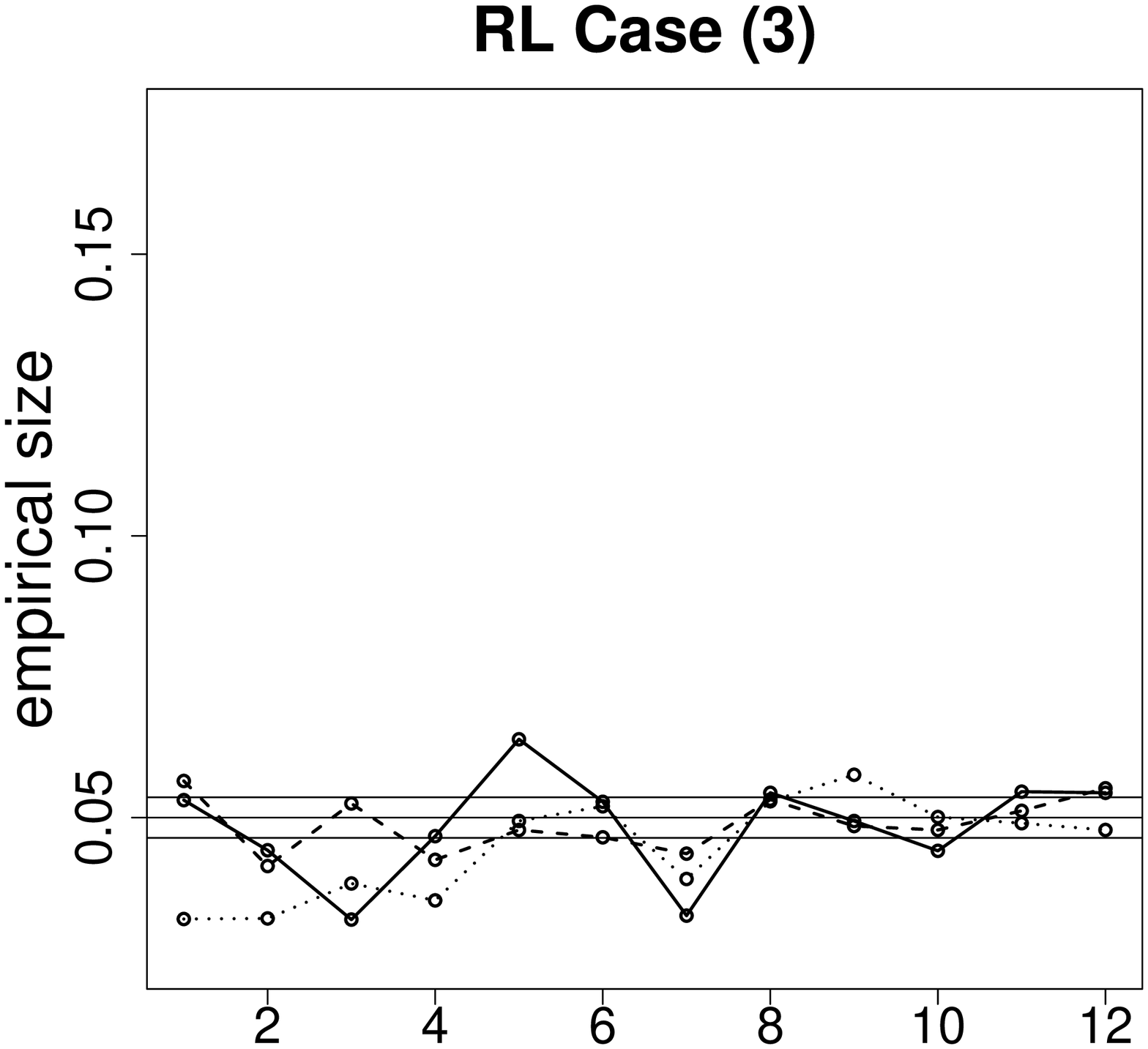} }}

\caption{
\label{fig:Emp-Size-NNCT-RL}
The empirical size estimates of the NNCT-tests based on 10000 Monte Carlo replications
under RL Cases (1)-(3) for various sample size combinations.
The horizontal lines are as in Figure \ref{fig:Emp-Size-NNCT-CSR}.
The numbers in the horizontal axis labels represent sample (i.e., class) size combinations:
1=(10,10), 2=(10,30), 3=(10,50), 4=(30,10), 5=(30,30),
6=(30,50), 7=(50,10), 8=(50,30), 9=(50,50), 10=(50,100), 11=(100,50), 12=(100,100).
The empirical size labeling is as in Table \ref{tab:MC-emp-sig-level-NNCT}.
}
\end{figure}

\begin{figure}
\centering
\rotatebox{-90}{ \resizebox{2.1 in}{!}{\includegraphics{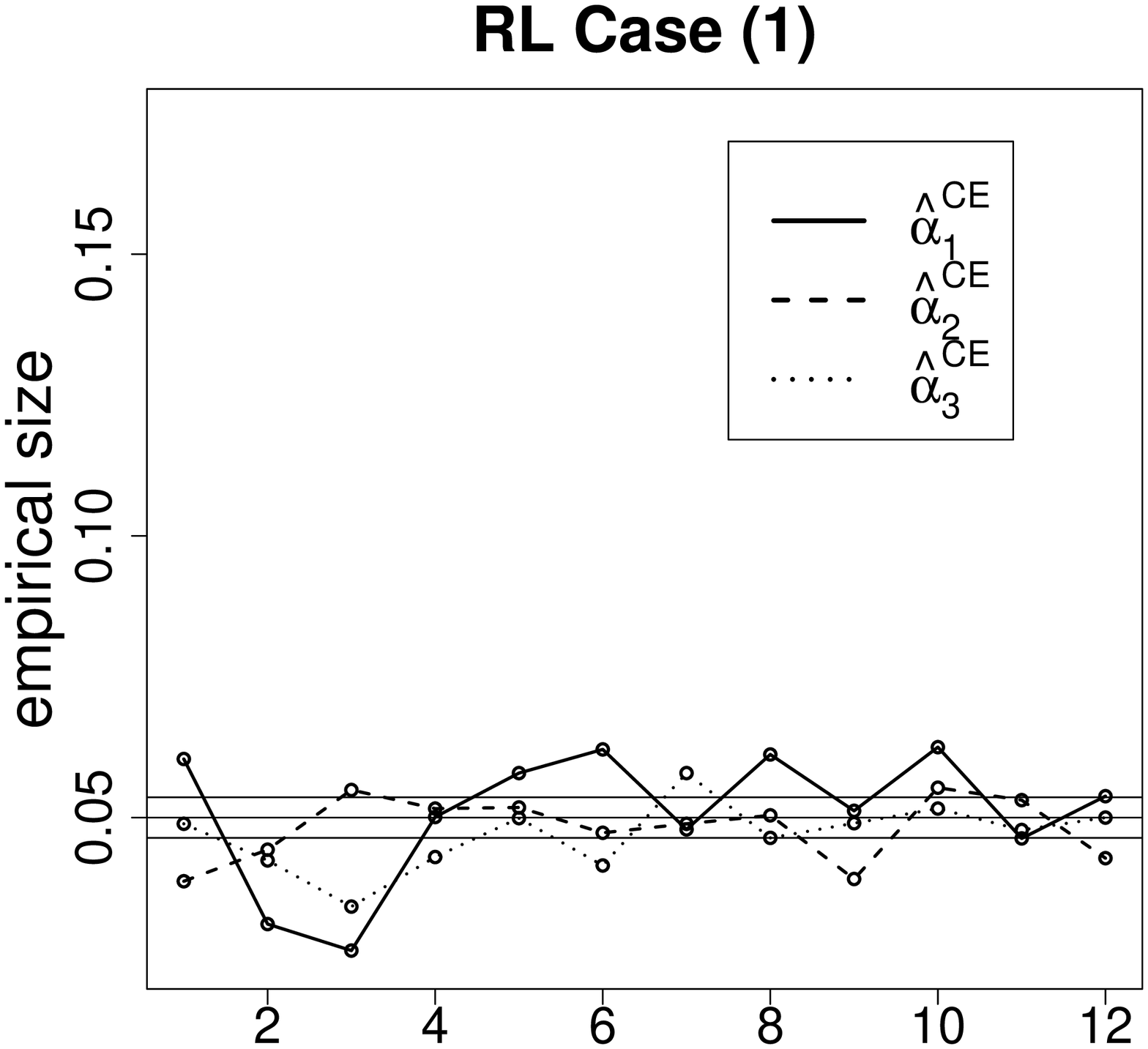} }}
\rotatebox{-90}{ \resizebox{2.1 in}{!}{\includegraphics{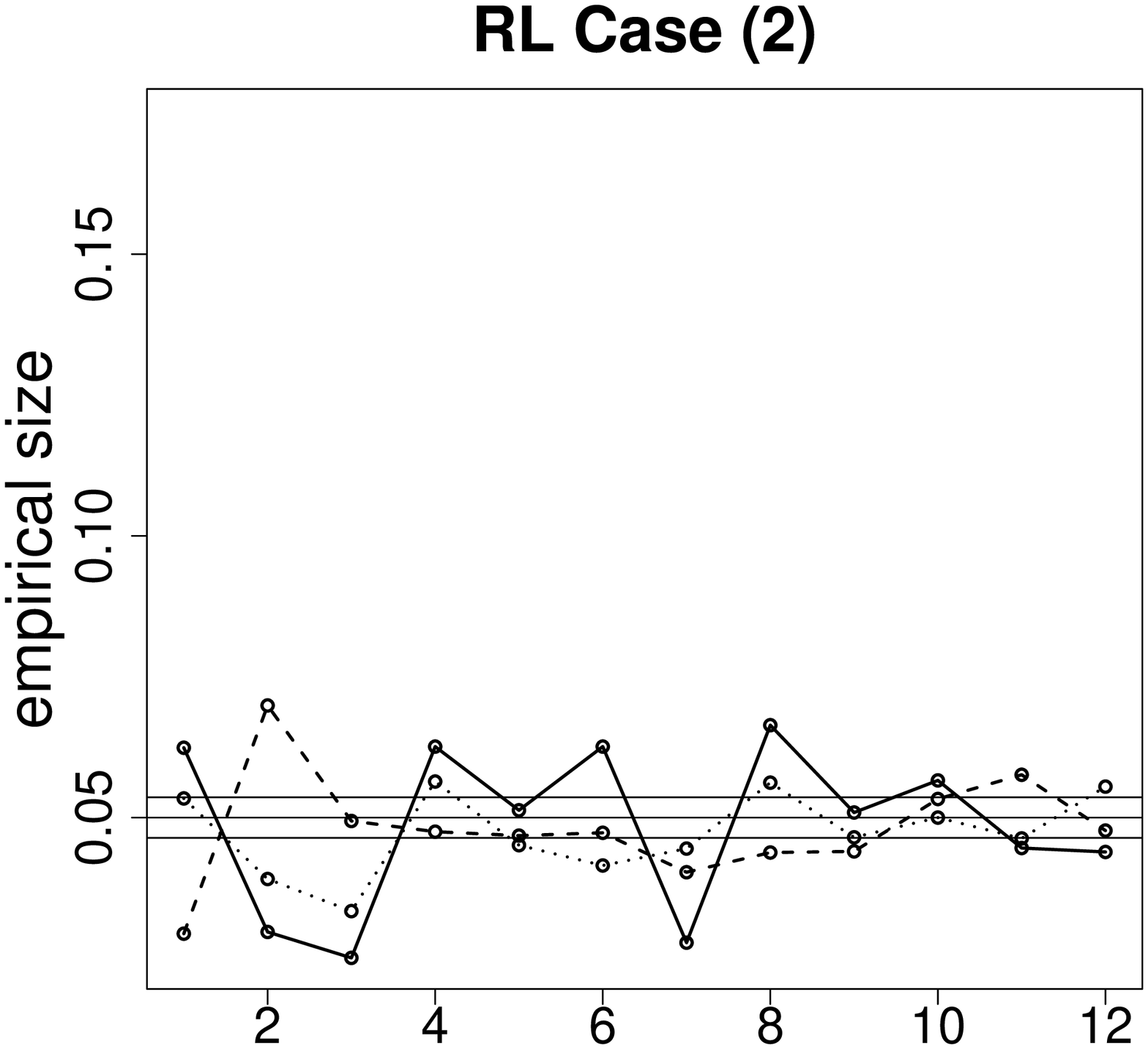} }}
\rotatebox{-90}{ \resizebox{2.1 in}{!}{\includegraphics{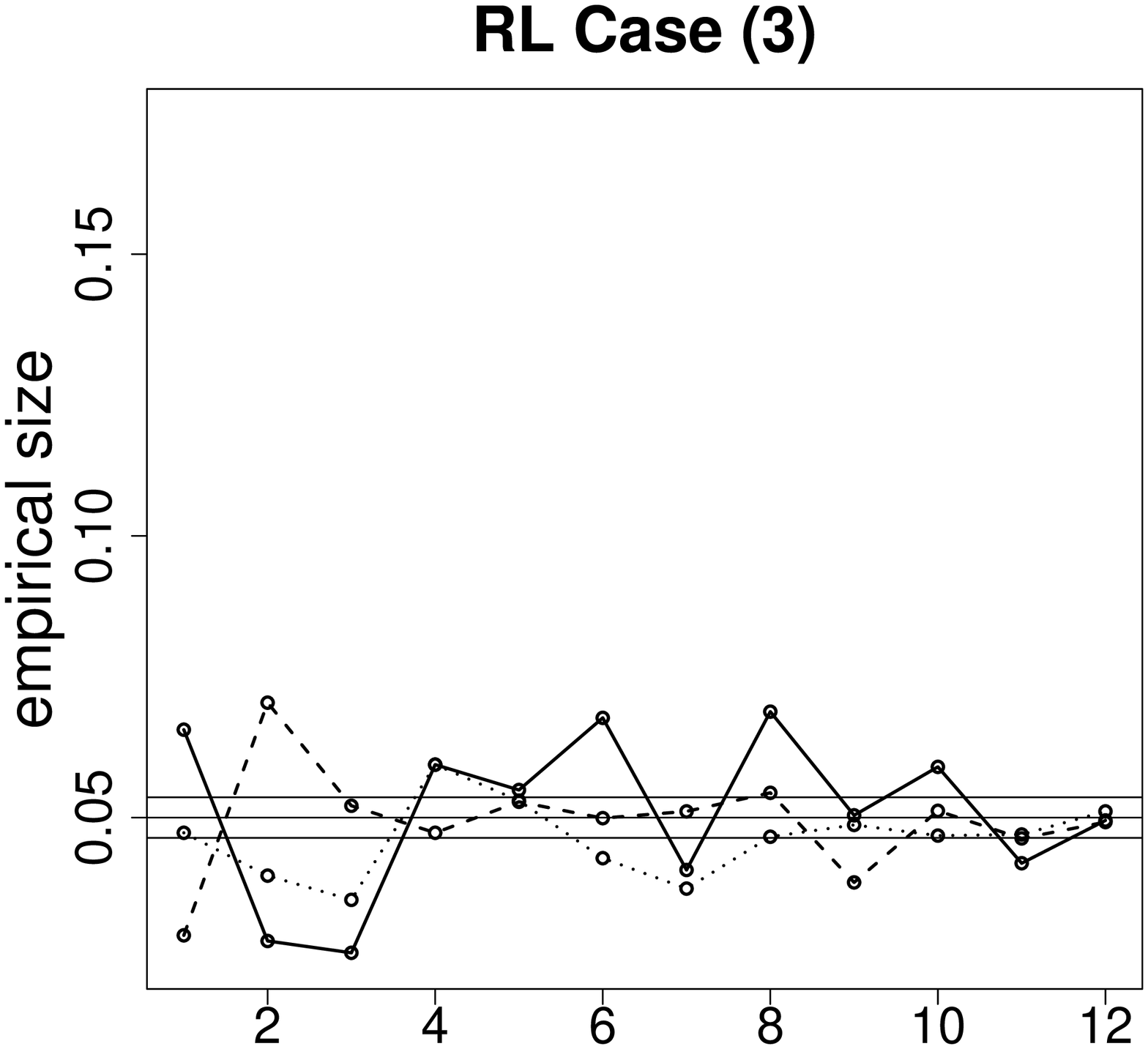} }}

\rotatebox{-90}{ \resizebox{2.1 in}{!}{\includegraphics{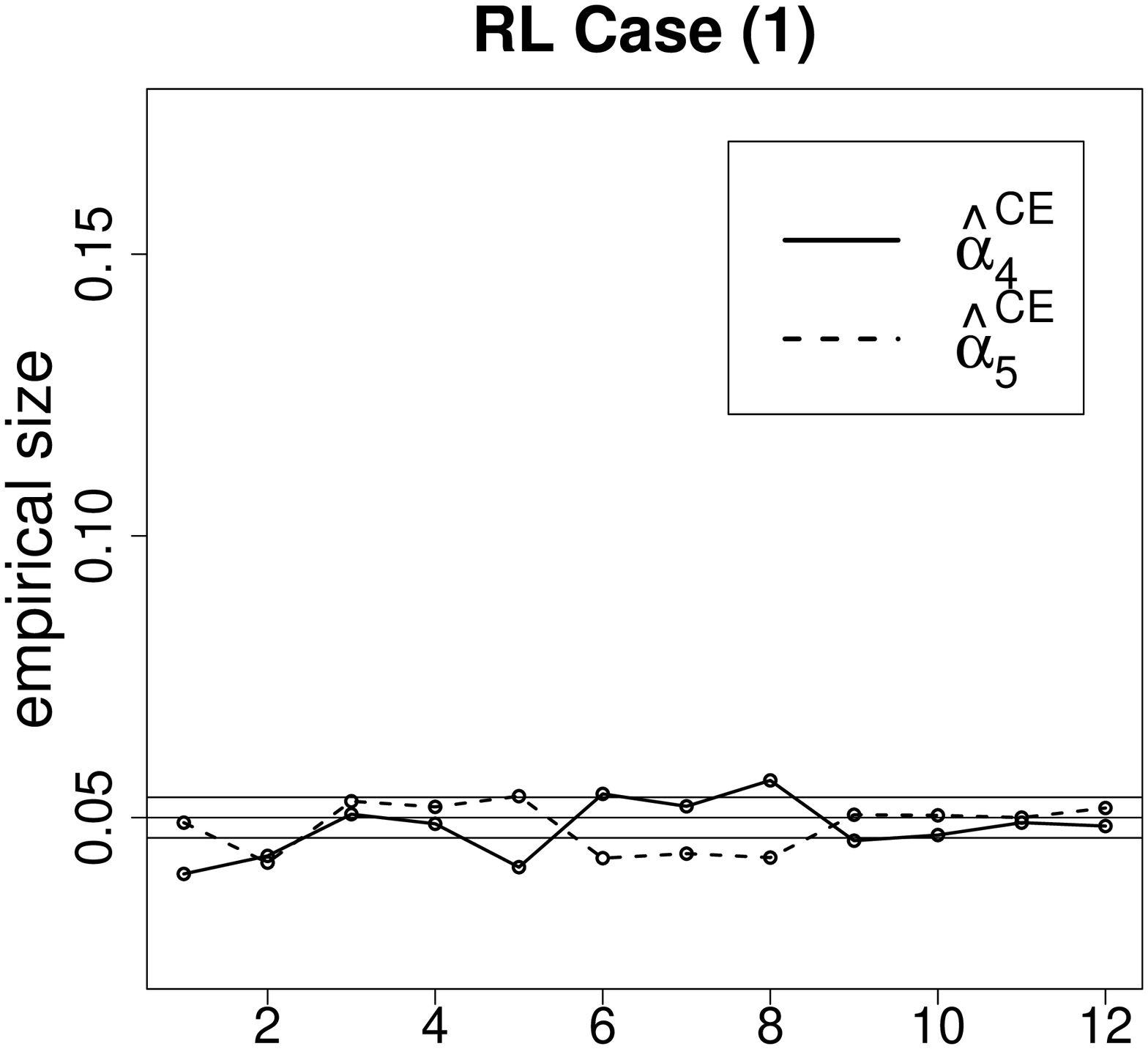} }}
\rotatebox{-90}{ \resizebox{2.1 in}{!}{\includegraphics{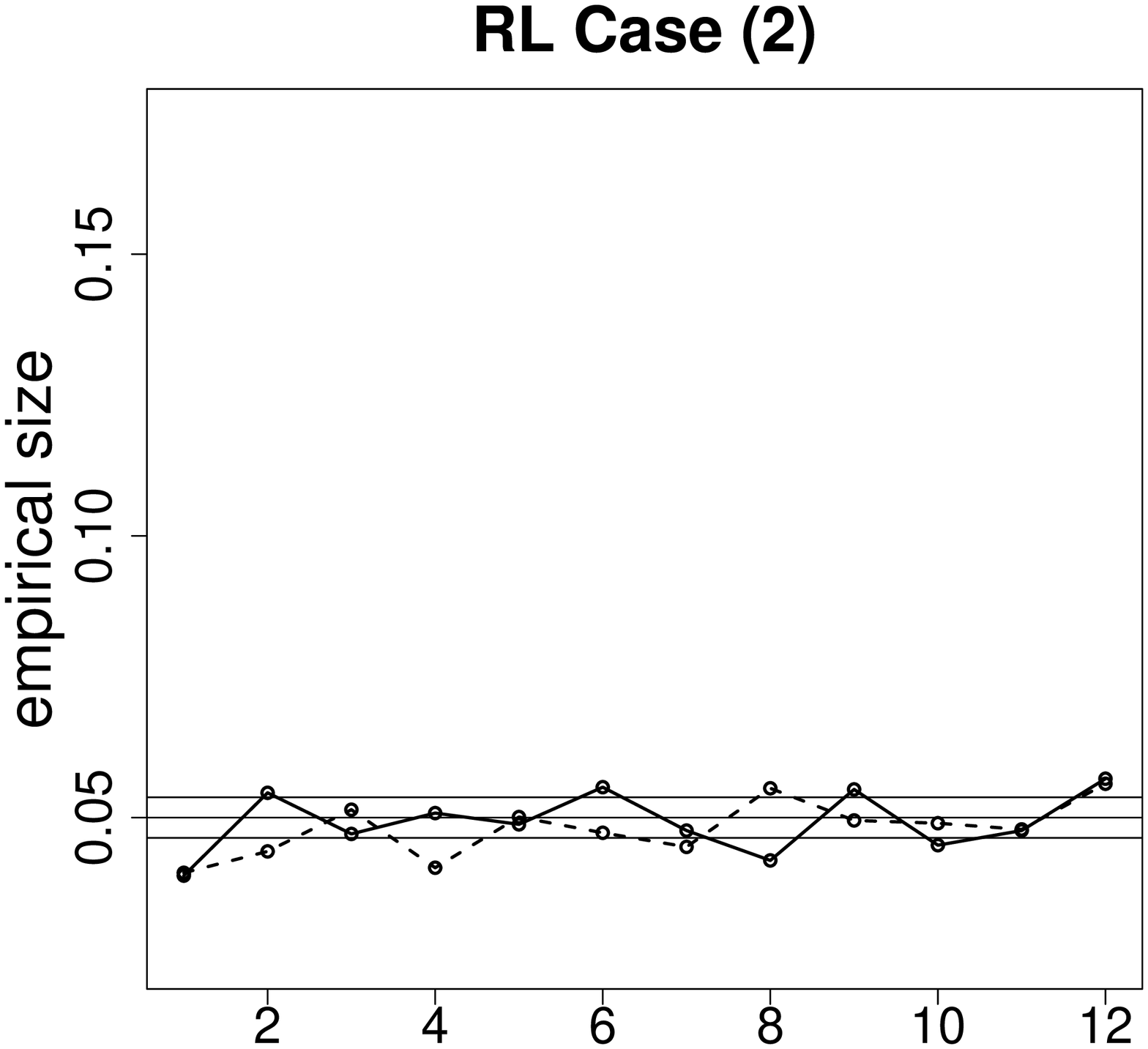} }}
\rotatebox{-90}{ \resizebox{2.1 in}{!}{\includegraphics{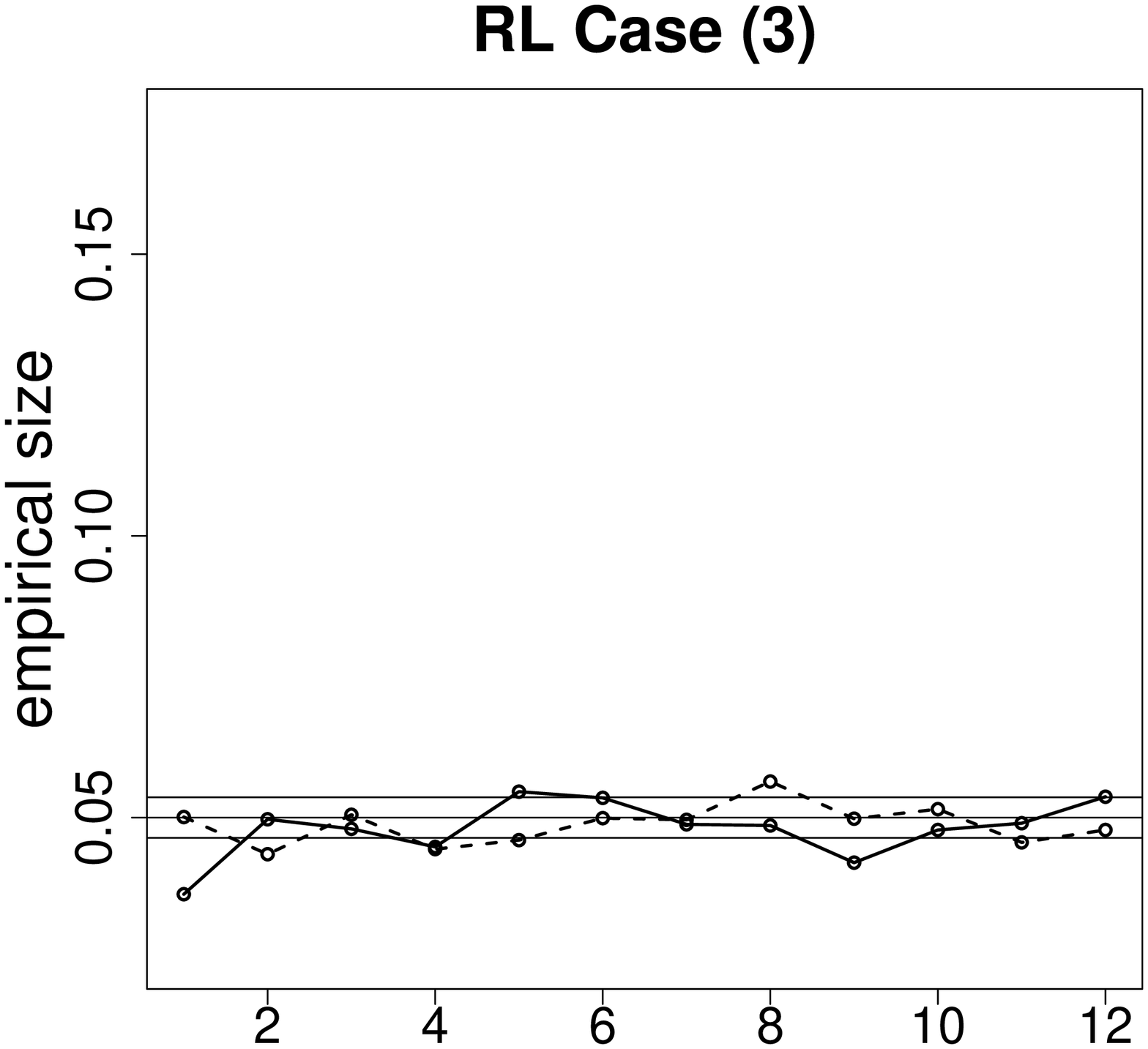} }}

\rotatebox{-90}{ \resizebox{2.1 in}{!}{\includegraphics{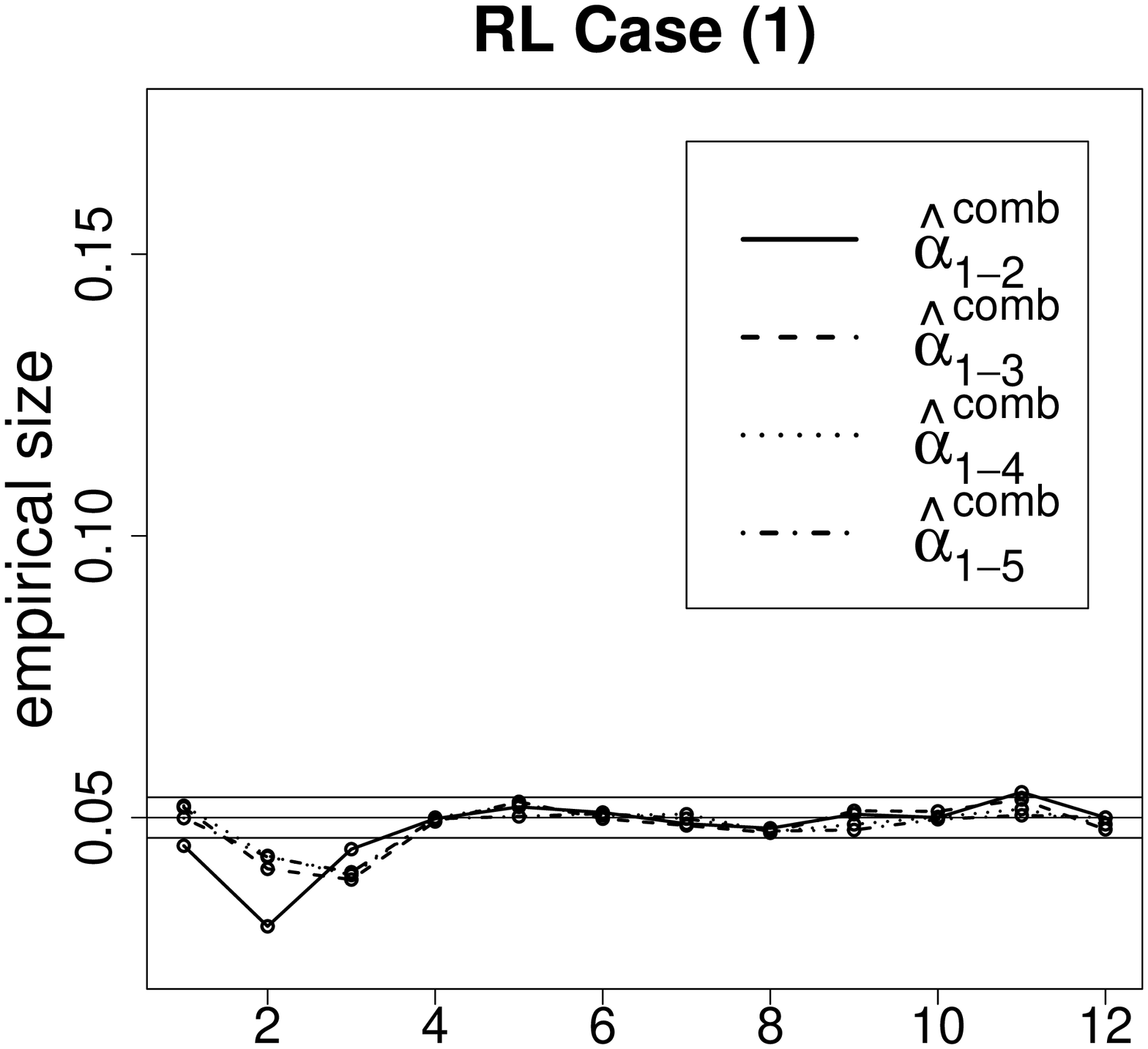} }}
\rotatebox{-90}{ \resizebox{2.1 in}{!}{\includegraphics{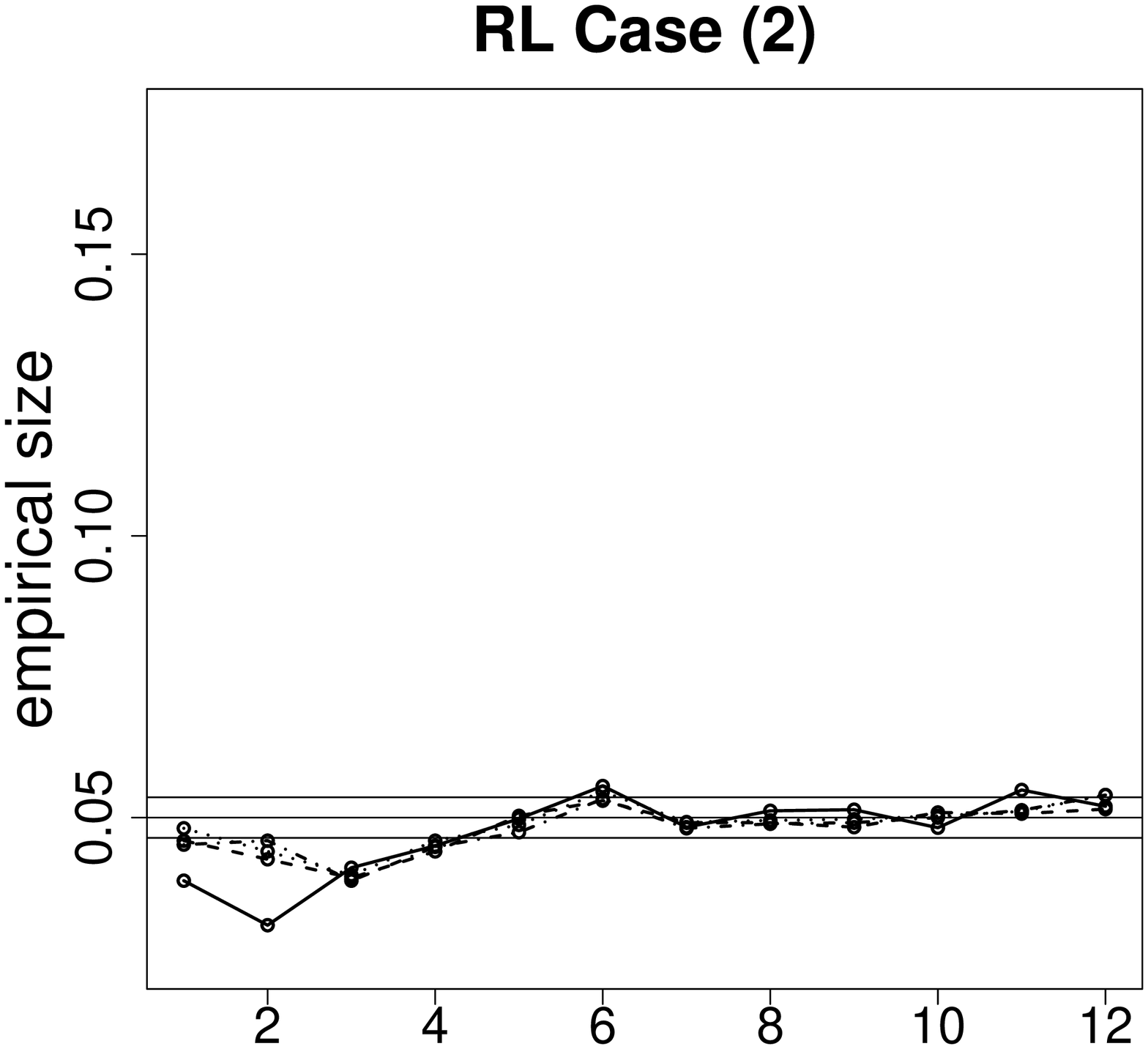} }}
\rotatebox{-90}{ \resizebox{2.1 in}{!}{\includegraphics{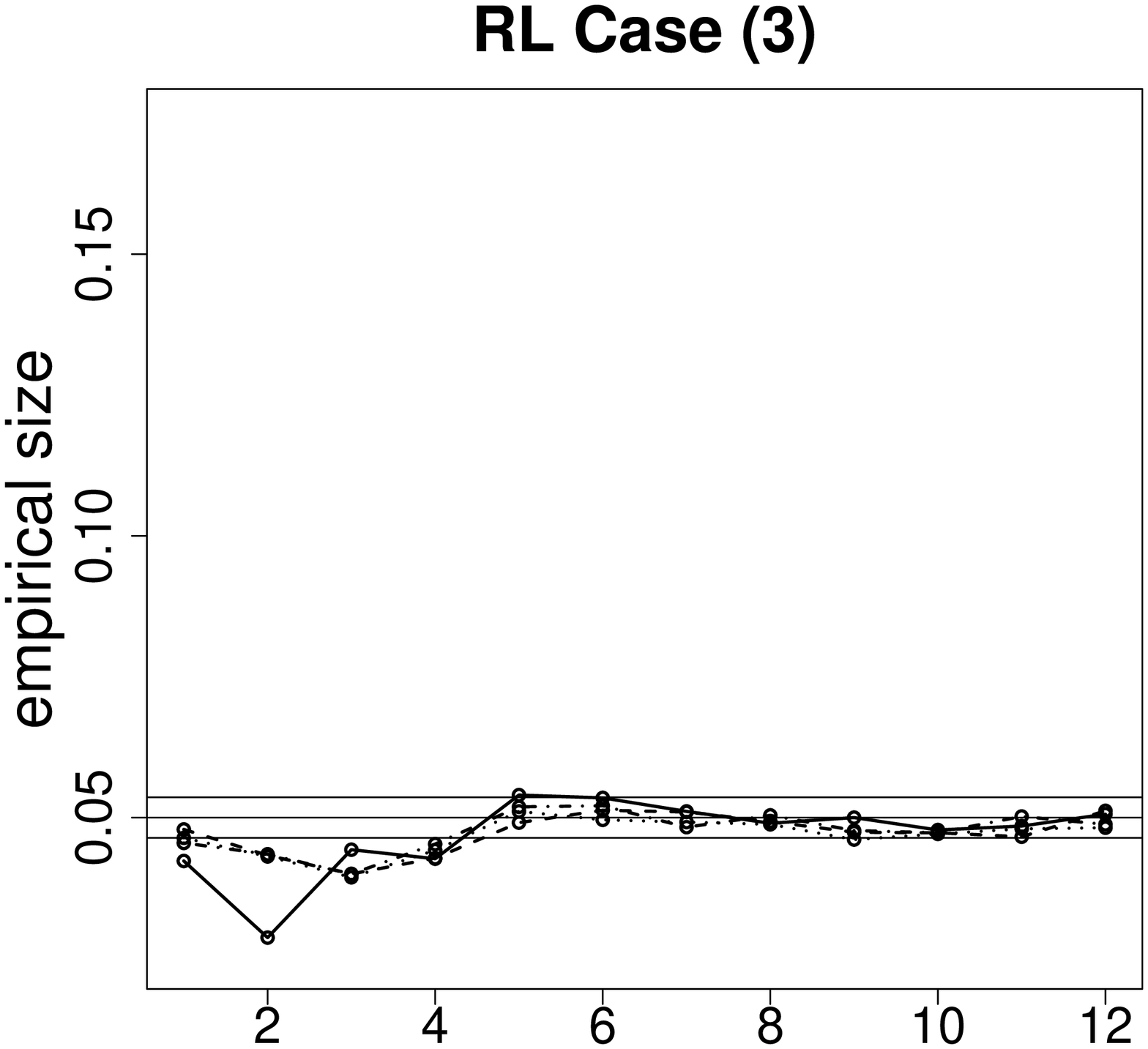} }}

\caption{
\label{fig:Emp-Size-CE-RL}
The empirical size estimates of Cuzick-Edwards $k$-NN and combined tests
based on 10000 Monte Carlo replications under RL Cases (1)-(3)
for various sample size combinations.
The horizontal lines are as in Figure \ref{fig:Emp-Size-NNCT-CSR},
the horizontal axis labeling is as in Figure \ref{fig:Emp-Size-NNCT-RL},
and the empirical size labeling is as in Table \ref{tab:MC-emp-sig-level-CE}.}
\end{figure}

\begin{remark}
\label{rem:MC-emp-size-RL}
\textbf{Main Result of Monte Carlo Simulations under RL:}
Based on the simulation results under RL,
we reach the same conclusions as in Remark \ref{rem:MC-emp-size-CSR} for NNCT-tests under RL.
That is, we recommend the disuse of Pielou's test;
and when sample sizes are small
(hence the corresponding cell counts are $\leq 5$),
we recommend using the Monte Carlo randomization.

Among the NNCT-tests, Dixon's test and version II of the new tests
have the best size performance under RL.
On the other hand, among Cuzick-Edward's tests,
$T_5$ and $T^{comb}_S$ have better size performance under RL,
and they have about the same size performance as Dixon's test
and version II of the new tests.
$\square$
\end{remark}

\section{Empirical Power Analysis}
\label{sec:emp-power}
To evaluate the power performance of the clustering tests,
we only consider alternatives against the CSR independence pattern.
That is, the points are generated in such a way that
they are from an inhomogeneous Poisson process ---conditional on the number of points---
in a region of interest (unit square in the simulations) for at least one class.
We avoid the alternatives against the RL pattern;
i.e., we do not consider non-random labeling of a fixed set of points
that would result in segregation or association.

\subsection{Empirical Power Analysis under the Segregation Alternatives}
\label{sec:emp-power-seg}
For the segregation alternatives (against the CSR independence pattern), three cases are considered.
We generate $X_i \stackrel{iid}{\sim} \U((0,1-s)\times(0,1-s))$ for $i=1,2,\ldots,n_1$
and $Y_j \stackrel{iid}{\sim} \U((s,1)\times(s,1))$ for $j=1,2,\ldots,n_2$.
In the pattern generated, appropriate choices of
$s$ will imply that $X_i$ and $Y_j$ are more segregated than expected under CSR independence.
That is, it will be more likely to have $(X,X)$
NN pairs than mixed NN pairs (i.e., $(X,Y)$ or $(Y,X)$ pairs).
The three values of $s$ we consider constitute
the three segregation alternatives:
\begin{equation}
\label{eqn:seg-alt}
H_S^{I}: s=1/6,\;\;\; H_S^{II}: s=1/4, \text{ and } H_S^{III}: s=1/3.
\end{equation}

Observe that, from $H_S^I$ to $H_S^{III}$ (i.e., as $s$ increases), the segregation gets stronger
in the sense that $X$ and $Y$ points tend to form one-class clumps or clusters.
By construction, the points are uniformly generated, hence exhibit homogeneity
with respect to their supports for each class,
but with respect to the unit square these alternative patterns are examples of
departures from first-order homogeneity which implies
segregation of the classes $X$ and $Y$.
The simulated segregation patterns are symmetric in the sense that,
$X$ and $Y$ classes are generated to be equally segregated (or clustered) from each other.
Hence, although class $X$ stands for the ``cases" in Cuzick-Edward's tests,
the results would be similar if class $Y$ is chosen instead.

\begin{figure}[ht]
\centering
\rotatebox{-90}{ \resizebox{2.1 in}{!}{\includegraphics{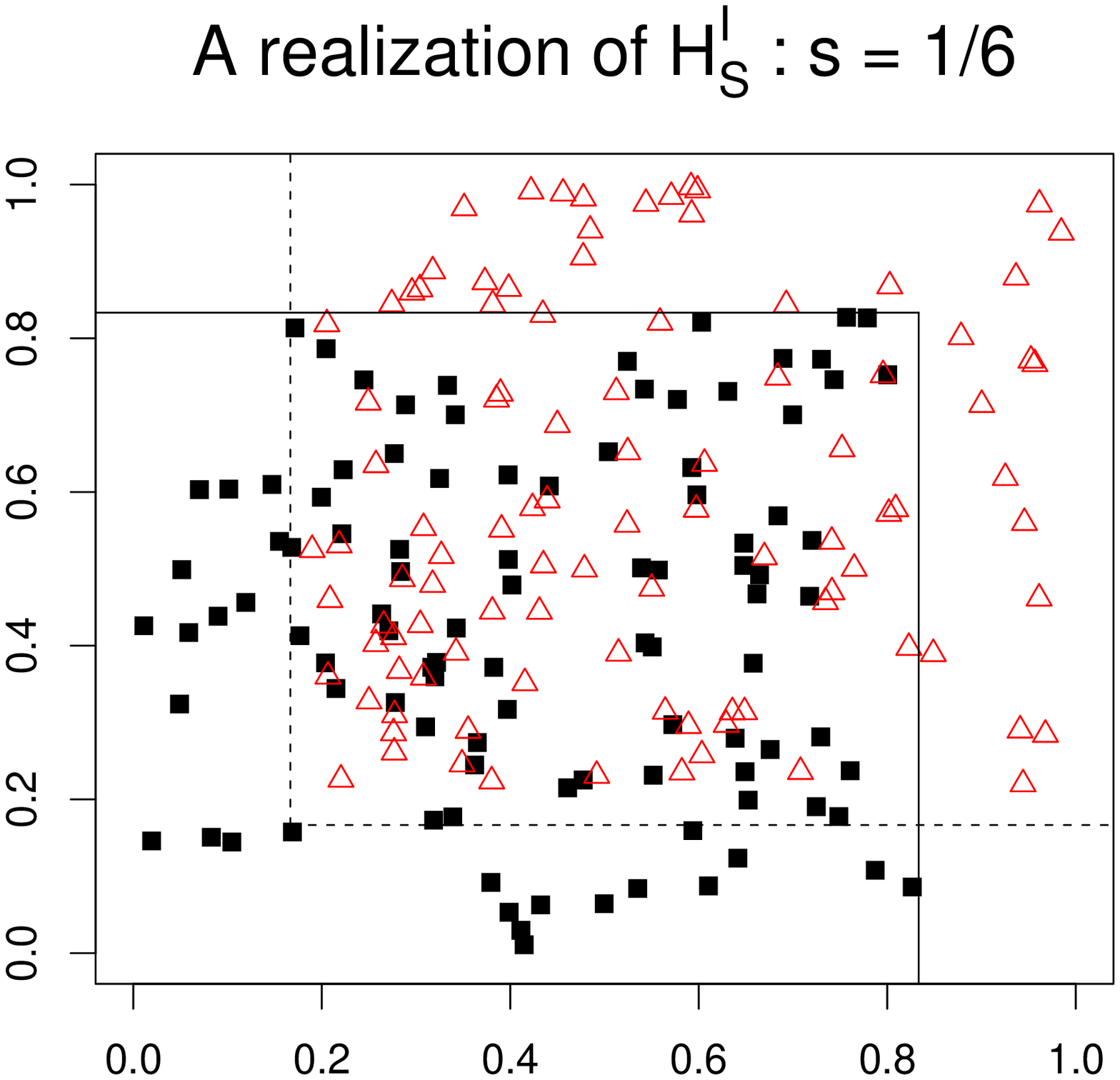} }}
\rotatebox{-90}{ \resizebox{2.1 in}{!}{\includegraphics{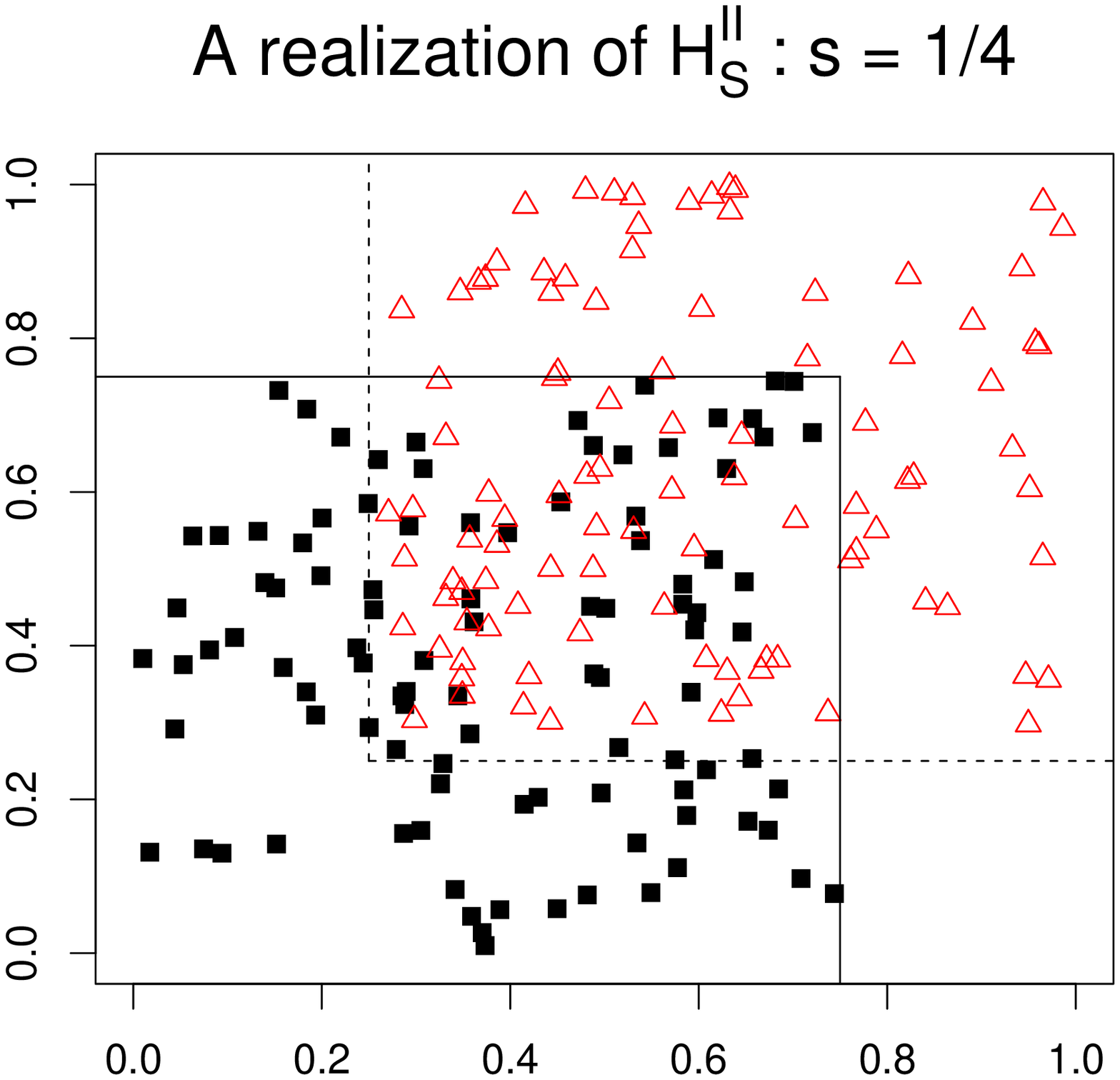} }}
\rotatebox{-90}{ \resizebox{2.1 in}{!}{\includegraphics{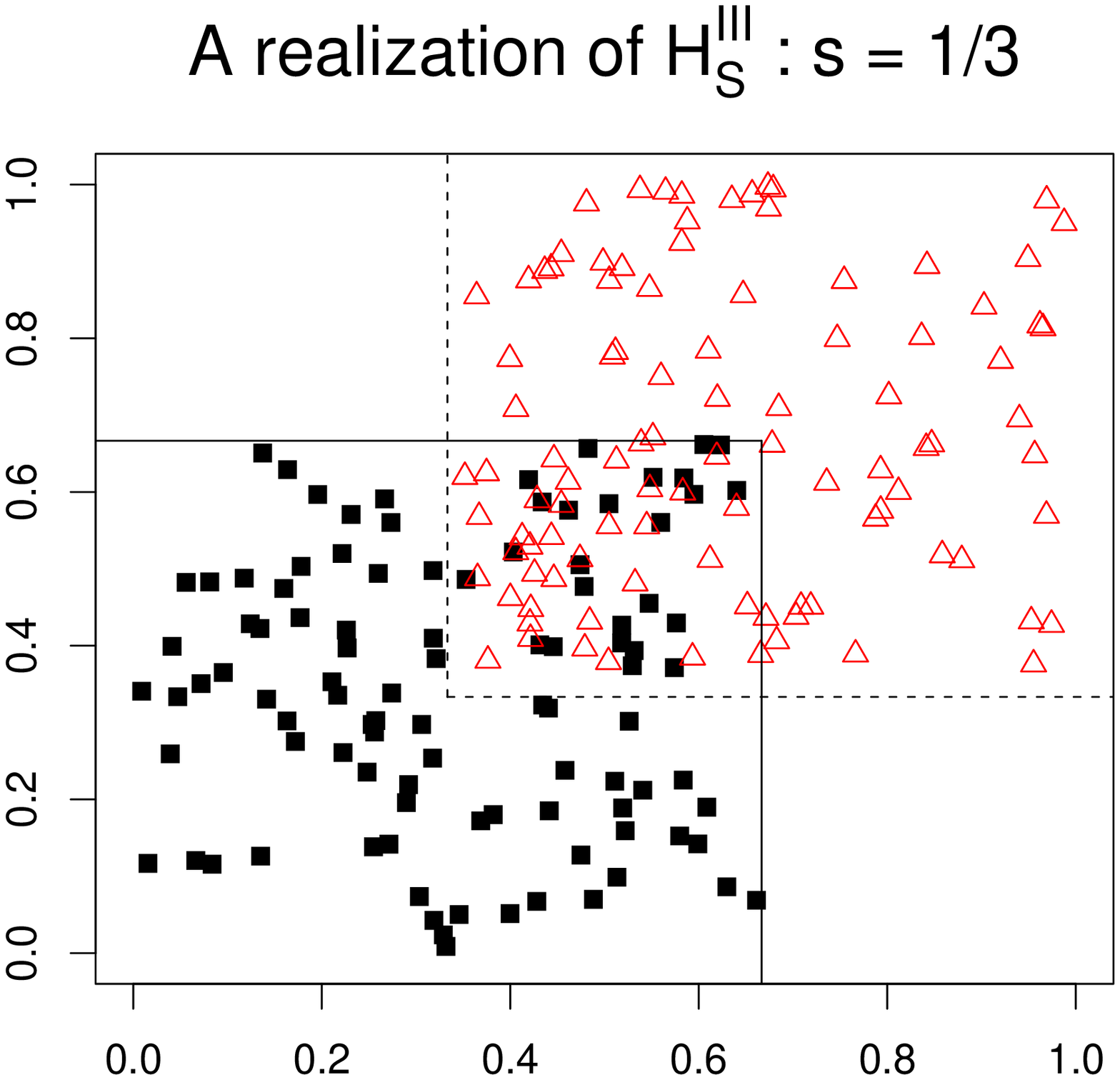} }}
 \caption{
\label{fig:SegAlt}
Three realizations for $H_S^{I}: s=1/6$ (left), $H_S^{II}: s=1/4$ (middle),
and $H_S^{III}: s=1/3$ (right) with $n_1=100$ $X$ points (solid squares $\blacksquare$)
and $n_2=100$ $Y$ points (triangles $\triangle$).}
\end{figure}

\begin{table}[ht]
\centering
\begin{tabular}{|c|c|c|c|c|c|c|c|}
\hline
\multicolumn{7}{|c|}{Empirical power estimates under} \\
\multicolumn{7}{|c|}{the segregation alternatives} \\
\hline
 & $(n_1,n_2)$  & $\bh_D$ & $\bh_I$ & $\bh_{II}$ & $\bh_{III}$ & $\bh_{P,mc}$ \\
\hline
\hline

 & $(10,10)$ & .0775 & .0881 & .0481 & .1026 & .0890\\
\cline{2-7}
 & $(10,30)$ & .1414 & .1737 & .1192 & .1983 & .1711\\
\cline{2-7}
 & $(10,50)$ & .2193 & .2187 & .1926 & .2491 & .2246\\
\cline{2-7} \raisebox{4.ex}[0pt]{$H_S^I$}
 & $(30,30)$ & .2904 & .3688 & .2456 & .3837 & .3717\\
\hline
\hline

 & $(10,10)$ & .2305 & .2830 & .1523 & .3203 & .2842\\
\cline{2-7}
 & $(10,30)$ & .4555 & .5344 & .4063 & .5734 & .5349\\
\cline{2-7}
 & $(10,50)$ & .6174 & .6413 & .5796 & .6794 & .6491\\
\cline{2-7} \raisebox{4.ex}[0pt]{$H_S^{II}$} &
$(30,30)$    & .8141 & .8847 & .7728 & .8917 & .8858\\
\hline
\hline

 & $(10,10)$ & .5817 & .6875 & .4697 & .7257 & .6890\\
\cline{2-7}
 & $(10,30)$ & .8787 & .9248 & .8467 & .9406 & .9245\\
\cline{2-7}
 & $(10,50)$ & .9528 & .9617 & .9395 & .9711 & .9627\\
\cline{2-7} \raisebox{4.ex}[0pt]{$H_S^{III}$}
 & $(30,30)$ & .9969 & .9988 & .9947 & .9990 & .9989\\
\hline

\end{tabular}
\caption{
\label{tab:emp-power-seg}
The empirical power estimates for the tests under the segregation alternatives,
$H_S^I-H_S^{III}$
with $N_{mc}=10000$, for some combinations of $n_1,n_2 \in
\{10,30,50\}$ at $\alpha=.05$.
$\bh_D$ stands for Dixon's test, $\bh_I$, $\bh_{II}$, and $\bh_{III}$
for versions I, II, and III of the new tests, respectively,
and $\bh_{P,mc}$ for Monte Carlo corrected version of Pielou's test.}
\end{table}

The empirical power estimates for NNCT-tests for $(n_1,n_2)\in \{(10,10),(10,30),(10,50),(30,30)\}$
are provided in Table \ref{tab:emp-power-seg}.
The power estimates against the sample size combinations for all the tests considered
are presented in Figure \ref{fig:Power-Est-Seg},
where $\bh_D$ is for Dixon's test, $\bh_I$, $\bh_{II}$, and $\bh_{III}$
are for versions I, II, and III of the new tests, respectively,
and $\bh_{P,mc}$ is for Monte Carlo corrected version of Pielou's test,
$\bh^{CE}_k$ is for Cuzick-Edwards $k$-NN test for $k=1,2,\ldots,5$,
and $\bh^{comb}_{1-j}$ is for Cuzick-Edwards $T^{comb}_{1-j}$ for $j=1,2,3,4$
(the empirical power estimate for Pielou's test is not presented
as it is misleading, see Remarks \ref{rem:MC-emp-size-CSR} and \ref{rem:MC-emp-size-RL}).
Observe that, as $n=(n_1+n_2)$ gets larger,
the power estimates get larger.
For the same $n=(n_1+n_2)$ values, the power estimate is larger
for classes with similar sample sizes.
Furthermore, as the segregation gets stronger,
the power estimates get larger.
The new version II test $\X^2_{II}$ has the lowest power
estimates for each sample size combination.
On the other hand the new versions,
$\X^2_I$, $\X^2_{III}$, and $\X^2_{P,mc}$ have about the
same power estimates that are larger than those of Dixon's test
with version III of the new tests having the highest power estimate
at each sample size combination.
Considering the empirical significance levels and power estimates,
we recommend version III of the new tests for testing against this type of segregation,
as $\X^2_{III}$ is at the correct significance level for similar sample sizes,
mildly conservative for very different sample sizes.
Additionally, $\X^2_{III}$ has the highest power for all sample size combinations.

\begin{figure}[]
\centering
Empirical power estimates for the NNCT-tests under $H_S$\\
\rotatebox{-90}{ \resizebox{2.1 in}{!}{\includegraphics{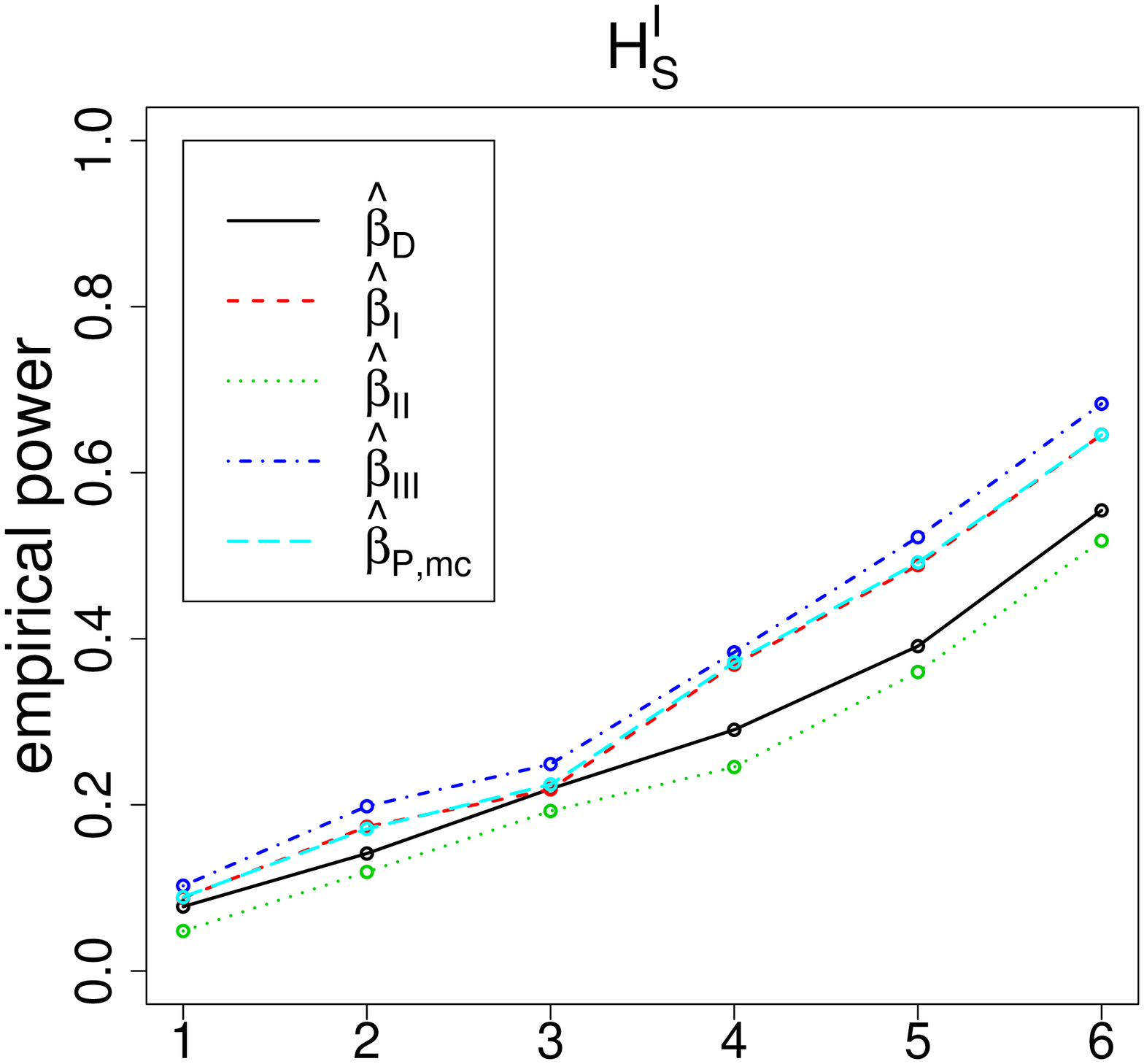} }}
\rotatebox{-90}{ \resizebox{2.1 in}{!}{\includegraphics{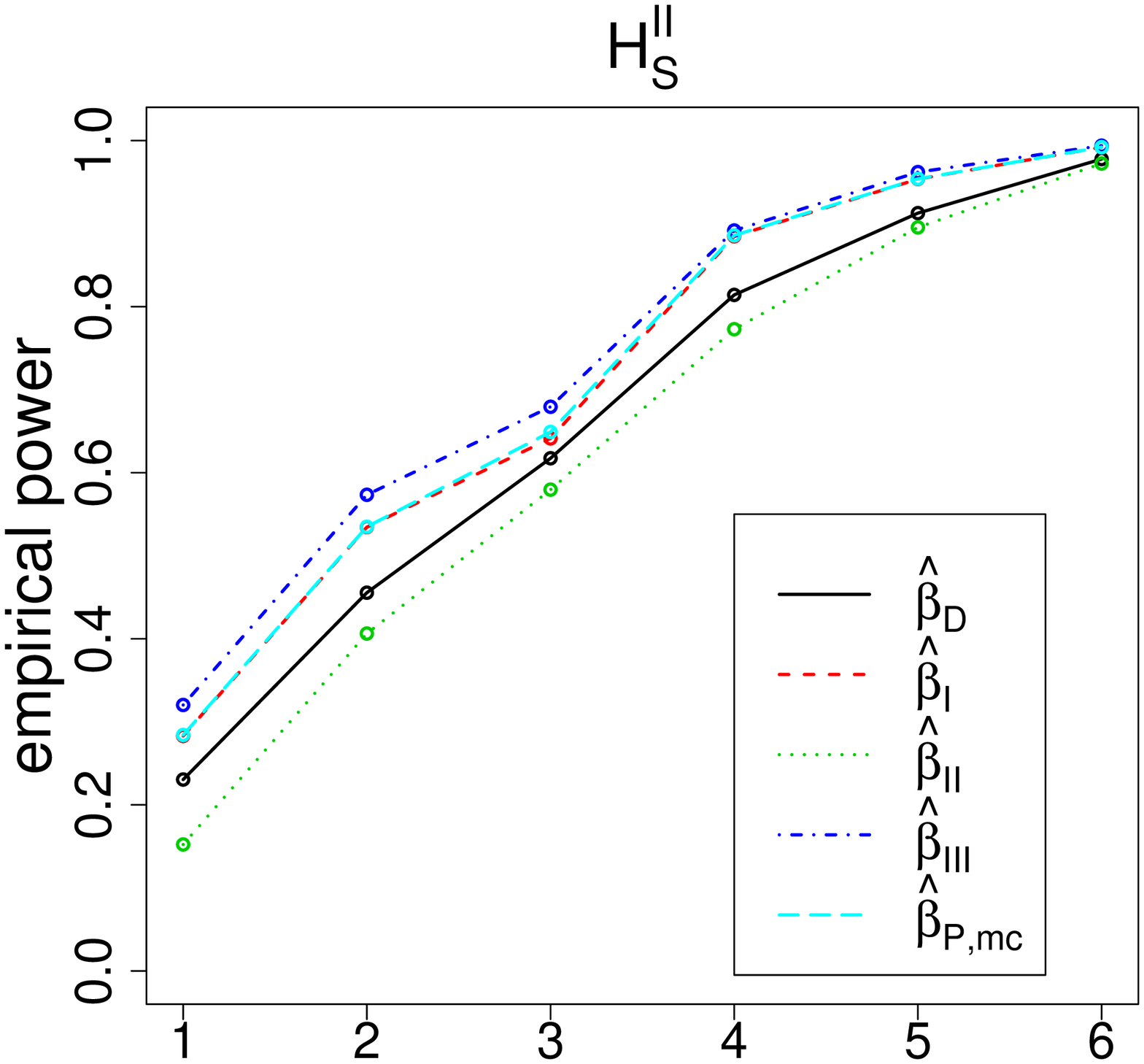} }}
\rotatebox{-90}{ \resizebox{2.1 in}{!}{\includegraphics{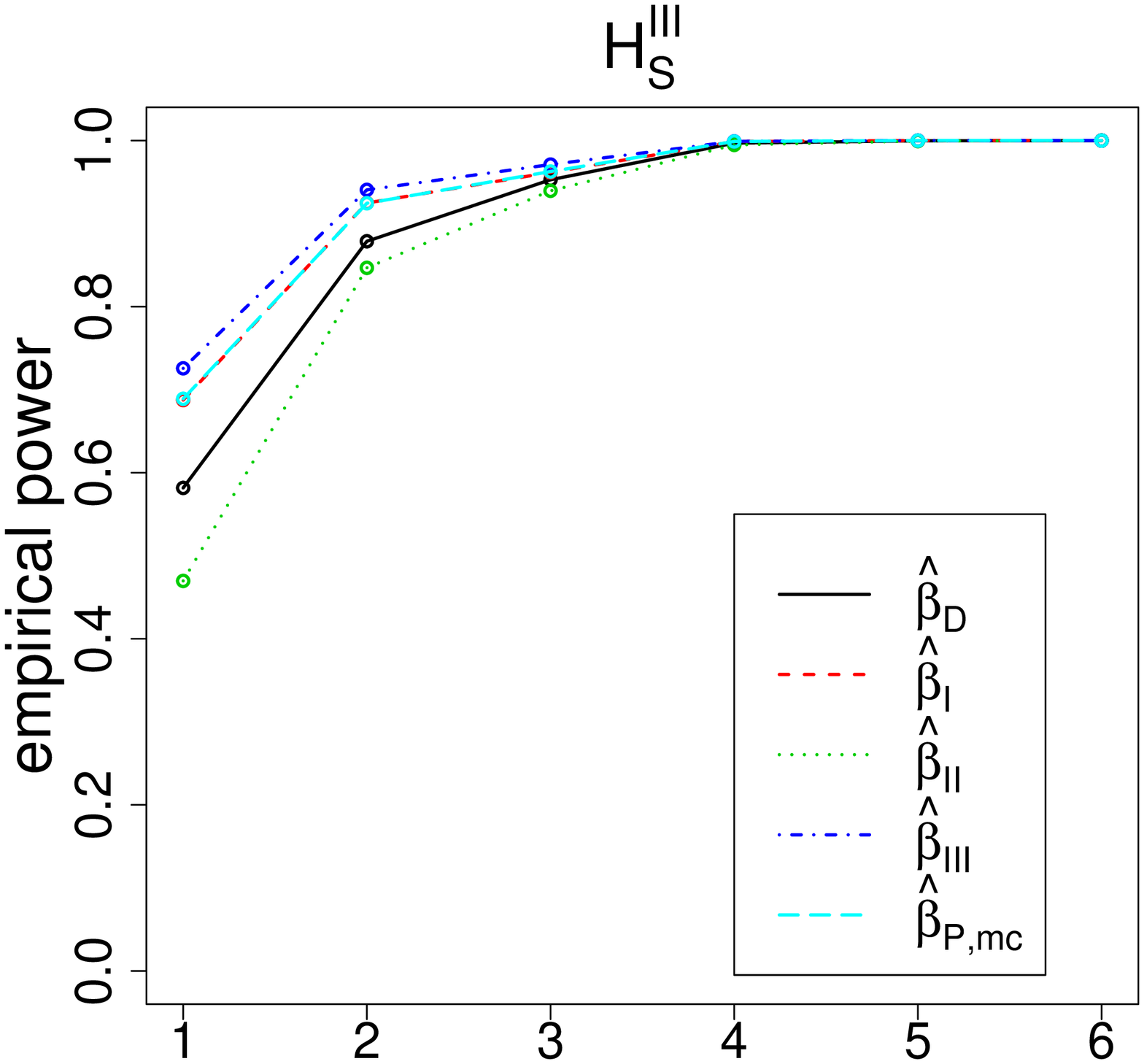} }}
Empirical power estimates for Cuzick-Edward's $k$-NN tests under $H_S$\\
\rotatebox{-90}{ \resizebox{2.1 in}{!}{\includegraphics{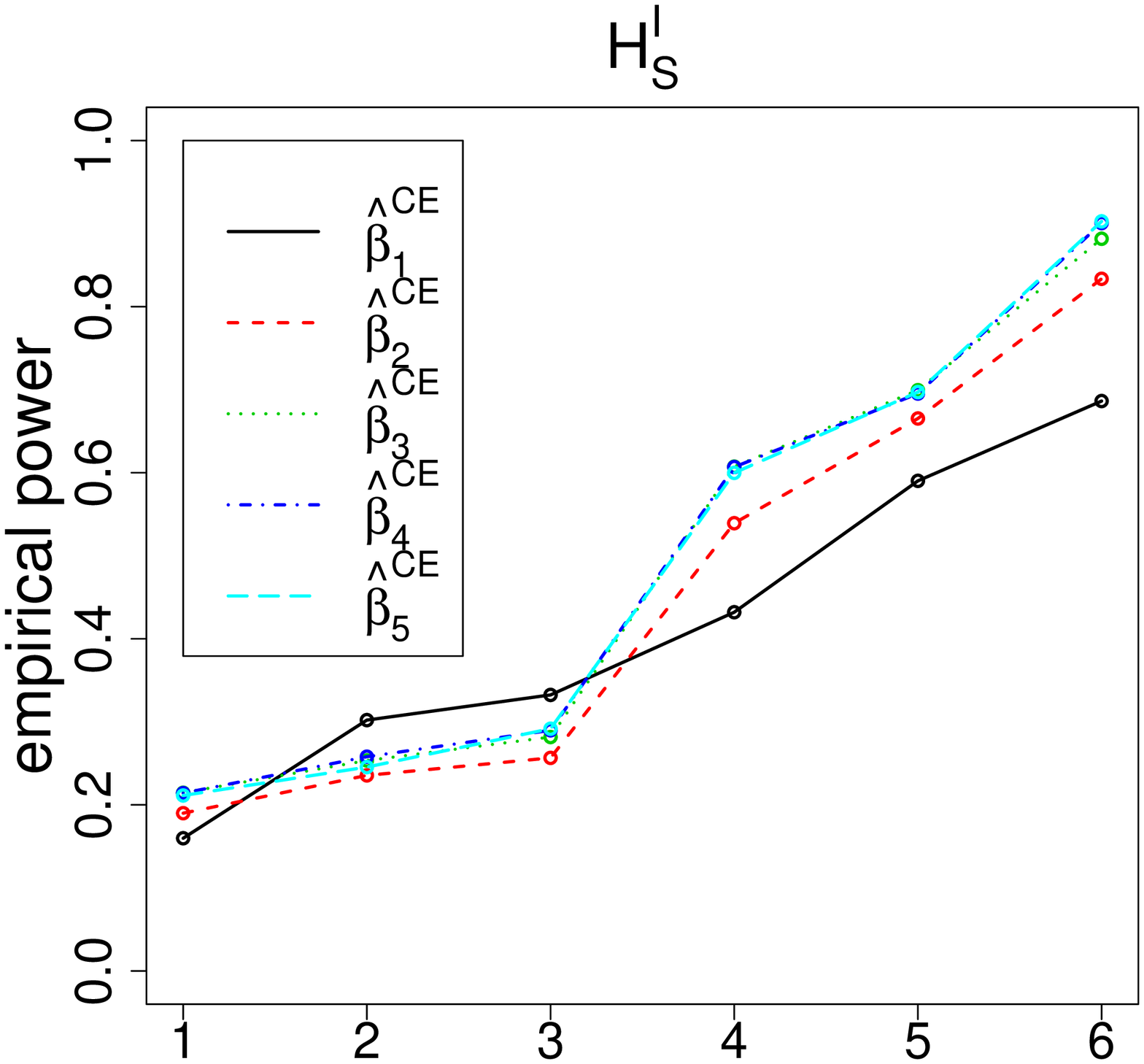} }}
\rotatebox{-90}{ \resizebox{2.1 in}{!}{\includegraphics{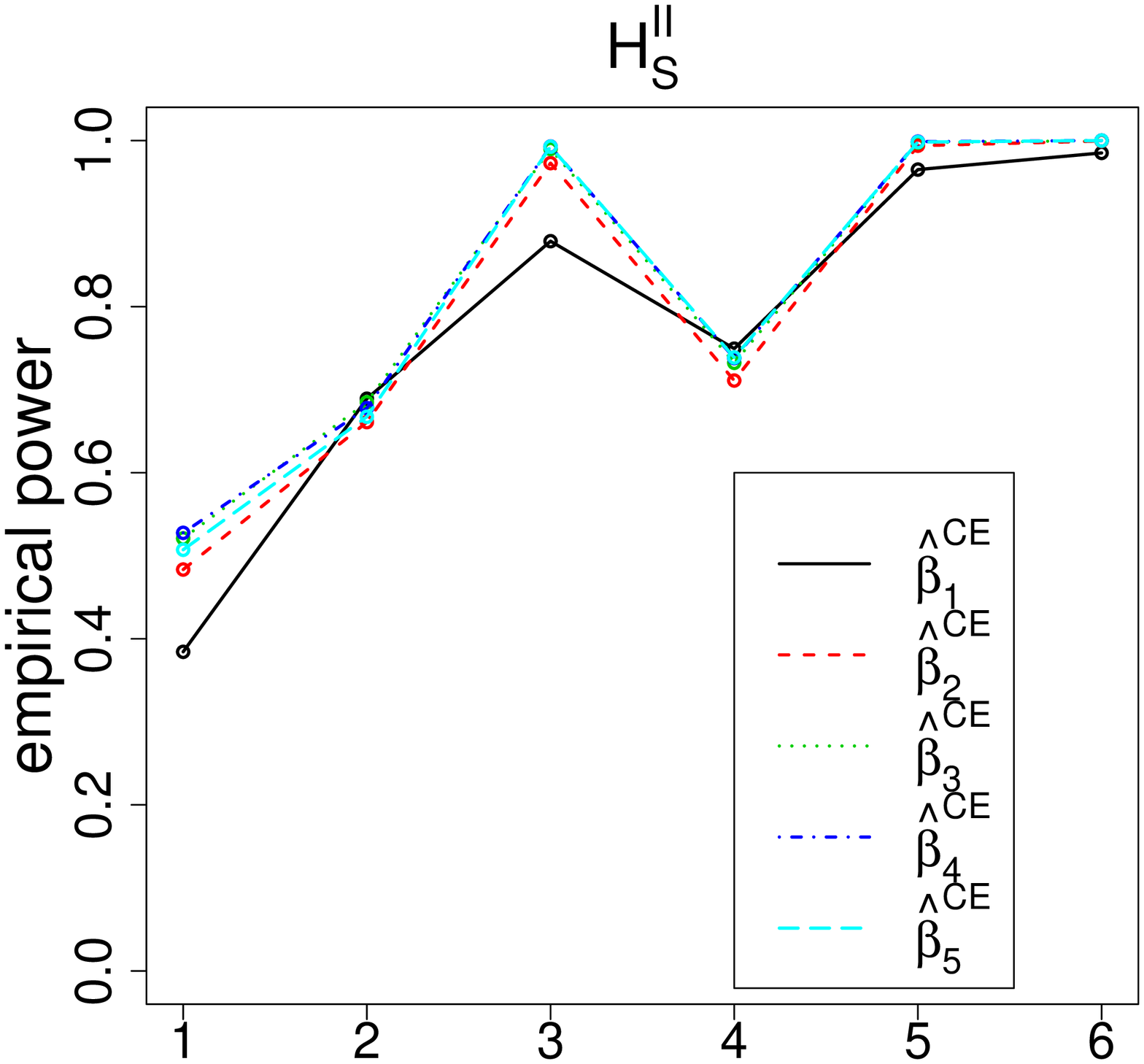} }}
\rotatebox{-90}{ \resizebox{2.1 in}{!}{\includegraphics{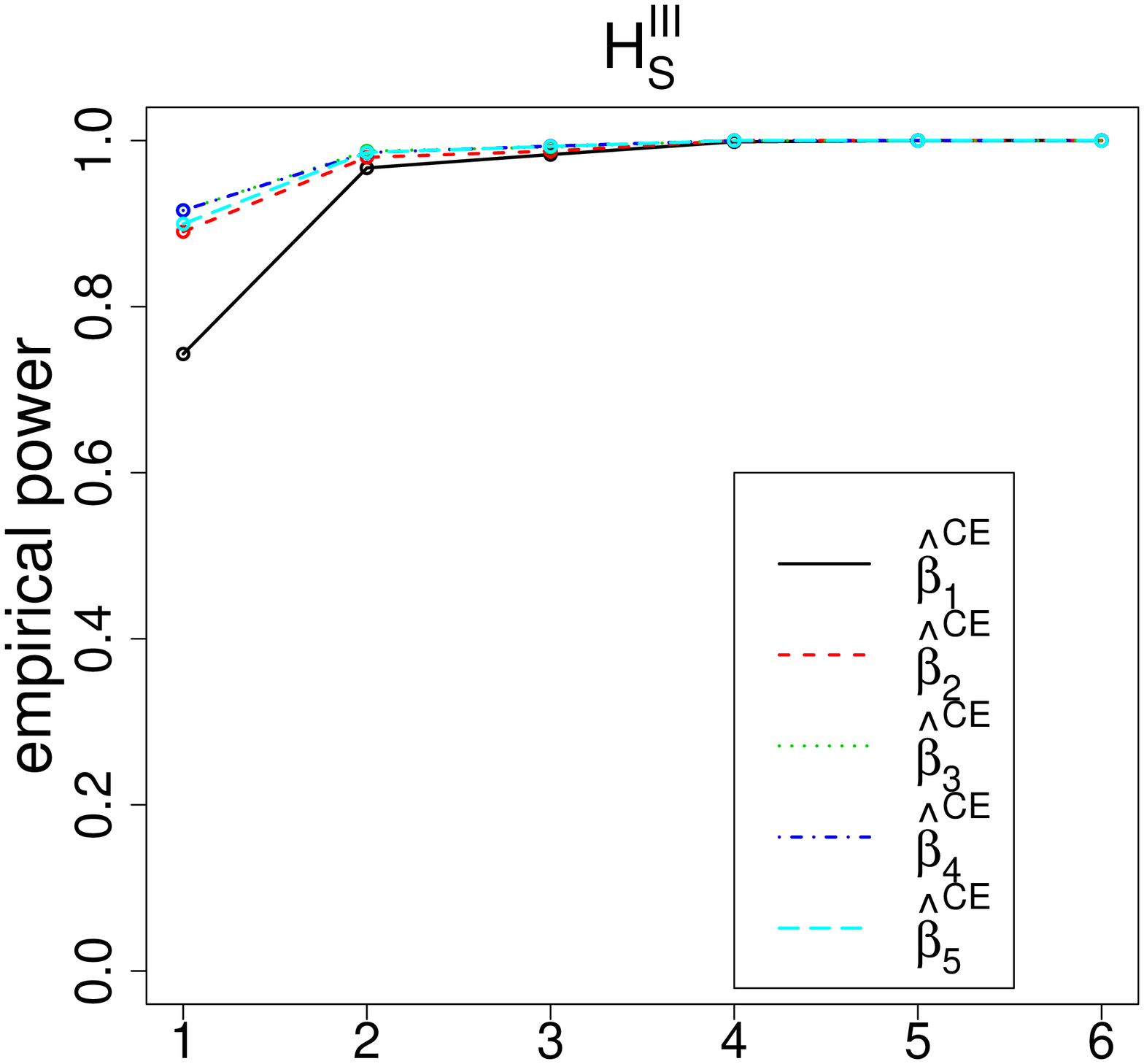} }}
Empirical power estimates for Cuzick-Edward's combined tests under $H_S$\\
\rotatebox{-90}{ \resizebox{2.1 in}{!}{\includegraphics{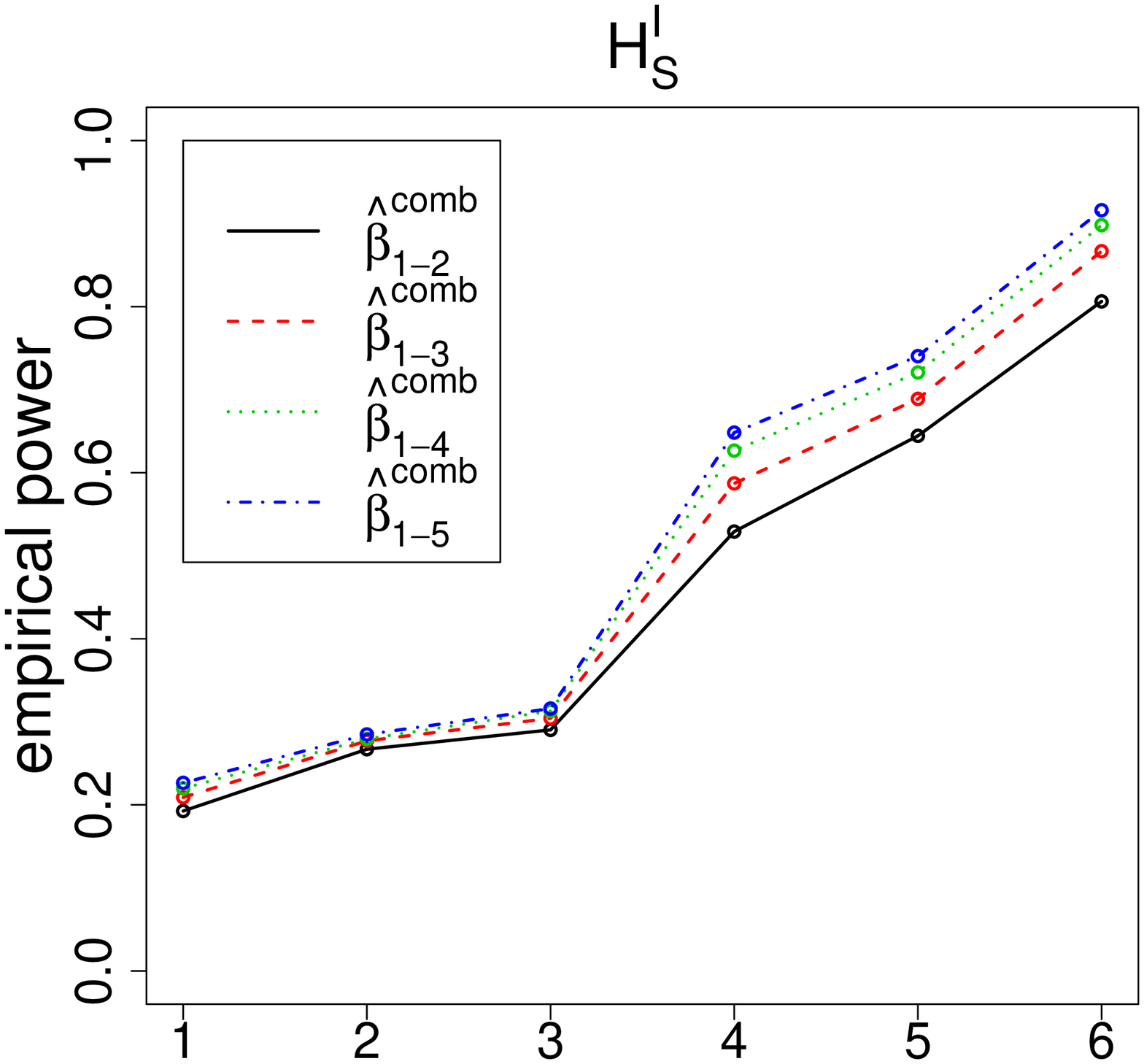} }}
\rotatebox{-90}{ \resizebox{2.1 in}{!}{\includegraphics{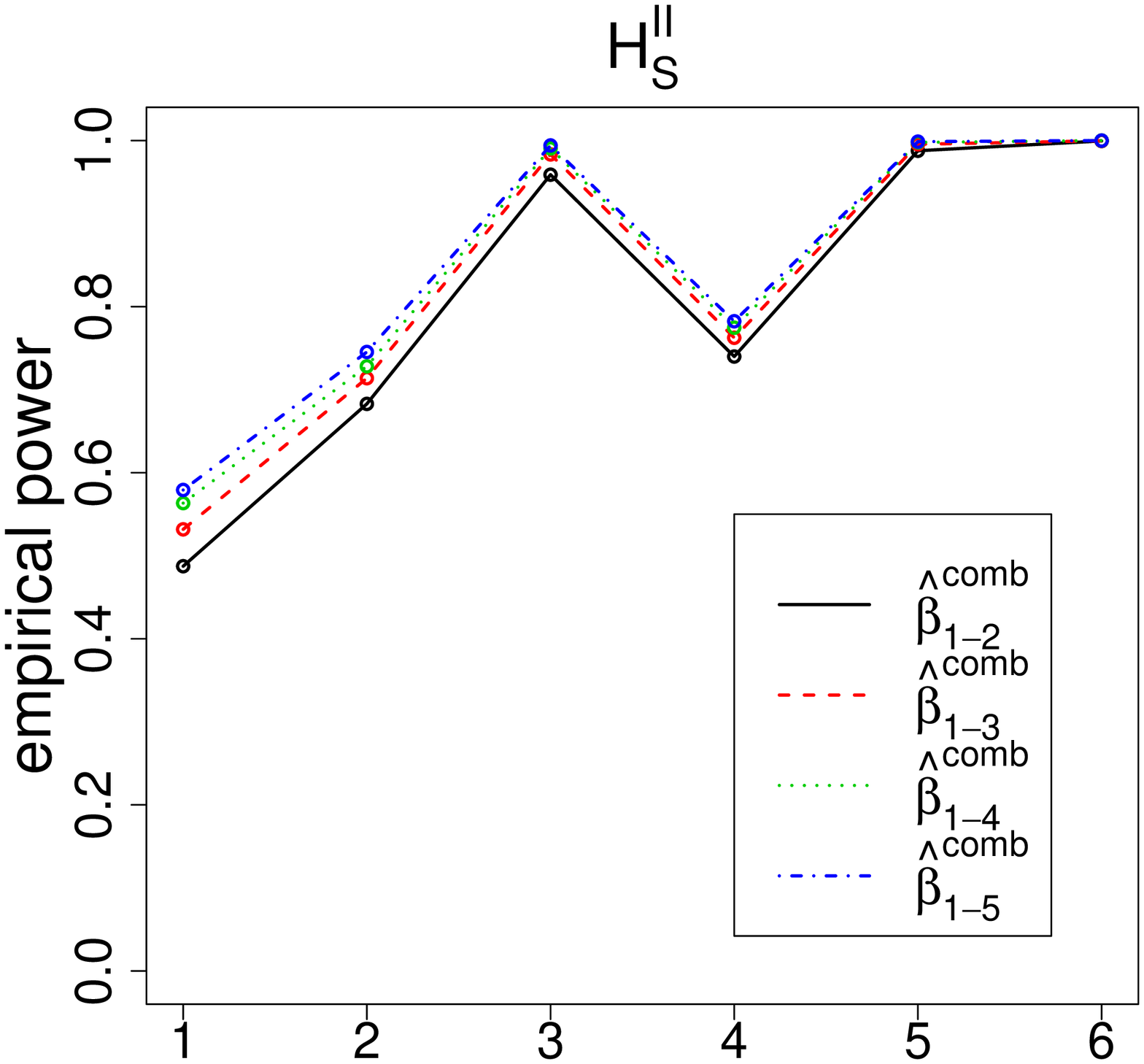} }}
\rotatebox{-90}{ \resizebox{2.1 in}{!}{\includegraphics{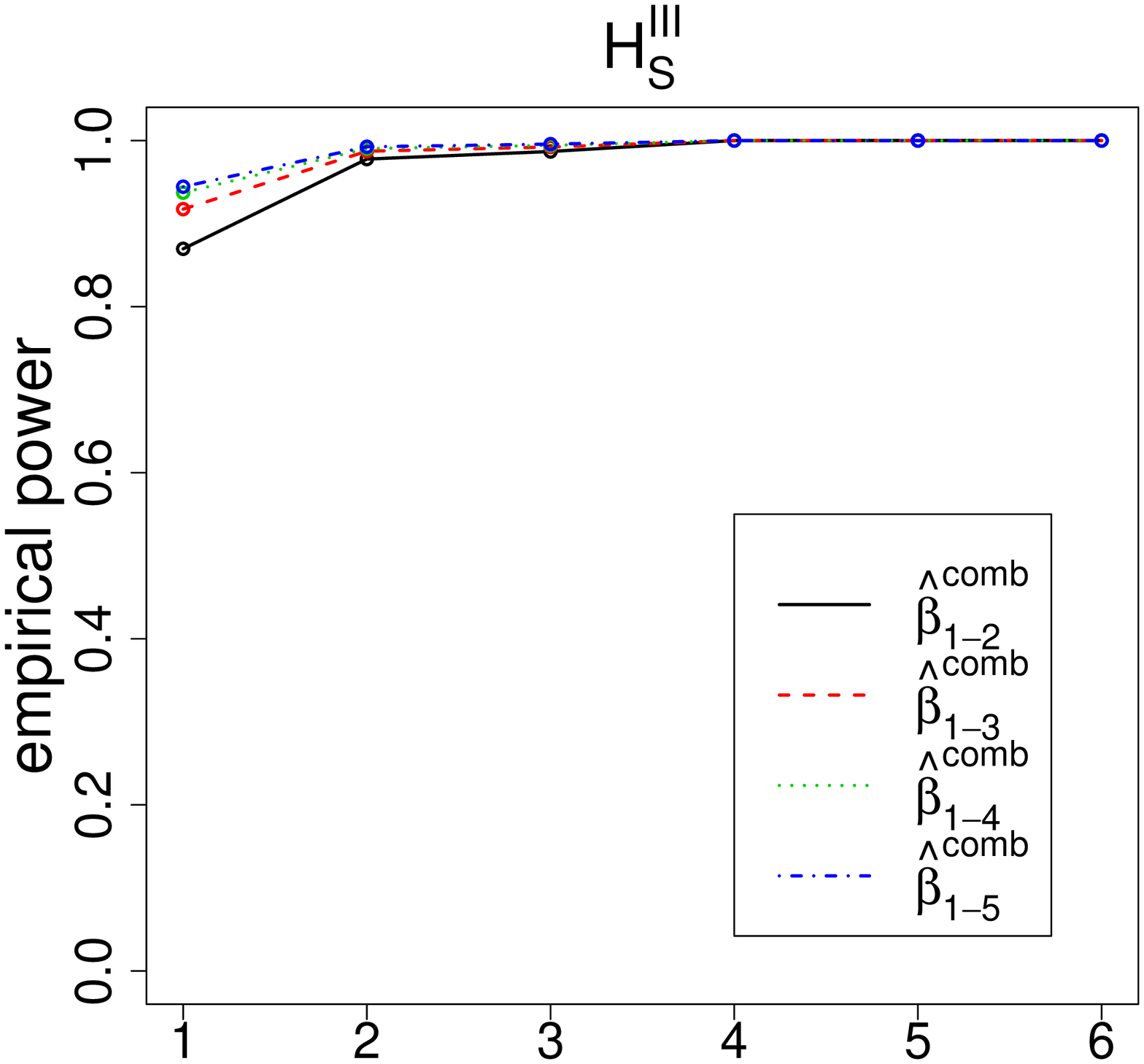} }}
\caption{
\label{fig:Power-Est-Seg}
Empirical power estimates for the NNCT-tests, Cuzick-Edward's $k$-NN tests for $k=1,2,\ldots,5$,
and $T^{comb}_{1-j}$ tests for $j=1,2,3,4$
based on 10000 Monte Carlo replications under the segregation alternatives.
The numbers in the horizontal axis labels represent sample (i.e., class) size combinations:
1=(10,10), 2=(10,30), 3=(10,50), 4=(30,30), 5=(30,50), 6=(50,50).
$\bh_D$, $\bh_I$, $\bh_{II}$, $\bh_{III}$, and $\bh_{P,mc}$
are as in Table \ref{tab:emp-power-seg}.
$\bh^{CE}_k$ stands for Cuzick-Edwards $k$-NN test for $k=1,2,\ldots,5$,
and $\bh^{comb}_{1-j}$ for Cuzick-Edwards $T^{comb}_{1-j}$ for $j=1,2,3,4$.}
\end{figure}

Among Cuzick-Edwards $k$-NN tests,
$T_k$ with $k>1$ have about the same power which is larger
than that of $T_1$ for large samples.
In particular, $T_4$ and $T_5$ have the highest power estimates
which are virtually indistinguishable.
The power estimates for $T_k$ tests seem to be higher
than those of NNCT-tests presented here.
As for the $T^{comb}_S$ tests,
their power estimates are higher than the individual $T_k$ tests,
and the power estimate increases as $j$ increases in $T^{comb}_{1-j}$,
i.e., the more successive $T_k$ tests are combined from $1,2,\ldots,k$,
the higher the power estimates for $T^{comb}_S$.

Considering the power estimates and empirical size performances,
$T^{comb}_{1-4}$ or $T^{comb}_{1-5}$ have the best performance,
hence either can be recommended against the segregation alternatives.
However given the computational cost of $T^{comb}_S$ tests for
larger $k$ values,
we recommend $T_k$ with $k=4$ or $5$ for the segregation alternatives.

\subsection{Empirical Power Analysis under the Association Alternatives}
\label{sec:emp-power-assoc}
For the association alternatives (against the CSR independence pattern), we also consider three cases.
First, we generate $X_i \stackrel{iid}{\sim} \U((0,1)\times(0,1))$ for $i=1,2,\ldots,n_1$.
Then we generate $Y_j$ for $j=1,2,\ldots,n_2$ as follows.
For each $j$, we pick an $i$ randomly, then generate $Y_j$ as
$X_i+R_j\,(\cos T_j, \sin T_j)'$ where
$R_j \stackrel{iid}{\sim} \U(0,r)$ with $r \in (0,1)$ and
$T_j \stackrel{iid}{\sim} \U(0,2\,\pi)$.
In the pattern generated, appropriate choices of
$r$ will imply $Y_j$ and $X_i$ are more associated than expected.
That is, it will be  more likely to have $(X,Y)$
NN pairs than self NN pairs (i.e., $(X,X)$ or $(Y,Y)$).
The three values of $r$ we consider constitute
the three association alternatives:
\begin{equation}
\label{eqn:assoc-alt}
H_A^{I}: r=1/4,\;\;\; H_A^{II}: r=1/7, \text{ and } H_A^{III}: r=1/10.
\end{equation}
Observe that, from $H_A^I$ to $H_A^{III}$ (i.e., as $r$ decreases),
the association gets stronger
in the sense that $X$ and $Y$ points tend to occur
together more and more frequently.
By construction,
$X$ points are from a homogeneous Poisson process with respect to the unit square,
while $Y$ points exhibit inhomogeneity in the same region.
Furthermore, these alternative patterns are examples of
departures from second-order homogeneity which implies
association of the class $Y$ with class $X$.
The simulated association patterns are contrary to the case/control framework
of Cuzick-Edward's tests, since class $X$ is used for the case class and
class $Y$ points are clustered around $X$ points.
However, we still include Cuzick-Edward's tests
to evaluate their performance under this type of deviation from the CSR independence pattern.

\begin{figure}[ht]
\centering
\rotatebox{-90}{ \resizebox{2.1 in}{!}{\includegraphics{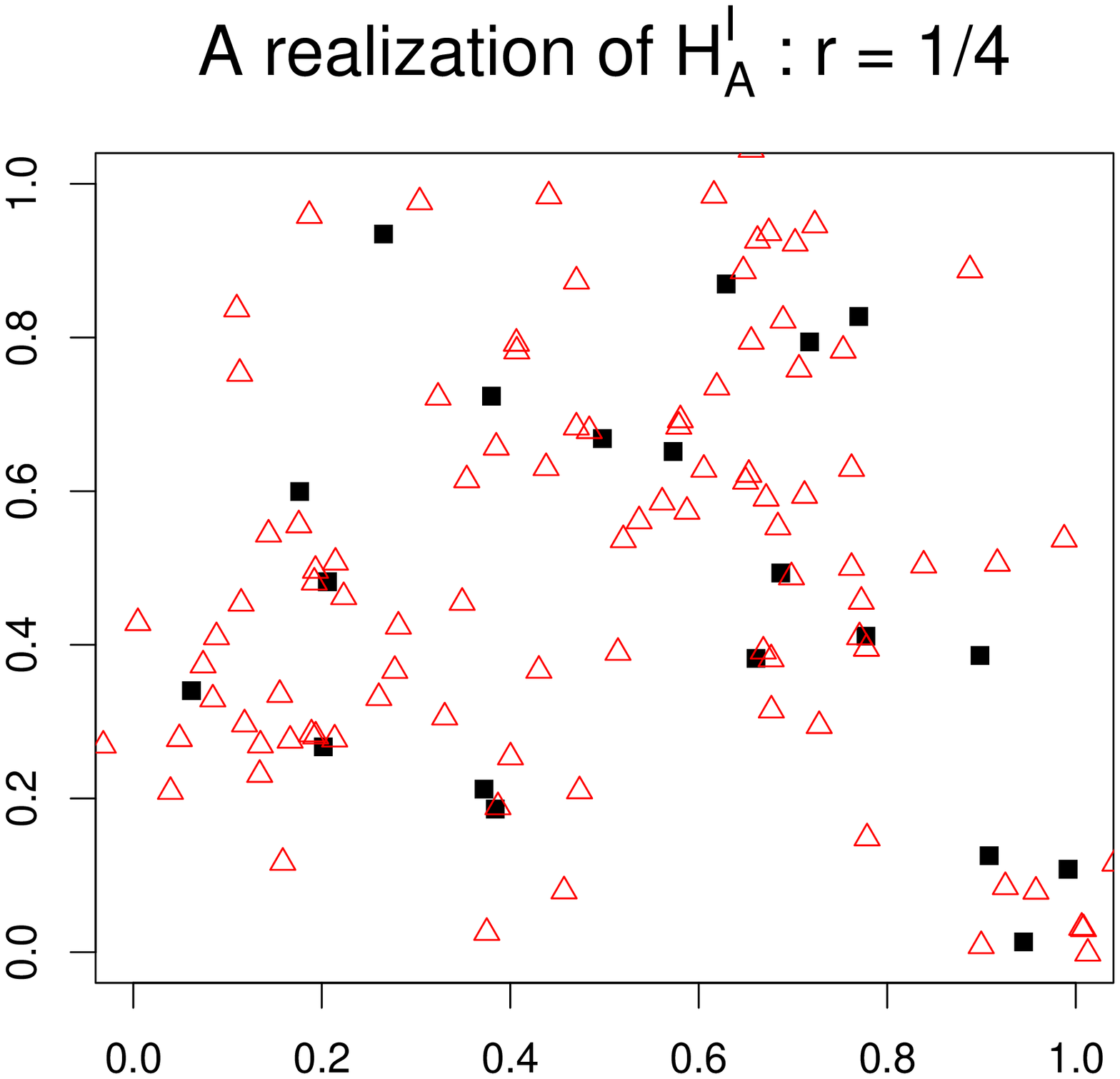} }}
\rotatebox{-90}{ \resizebox{2.1 in}{!}{\includegraphics{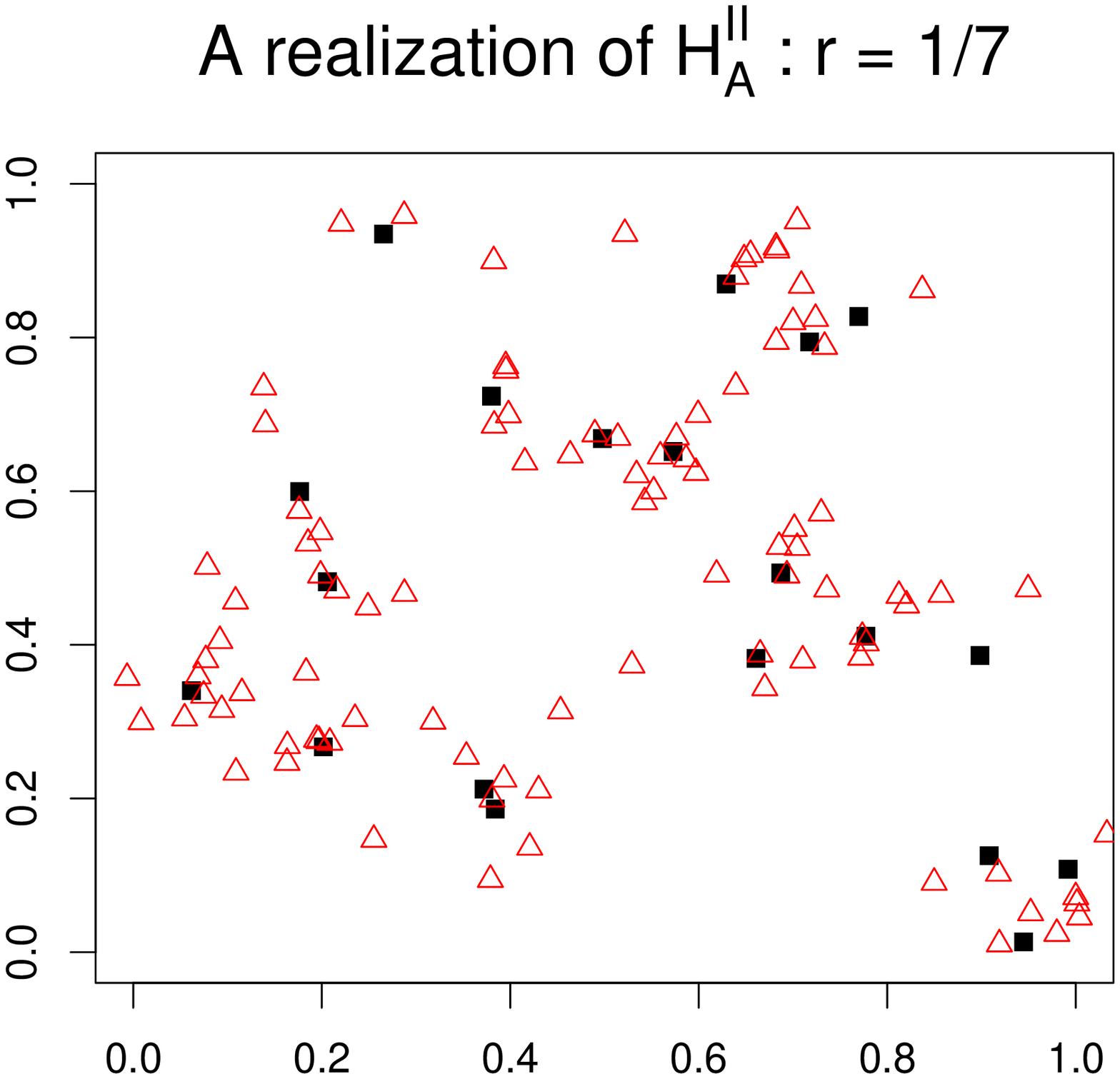} }}
\rotatebox{-90}{ \resizebox{2.1 in}{!}{\includegraphics{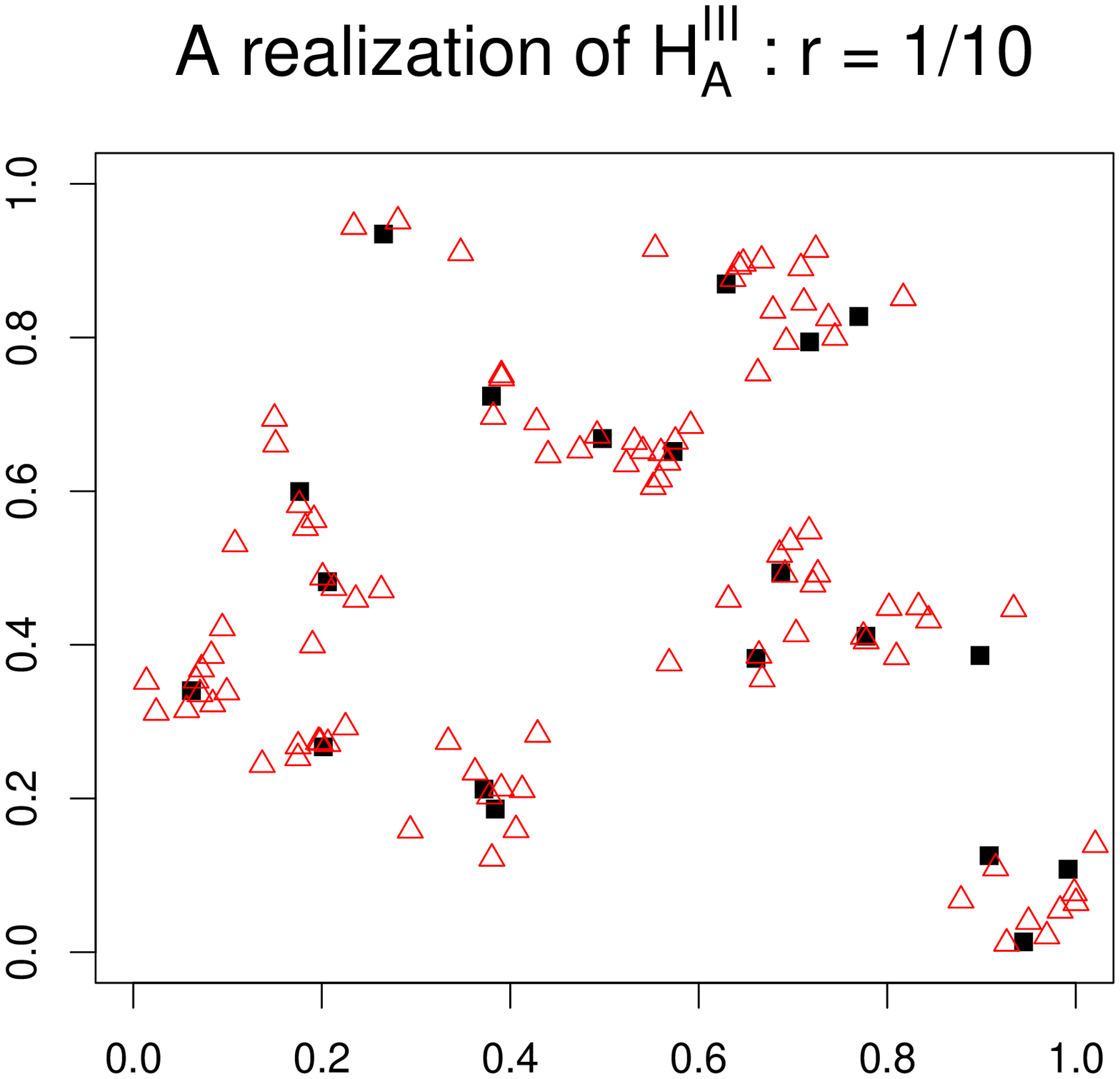} }}
 \caption{
\label{fig:AssocAlt}
Three realizations for $H_A^{I}: r=1/4$ (left), $H_A^{II}: s=1/7$ (middle),
and $H_A^{III}: r=1/10$ (right) with $n_1=20$ $X$ points (solid squares $\blacksquare$)
and $n_2=100$ $Y$ points (triangles $\triangle$).}
\end{figure}

\begin{table}[]
\centering
\begin{tabular}{|c|c|c|c|c|c|c|}
\hline
\multicolumn{7}{|c|}{Empirical power estimates under} \\
\multicolumn{7}{|c|}{the association alternatives} \\
\hline
 & $(n_1,n_2)$  & $\bh_D$ & $\bh_I$ & $\bh_{II}$ & $\bh_{III}$ & $\bh_{P,mc}$ \\
\hline
\hline

 & $(10,10)$ & .1105 & .2638 & .1670 & .1663 & .2690\\
\cline{2-7}
 & $(10,30)$ & .3007 & .3639 & .3532 & .2602 & .2497\\
\cline{2-7}
 & $(10,50)$ & .3318 & .1655 & .3656 & .0870 & .0026\\
\cline{2-7} \raisebox{4.ex}[0pt]{$H_A^I$}
 & $(30,30)$ & .1697 & .2643 & .2021 & .2082 & .2663\\
\hline
\hline

 & $(10,10)$ & .1834 & .4362 & .2596 & .2871 & .4422\\
\cline{2-7}
 & $(10,30)$ & .4956 & .6155 & .5782 & .4964 & .4847\\
\cline{2-7}
 & $(10,50)$ & .5500 & .3645 & .5820 & .2352 & .0112\\
\cline{2-7} \raisebox{4.ex}[0pt]{$H_A^{II}$} &
$(30,30)$    & .4141 & .6037 & .4663 & .5243 & .6070\\
\hline
\hline

 & $(10,10)$ & .2222 & .5068 & .2988 & .3448 & .5138\\
\cline{2-7}
 & $(10,30)$ & .6003 & .7232 & .6811 & .6148 & .6037\\
\cline{2-7}
 & $(10,50)$ & .6512 & .4753 & .6776 & .3267 & .0203\\
\cline{2-7} \raisebox{4.ex}[0pt]{$H_A^{III}$}
 & $(30,30)$ & .6157 & .7912 & .6667 & .7283 & .7962\\
\hline

\end{tabular}
\caption{
\label{tab:emp-power-assoc}
The empirical power estimates for the tests under the association alternatives
$H_A^I-H_A^{III}$
with $N_{mc}=10000$, for some combinations of $n_1,n_2 \in
\{10,30,50\}$ at $\alpha=.05$.
The empirical power labeling is as in Table \ref{tab:emp-power-seg}.}
\end{table}

The empirical power estimates for NNCT-tests for $(n_1,n_2)\in \{(10,10),(10,30),(10,50),(30,30)\}$
are provided in Table \ref{tab:emp-power-assoc}.
The power estimates under the association alternatives
are presented in Figure \ref{fig:Power-Est-Assoc},
where labeling is as in Figure \ref{fig:Power-Est-Seg}.
Observe that, for similar sample sizes as $n=(n_1+n_2)$ gets larger,
the power estimates get larger at each association alternative.
Furthermore, as the association gets stronger,
the power estimates get larger at each sample size combination.
Considering the NNCT-tests, for most sample size combinations,
version I of the new tests has the highest power estimate
(except for $(n_1,n_2)=(10,50)$ in which case, version II of the new tests has the highest power
and most tests perform very poorly, with Monte Carlo corrected version of
Pielou's test being the worst).
Hence considering the empirical size and power estimates,
we recommend version I of the new tests for large samples,
and Monte Carlo randomization for the NNCT-tests for small samples.

\begin{figure}[]
\centering
Empirical power estimates for the NNCT-tests under $H_A$\\
\rotatebox{-90}{ \resizebox{2.1 in}{!}{\includegraphics{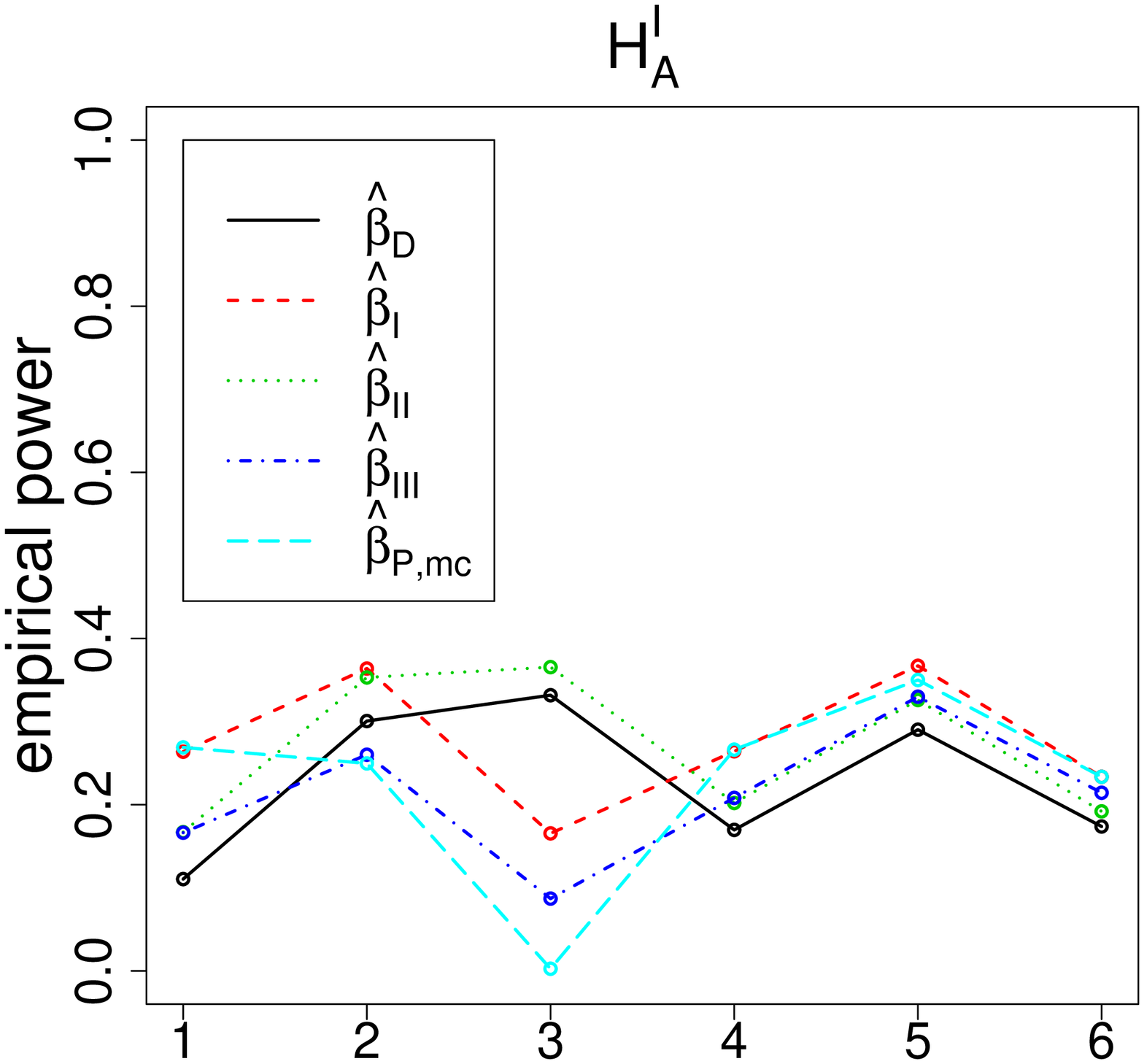} }}
\rotatebox{-90}{ \resizebox{2.1 in}{!}{\includegraphics{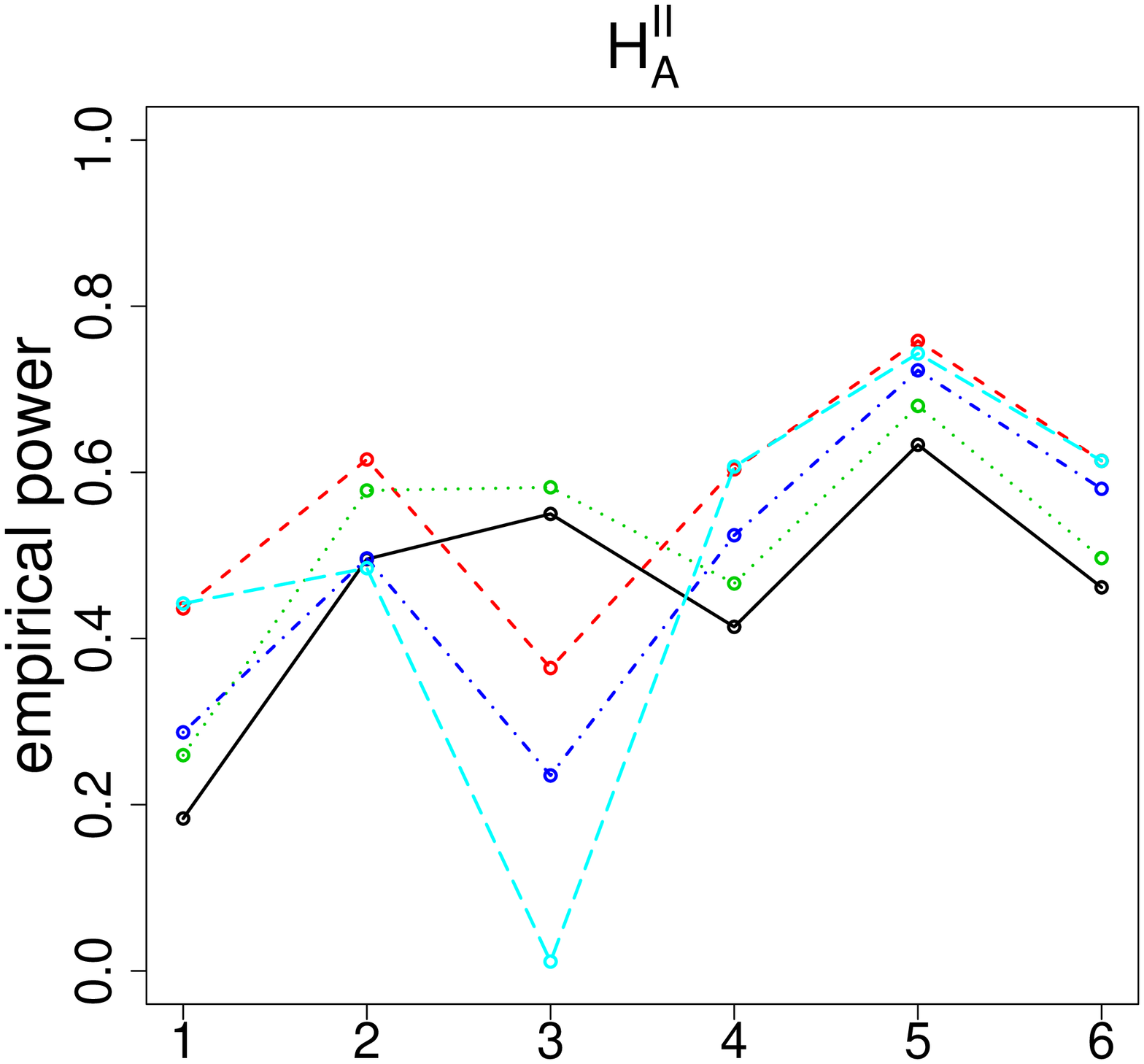} }}
\rotatebox{-90}{ \resizebox{2.1 in}{!}{\includegraphics{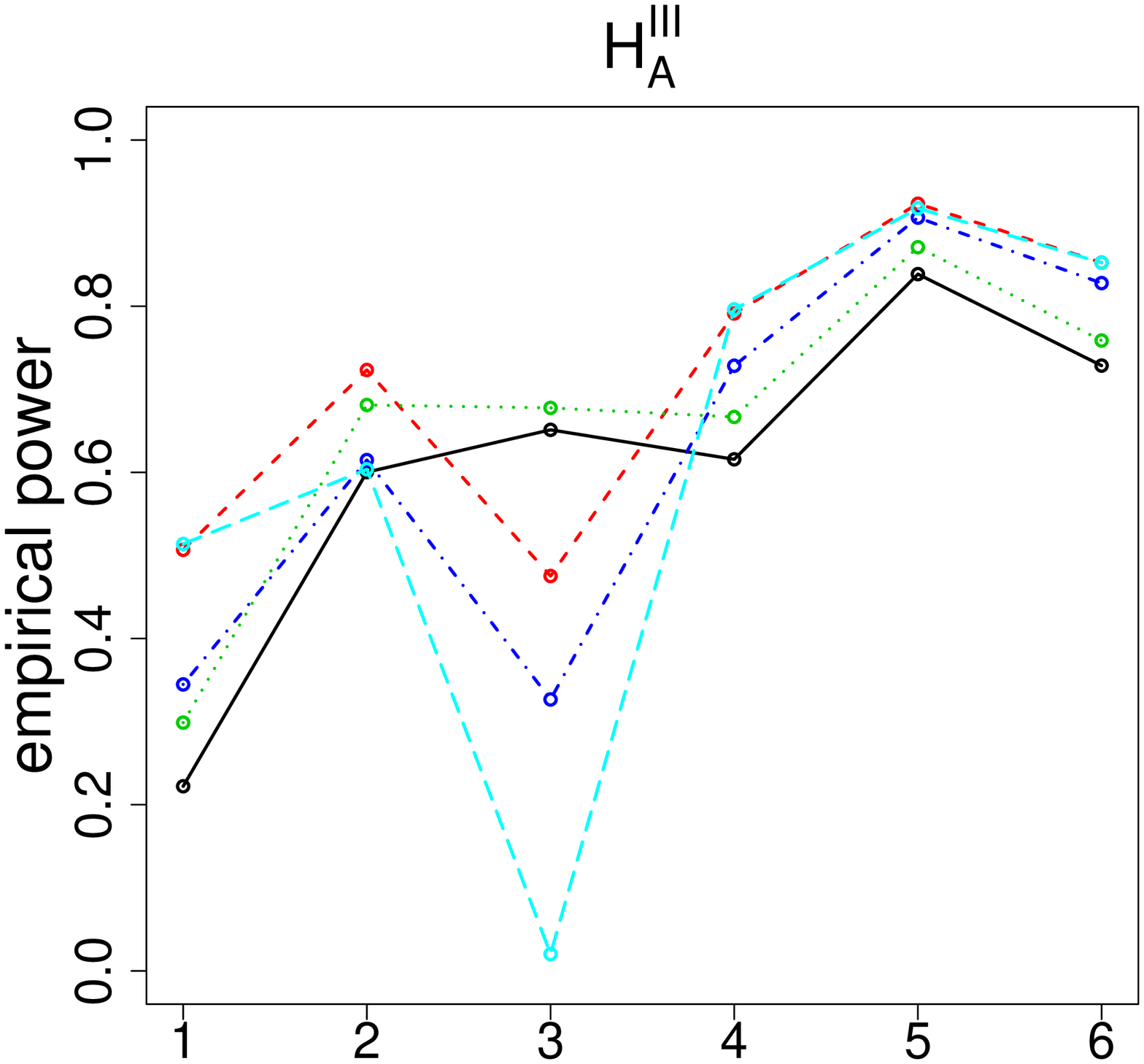} }}
Empirical power estimates for Cuzick-Edward's $k$-NN tests under $H_A$\\
\rotatebox{-90}{ \resizebox{2.1 in}{!}{\includegraphics{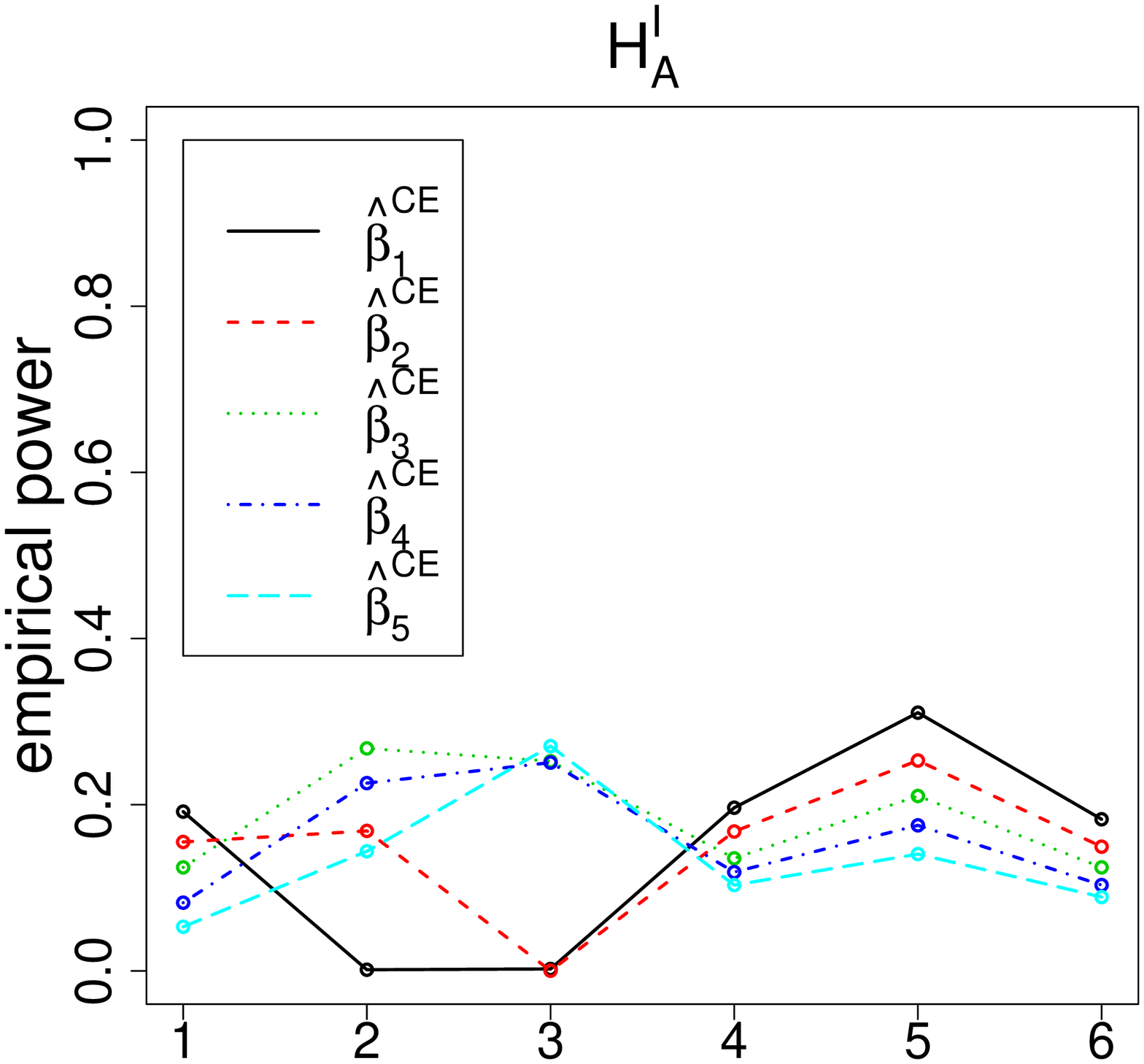} }}
\rotatebox{-90}{ \resizebox{2.1 in}{!}{\includegraphics{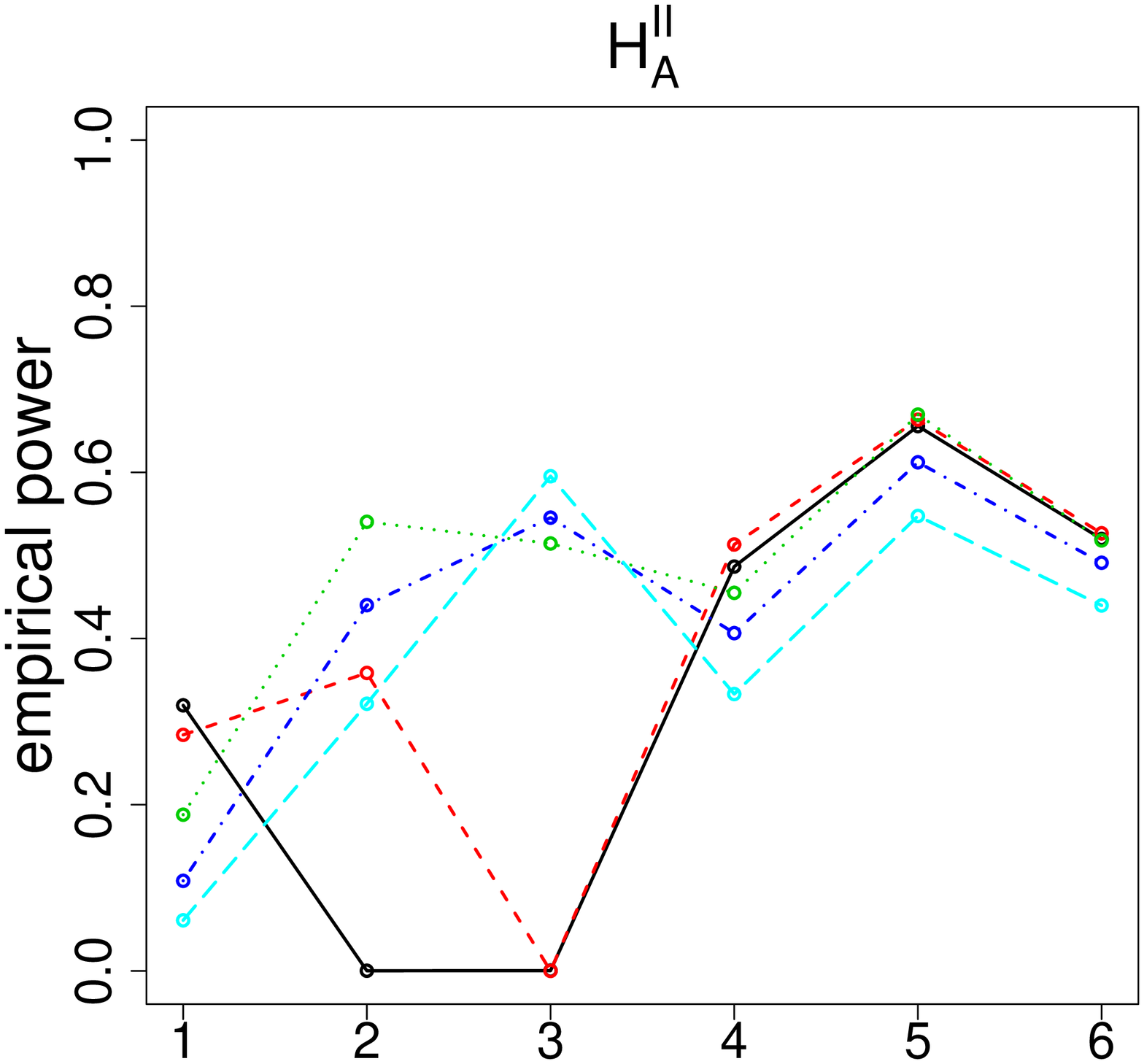} }}
\rotatebox{-90}{ \resizebox{2.1 in}{!}{\includegraphics{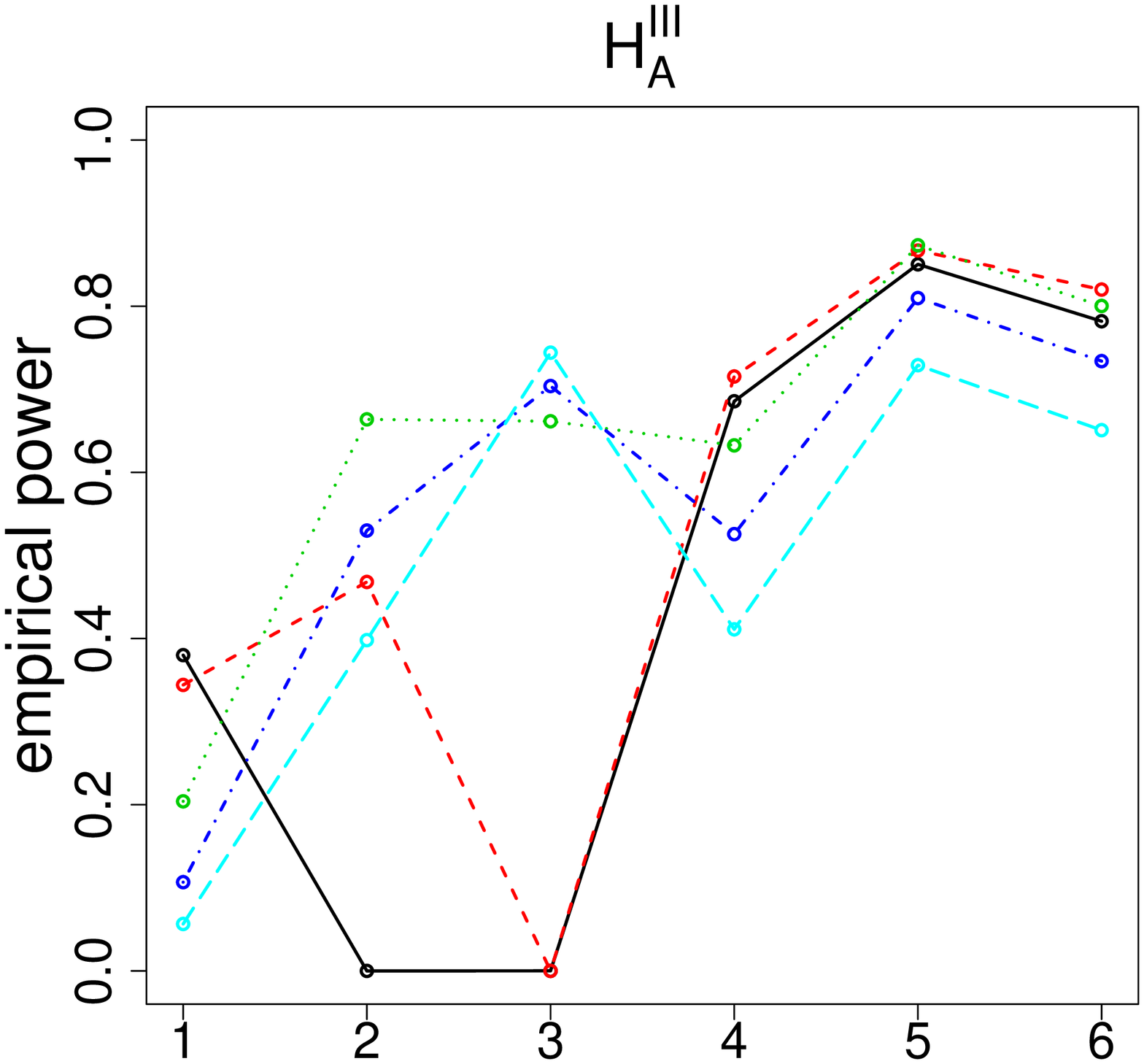} }}
Empirical power estimates for Cuzick-Edward's combined tests under $H_A$\\
\rotatebox{-90}{ \resizebox{2.1 in}{!}{\includegraphics{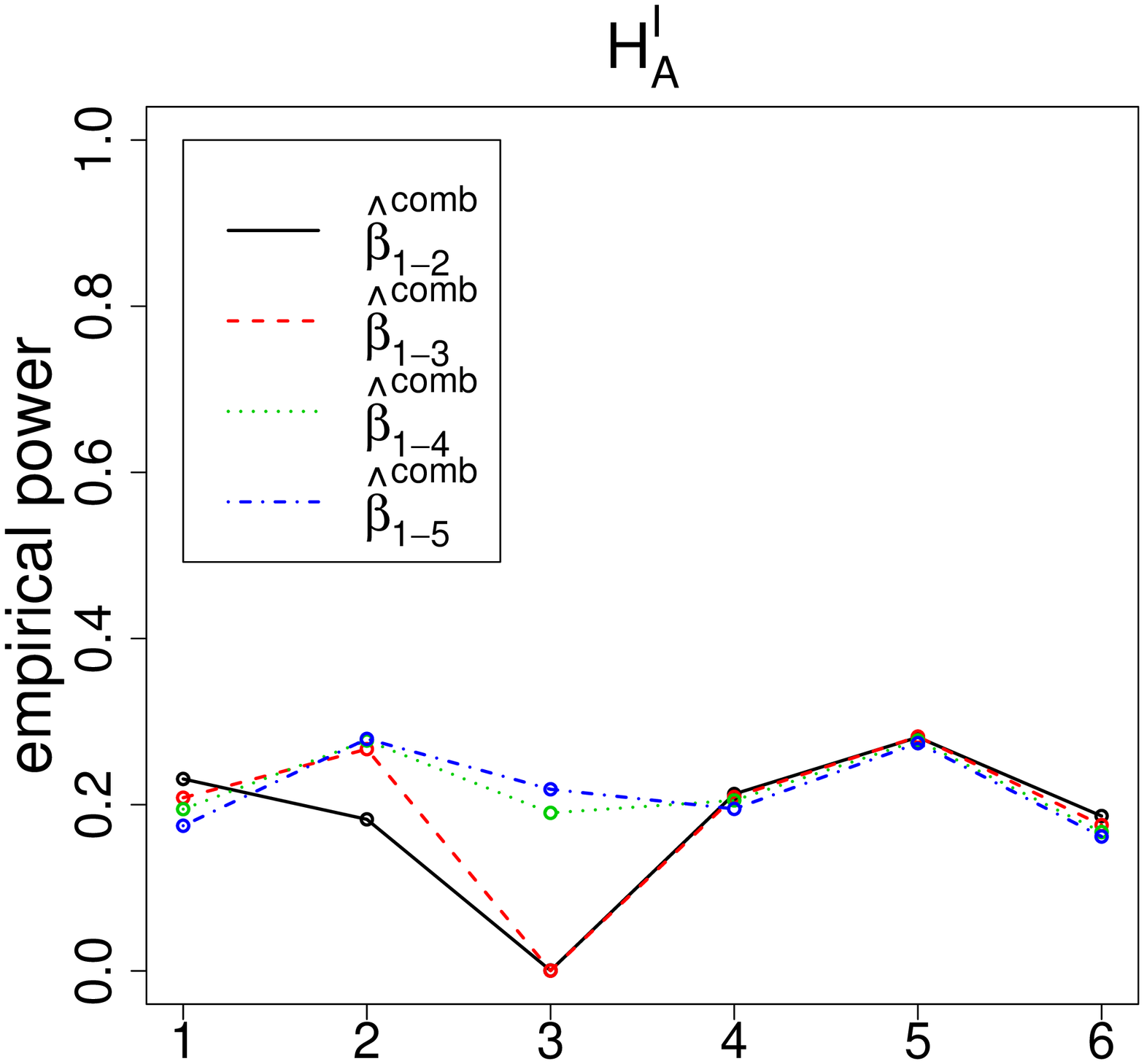} }}
\rotatebox{-90}{ \resizebox{2.1 in}{!}{\includegraphics{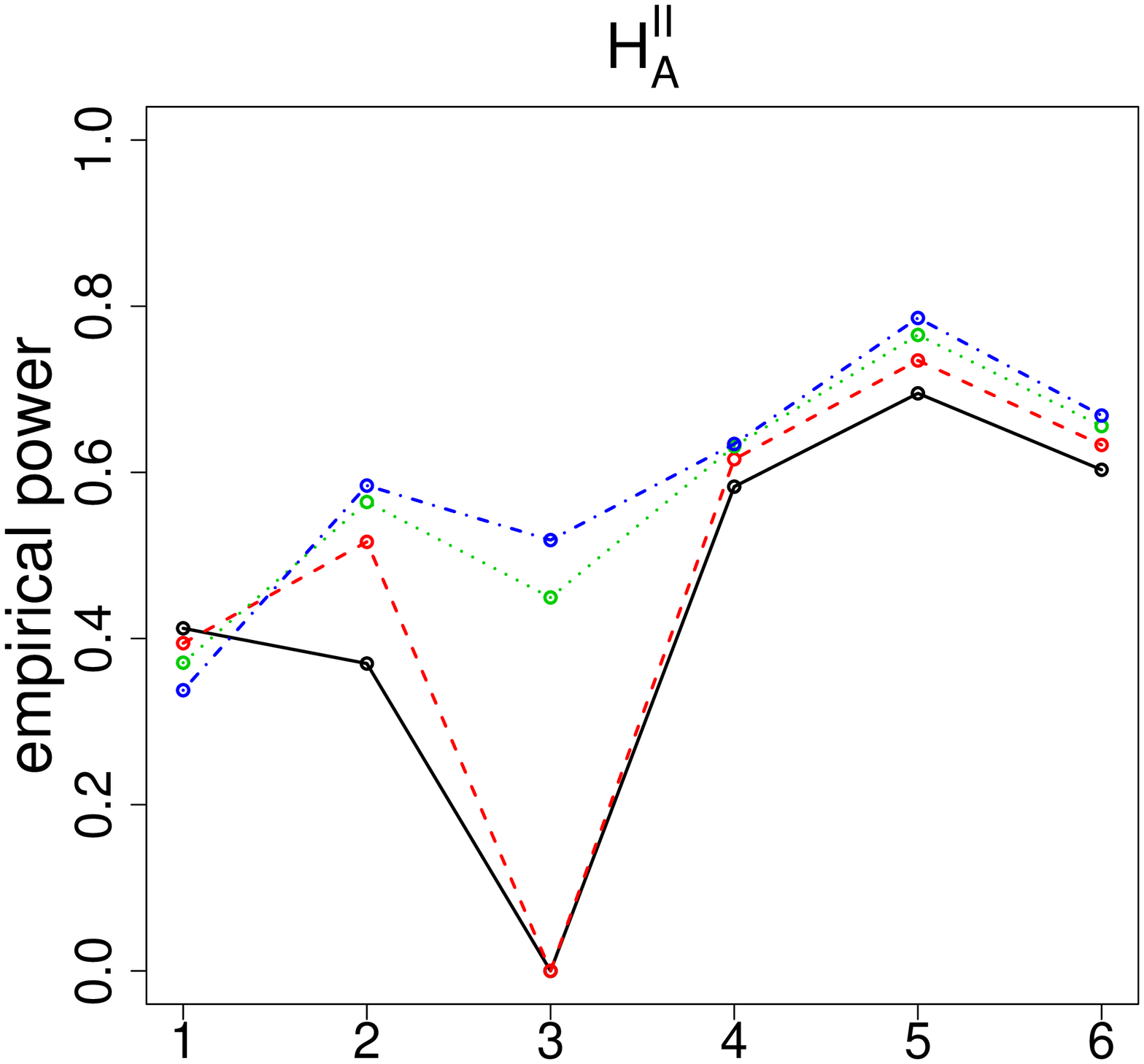} }}
\rotatebox{-90}{ \resizebox{2.1 in}{!}{\includegraphics{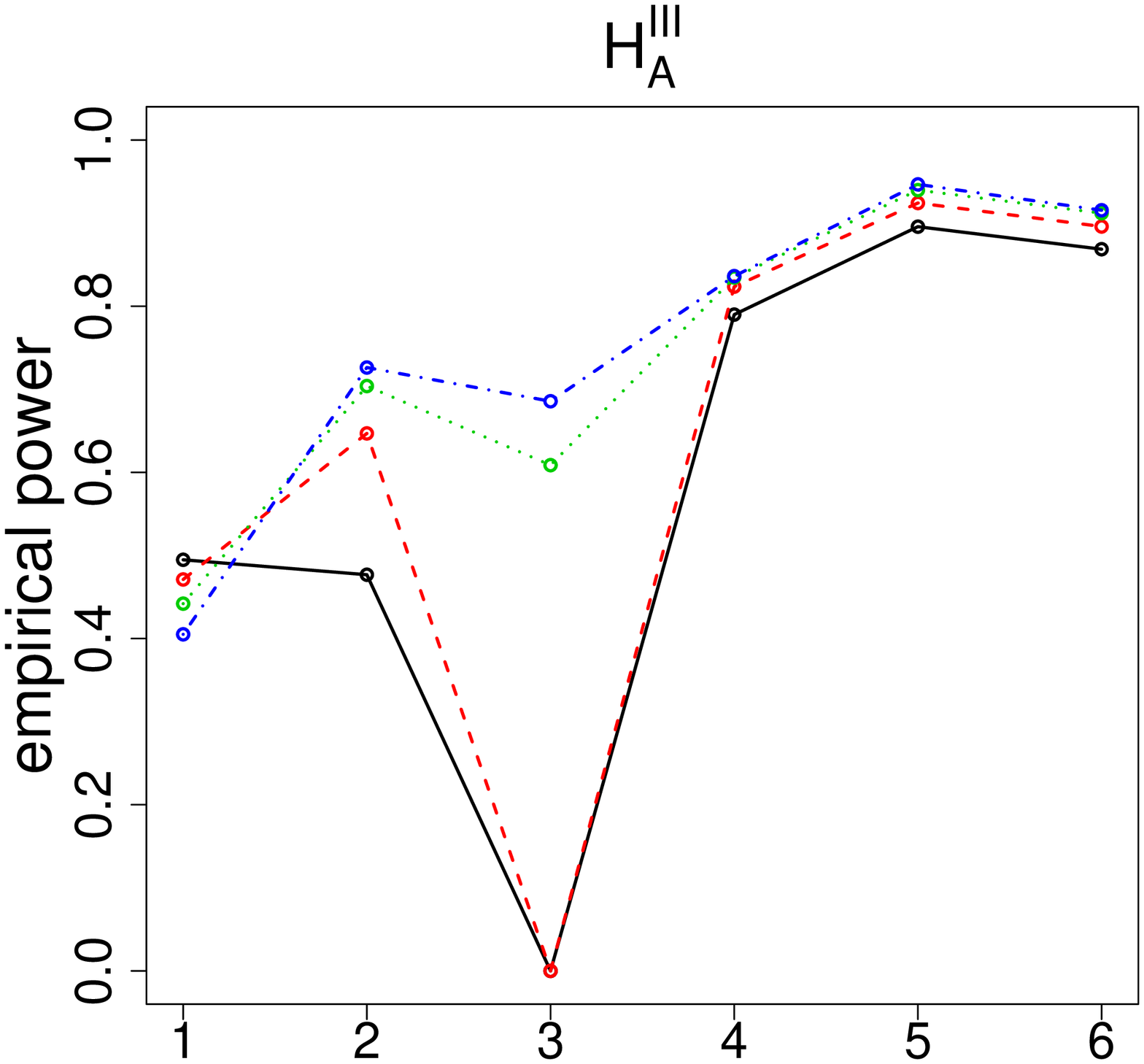} }}
\caption{
\label{fig:Power-Est-Assoc}
Empirical power estimates for the NNCT-tests, Cuzick-Edward's $k$-NN tests
for $k=1,2,\ldots,5$, and $T^{comb}_{1-j}$ tests for $j=1,2,3,4$
under the association alternatives.
The numbers in the horizontal axis labels represent sample (i.e., class) size combinations:
1=(10,10), 2=(10,30), 3=(10,50), 4=(30,30), 5=(30,50), 6=(50,50).
The empirical power labeling is as in Figure \ref{fig:Power-Est-Seg}.}
\end{figure}

Considering Cuzick-Edwards $k$-NN tests,
it is seen that $T_1$ has virtually no power for $(n_1,n_2)=(10,30)$ or $(10,50)$
and $T_2$ has virtually no power for $(n_1,n_2)=(10,50)$.
Under $H_A^I$, $T_1$ has the highest power estimate,
while under other association alternatives, $T_2$ has
higher power estimates for larger sample sizes.
For smaller samples either $T_3$ or Monte Carlo randomization
of the tests can be used.
Considering Cuzick-Edwards $k$-NN tests with NNCT-tests together,
observe that version III of the new overall tests has the best power performance
under the association alternatives.

Among Cuzick-Edwards combined tests,
$T^{comb}_{1-2}$ and $T^{comb}_{1-3}$ have virtually no power for $(n_1,n_2)=(10,50)$.
For almost all sample size combinations, $T^{comb}_{1-5}$ has the highest power estimates.
Considering all tests together, still $T^{comb}_{1-5}$ has the best power performances
under the association alternatives,
hence can be recommended for use against this type of association.
However, given the computational cost of combined tests,
one might prefer version III of the new tests under the association alternatives for larger samples,
as its power is very close to that of $T^{comb}_{1-5}$.
For smaller samples, either Monte Carlo randomization for NNCT-tests,
or asymptotic approximation for $T^{comb}_{1-4}$ or $T^{comb}_{1-5}$ can be used.

\begin{remark}
\label{rem:edge-correct}
\textbf{Edge Correction for NNCT-Tests:}
Edge (or boundary) effects are not a
concern for testing against the RL pattern.
However, the CSR independence pattern assumes that the study region
is unbounded for the analyzed pattern,
which is not the case in practice.
So edge effects are a constant problem in the analysis of
empirical (bounded) data sets
if the null pattern is the CSR independence
and much effort has gone into the
development of edge corrections methods
(\cite{yamada:2003} and \cite{dixon:EncycEnv2002}).

Two correction methods to mitigate the edge effects on NNCT-tests,
namely, buffer zone correction and toroidal correction,
are investigated in (\cite{ceyhan:segreg-edge-correctAS2007,ceyhan:overall}) where it is shown that the empirical sizes of the
NNCT-tests are not affected by the toroidal edge correction under CSR independence.
On the other hand, the toroidal
correction has a mild influence on the results provided that there
are no clusters around the edges.
Furthermore, toroidal correction
(slightly) improves the results of some of the segregation tests based on NNCTs.
However, toroidal correction is biased for non-CSR patterns.
In particular if the pattern outside the plot (which is often unknown) is not the same as that inside it
it yields questionable results (\cite{haase:1995} and \cite{yamada:2003}).
The bias is more severe especially when the edges cut through some cluster(s).
The (outer) buffer zone edge correction method
seems to have slightly stronger influence on the tests compared to toroidal correction.
But for these tests, buffer zone correction
does not change the sizes significantly for most sample size combinations.
This is in agreement with the findings of \cite{barot:1999}
who say NN methods only require a small buffer area around the study region.
A large buffer area does not help much since one only
needs to be able to see far enough away from an event to find its NN.
Once the buffer area extends past the likely NN distances
(i.e., about the average NN distances),
it is not adding much helpful information for NNCTs.
Hence we recommend inner or outer buffer zone correction for NNCT-tests
with the width of the buffer area being about the average NN distance.
We do not recommend larger buffer areas,
since they are wasteful with little additional gain.
$\square$
\end{remark}

\section{Examples}
\label{sec:examples}
We illustrate the tests on three examples:
two ecological data sets, namely
Pielou's Douglas-fir/ponderosa pine data (\cite{pielou:1961})
and swamp tree data (\cite{good:1982}),
and an epidemiological data set, namely leukemia data set (\cite{diggle:2003}).

\subsection{Pielou's Data}
\label{sec:pielou-data}
Pielou used a completely mapped data set that is
comprised of ponderosa pine (\emph{Pinus Ponderosa})
and Douglas-fir trees (\emph{Pseudotsuga menziesii}
formerly \emph{P. taxifolia}) from a region in
British Columbia (\cite{pielou:1961}).
Her data are also used by
Dixon as an illustrative example (\cite{dixon:1994}).
Since the data consist of individuals of different species,
it is more reasonable to assume CSR independence as the underlying pattern
for the null hypothesis of randomness in NN structure.
Deviation from CSR independence implies that the two classes are a priori
the result of different processes.
The question of interest is whether the two tree species are segregated,
associated, or do not significantly deviate from CSR independence.
The corresponding NNCT and
the percentages are provided in Table \ref{tab:NNCT-pielou}.
The percentages for the cells are based on the size of each tree species.
For example, 86 \% of Douglas-firs have NNs from
Douglas firs, and remaining 15 \% have NNs are from ponderosa pines.
The row and column percentages are marginal percentages with respect to the
total sample size.
The percentage values in the diagonal cells are suggestive of segregation for both species.

\begin{table}[ht]
\centering
\begin{tabular}{cc|cc|c}
\multicolumn{2}{c}{}& \multicolumn{2}{c}{NN}& \\
\multicolumn{2}{c}{}&    D.F. &  P.P.   &   sum  \\
\hline
&D.F.&    137 (86 \%)  &   23 (15 \%)    &   160 (70 \%)  \\
\raisebox{1.5ex}[0pt]{base}
&P.P. &    38 (56 \%) &  30 (44 \%)   &   68 (30 \%)  \\
\hline
&sum     &    175 (77 \%)  & 53 (23 \%)             &  228 (100 \%) \\
\end{tabular}
\caption{
\label{tab:NNCT-pielou}
The NNCT for Pielou's data and the corresponding percentages (in parentheses).
D.F.= Douglas-fir, P.P.= ponderosa pine.}
\end{table}

The raw data are not available, but fortunately,
\cite{pielou:1961} provided $Q=162$ and $R=134$.
Hence, we could calculate the NNCT-test statistics
which are provided in Table \ref{tab:ex-NNCT-test-stat}
where $C_D$ stands for Dixon's test of segregation,
$\X^2_P$ for Pielou's test,
$\X^2_{P,mc}$ for Pielou's test by Monte Carlo simulations,
$\X_I^2$ for version I as in Equation (\ref{eqn:X-square-I}),
$\X_{II}^2$ as in Equation (\ref{eqn:X-square-II}),
and $\X_{III}^2$ as in Equation (\ref{eqn:X-square-III}).
The $p$-values are also provided below the test statistics in parentheses.
Observe that all of the tests are significant,
implying significant deviation from independence in NN structure (hence CSR independence),
and the percentages in the NNCT imply
that there is significant segregation for both species.
On the other hand, since the raw data is not available,
neither Cuzick-Edward's $k$-NN and combined tests
nor Ripley's $K$ or $L$-functions and pair correlation functions can be calculated.

\begin{table}[ht]
\centering
\begin{tabular}{|c||c|c||c|c|c|c|}
\hline
\multicolumn{7}{|c|}{NNCT-test statistics and the associated $p$-values} \\
\hline
Data & $\X^2_P$ & $C_D$ & $\X_I^2$ & $\X_{II}^2$ & $\X_{III}^2$ & $\X^2_{P,mc}$ \\
\hline
Pielou's & 23.66 & 19.67 & 12.73 & 19.29 & 13.09 & 14.41 \\
Data &($<.0001$) & (.0001) & (.0004) & (.0001) & (.0003) & (.0001) \\
\hline
Swamp Tree & 212.20 & 133.48 & 132.13 & 132.42 & 133.20 & 129.16 \\
Data & ($<.0001$) & ($<.0001$) & ($<.0001$) & ($<.0001$) & ($<.0001$) & ($<.0001$) \\
\hline
Leukemia & 3.31 & 2.25 & 1.98 & 2.10 & 2.13 & 2.02 \\
Data & (.0687) & (.3249) & (.1599) & (.3505) & (.1449) & (.1547) \\
\hline
\end{tabular}
\caption{ \label{tab:ex-NNCT-test-stat}
Test statistics and the associated $p$-values (in parentheses)
for NNCT-tests for the example data sets.
$C_D$ stands for Dixon's test of segregation,
$\X^2_P$ for Pielou's test,
$\X^2_{P,mc}$ for Pielou's test by Monte Carlo simulations,
$\X_I^2$ for version I as in Equation (\ref{eqn:X-square-I}),
$\X_{II}^2$ for version II as in Equation (\ref{eqn:X-square-II}),
and $\X_{III}^2$ for version III as in Equation (\ref{eqn:X-square-III}).
}
\end{table}

\subsection{Swamp Tree Data}
\label{sec:swamp-data}
\cite{good:1982} considered the spatial patterns of tree species
along the Savannah River, South Carolina, U.S.A.
From this data, \cite{dixon:NNCTEco2002} used a single 50m $\times$ 200m rectangular plot
to illustrate his tests.
All live or dead trees with 4.5 cm or more dbh (diameter at breast height)
were recorded together with their species.
Hence it is an example of a realization of a marked multi-variate point pattern.
The plot contains 13 different tree species,
four of which comprise over 90 \% of the 734 tree stems.
The remaining tree stems were categorized as ``other trees".
The plot consists of 215 water tupelo (\emph{Nyssa aquatica}),
205 black gum (\emph{Nyssa sylvatica}), 156 Carolina ash (\emph{Fraxinus caroliniana}),
98 bald cypress (\emph{Taxodium distichum}), and 60 stems of 8 additional species (i.e., other species).
We will only consider the three most frequent tree species in this data set
(i.e., water tupelos, black gums, and Carolina ashes).
So a $3 \times 3$ NNCT-analysis is conducted for this data set.
If segregation among the less frequent species were important,
a more detailed $5 \times 5$ or a $12 \times 12$ NNCT-analysis should be performed.
The locations of these trees in the study region are plotted in Figure \ref{fig:SwampTrees}
and the corresponding $3 \times 3$ NNCT together with percentages
based on row and grand sums are provided in Table \ref{tab:NNCT-swamp}.
For example, for black gum as the base species and Carolina ash as the NN species,
the cell count is 31 which is 15 \% of the 205 black gums (which is 36 \% of all trees).
Observe that the percentages and Figure \ref{fig:SwampTrees} are suggestive of segregation for
all three tree species since the observed percentage of species with themselves as the NN is much larger
than the row percentages.

\begin{figure}[ht]
\centering
\rotatebox{-90}{ \resizebox{3.6 in}{!}{\includegraphics{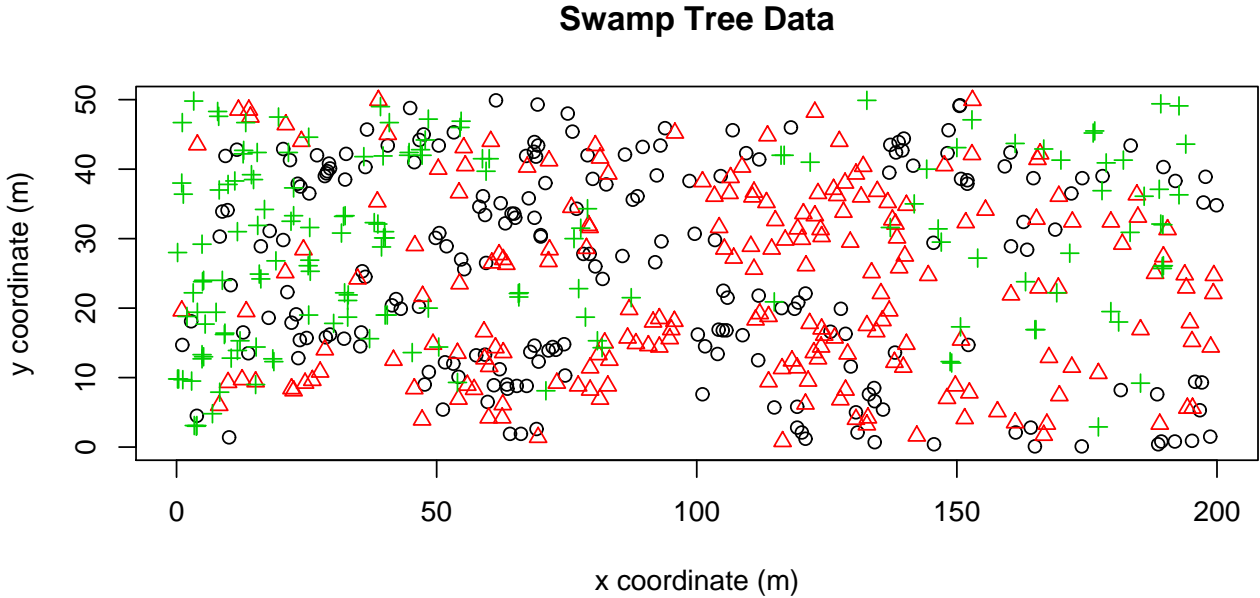} }}
 \caption{
\label{fig:SwampTrees}
The scatter plot of the locations of water tupelos (circles $\circ$),
black gum trees (triangles $\triangle$), and Carolina ashes (pluses $+$).}
\end{figure}

\begin{table}[ht]
\centering
\begin{tabular}{cc|ccc|c}
\multicolumn{2}{c}{}& \multicolumn{3}{c}{NN} & \\
\multicolumn{2}{c}{} & W.T. & B.G. & C.A. &  sum  \\
\hline
& W.T. &    134 (62 \%) &   47 (22 \%) &  34 (16 \%) & 215 (37 \%)\\
& B.G. &    47  (23 \%) &  128 (62 \%) &  31 (15 \%) & 206 (36 \%) \\
\raisebox{2.5ex}[0pt]{base}
& C.A. &    34  (22 \%) &   27 (17 \%) &  96 (61 \%)  & 157 (27 \%) \\
\hline
& sum   &   215  (37 \%) &  202 (35 \%) & 162 (28 \%) &  578 (100 \%)\\
\end{tabular}
\caption{ \label{tab:NNCT-swamp}
The NNCT for swamp tree data and the corresponding percentages (in parentheses),
where the cell percentages are with respect to the row sums
and marginal percentages are with respect to the total size.
W.T. = water tupelos, B.G. = black gums, and C.A. = Carolina ashes.}
\end{table}

\begin{table}[ht]
\centering
\begin{tabular}{|c||c|c|c|c|c|}
\hline
\multicolumn{6}{|c|}{Test statistics and the associated $p$-values for Swamp Tree Data} \\
\hline
\multicolumn{6}{|c|}{Cuzick-Edward's $k$-NN tests} \\
\hline
Data & $T_1$ & $T_2$ & $T_3$ & $T_4$ & $T_5$ \\
\hline
W.T. vs B.G. & 155 ($<.0001$) & 309 ($<.0001$) & 451 ($<.0001$) & 588 ($<.0001$) & 703 ($<.0001$) \\
\hline
B.G. vs W.T. & 149 ($<.0001$) & 279 ($<.0001$) & 411 ($<.0001$) & 529 ($<.0001$) & 650 ($<.0001$) \\
\hline
\hline
W.T. vs C.A. & 171 ($<.0001$) & 337 ($<.0001$) & 498 ($<.0001$) & 653 ($<.0001$) & 812 ($<.0001$) \\
\hline
C.A. vs W.T. & 108 ($<.0001$) & 213 ($<.0001$) & 297 ($<.0001$) & 378 ($<.0001$) & 455 ($<.0001$) \\
\hline
\hline
B.G. vs C.A. & 159 ($<.0001$) & 303 ($<.0001$) & 461 ($<.0001$) & 606 ($<.0001$) & 755 ($<.0001$) \\
\hline
C.A. vs B.G. & 115 ($<.0001$) & 216 ($<.0001$) & 315 ($<.0001$) & 410 ($<.0001$) & 511 ($<.0001$) \\
\hline
\end{tabular}
\begin{tabular}{|c||c|c|c|c|}
\hline
\multicolumn{5}{|c|}{Cuzick-Edward's combined tests} \\
\hline
Data & $T^{comb}_{1-2}$ & $T^{comb}_{1-3}$ & $T^{comb}_{1-4}$ & $T^{comb}_{1-5}$ \\
\hline
W.T. vs B.G. & 7.31 ($<.0001$) & 8.15 ($<.0001$) & 8.78 ($<.0001$) & 9.06 ($<.0001$) \\
\hline
B.G. vs W.T. & 7.27 ($<.0001$) & 7.95 ($<.0001$) & 8.36 ($<.0001$) & 8.70 ($<.0001$) \\
\hline
\hline
W.T. vs C.A. & 7.64 ($<.0001$) & 8.71 ($<.0001$) & 9.46 ($<.0001$) & 10.14 ($<.0001$) \\
\hline
C.A. vs W.T. & 7.29 ($<.0001$) & 7.93 ($<.0001$) & 8.27 ($<.0001$) & 8.56 ($<.0001$) \\
\hline
\hline
B.G. vs C.A. & 6.80 ($<.0001$) & 7.83 ($<.0001$) & 8.55 ($<.0001$) & 9.23 ($<.0001$) \\
\hline
C.A. vs B.G. & 7.96 ($<.0001$) & 8.83 ($<.0001$) & 9.41 ($<.0001$) & 9.57 ($<.0001$) \\
\hline
\end{tabular}
\caption{ \label{tab:swamp-tree-CE-test-stat}
Test statistics and the associated $p$-values (in parentheses)
for Cuzick-Edward's NN tests the swamp tree data.
$T_k$ stands for Cuzick-Edward's $k$-NN test for $k=1,2,3,4,5$
and $T^{comb}_{1-j}$ stands for the combined tests for $j=2,3,4,5$.}
\end{table}

The locations of the tree species can be viewed a priori resulting
from different processes so the more appropriate null hypothesis is the CSR independence pattern.
Hence our inference will be a conditional one (see Remark \ref{rem:QandR}).
We calculate $Q=472$ and $R=454$ for this data set.
We present the tests statistics and the associated $p$-values for NNCT-tests
in Table \ref{tab:ex-NNCT-test-stat}.
Based on the NNCT-tests, we find that the segregation between all species are significant,
since all the tests considered yield significant $p$-values
and the diagonal cells are larger than expected.

The swamp tree data have the null hypothesis as the CSR independence of three tree species,
hence do not fall in the generalized two-class
case/control framework of Cuzick-Edward's tests.
So we apply these tests on the swamp tree data for two species at a time.
As Cuzick-Edward's tests are more sensitive to detect the clustering
of the cases (i.e., the first class in the generalized framework),
they are not symmetric in the two species they are used for.
Hence, we apply these tests for each of the six different ordered pairs of tree species
and the resulting test statistics and the associated $p$-values
are presented in Table \ref{tab:swamp-tree-CE-test-stat}.
Based on Cuzick-Edwards $k$-NN tests (i.e., $T_k$),
water tupelos and black gums exhibit significant segregation
since all $T_k$ values are significant.
Likewise for water tupelos versus Carolina ashes
and bald cypresses versus Carolina ashes.

\begin{figure}[t]
\centering
\rotatebox{-90}{ \resizebox{2 in}{!}{\includegraphics{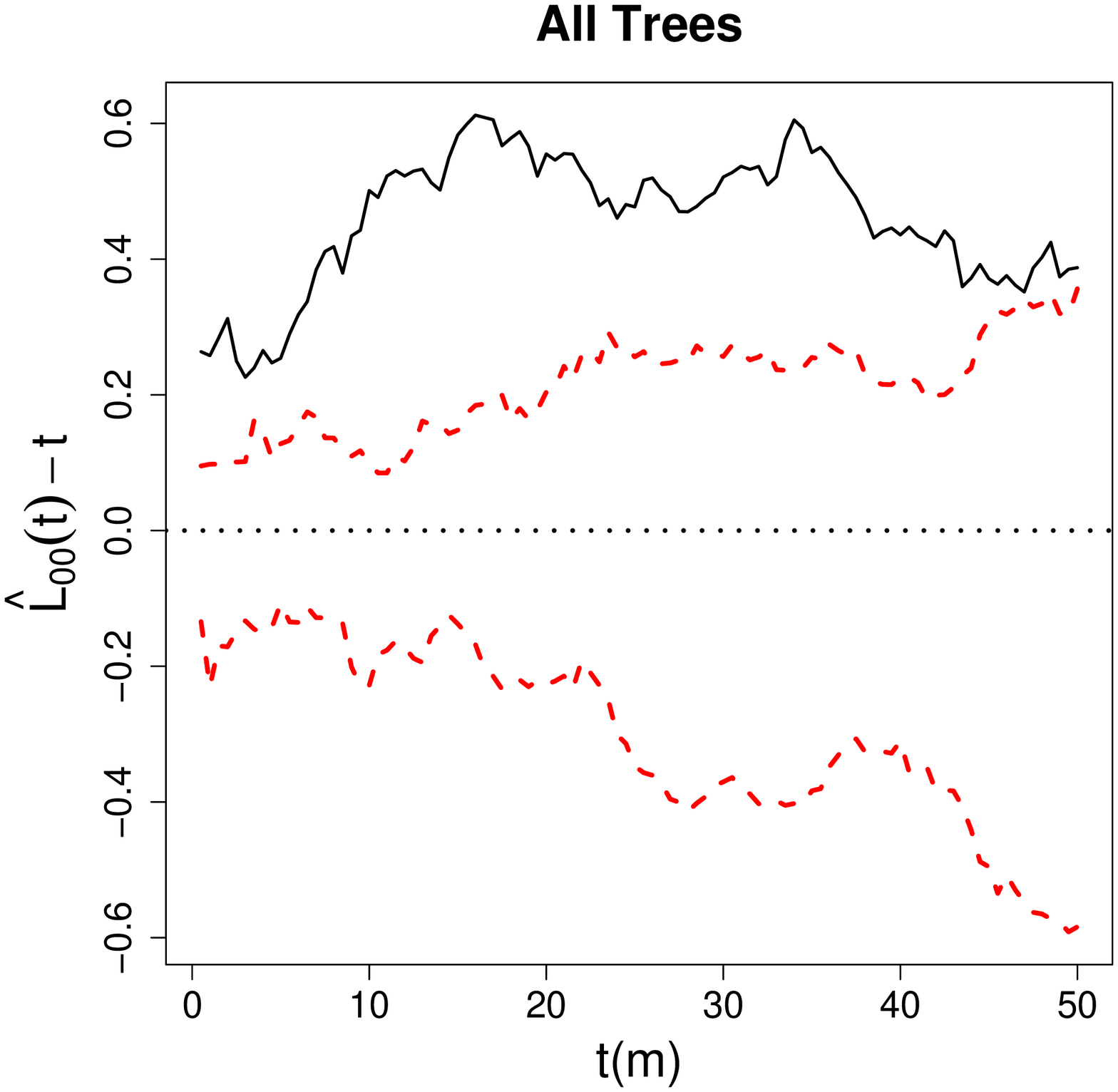} }}\\
\rotatebox{-90}{ \resizebox{2 in}{!}{\includegraphics{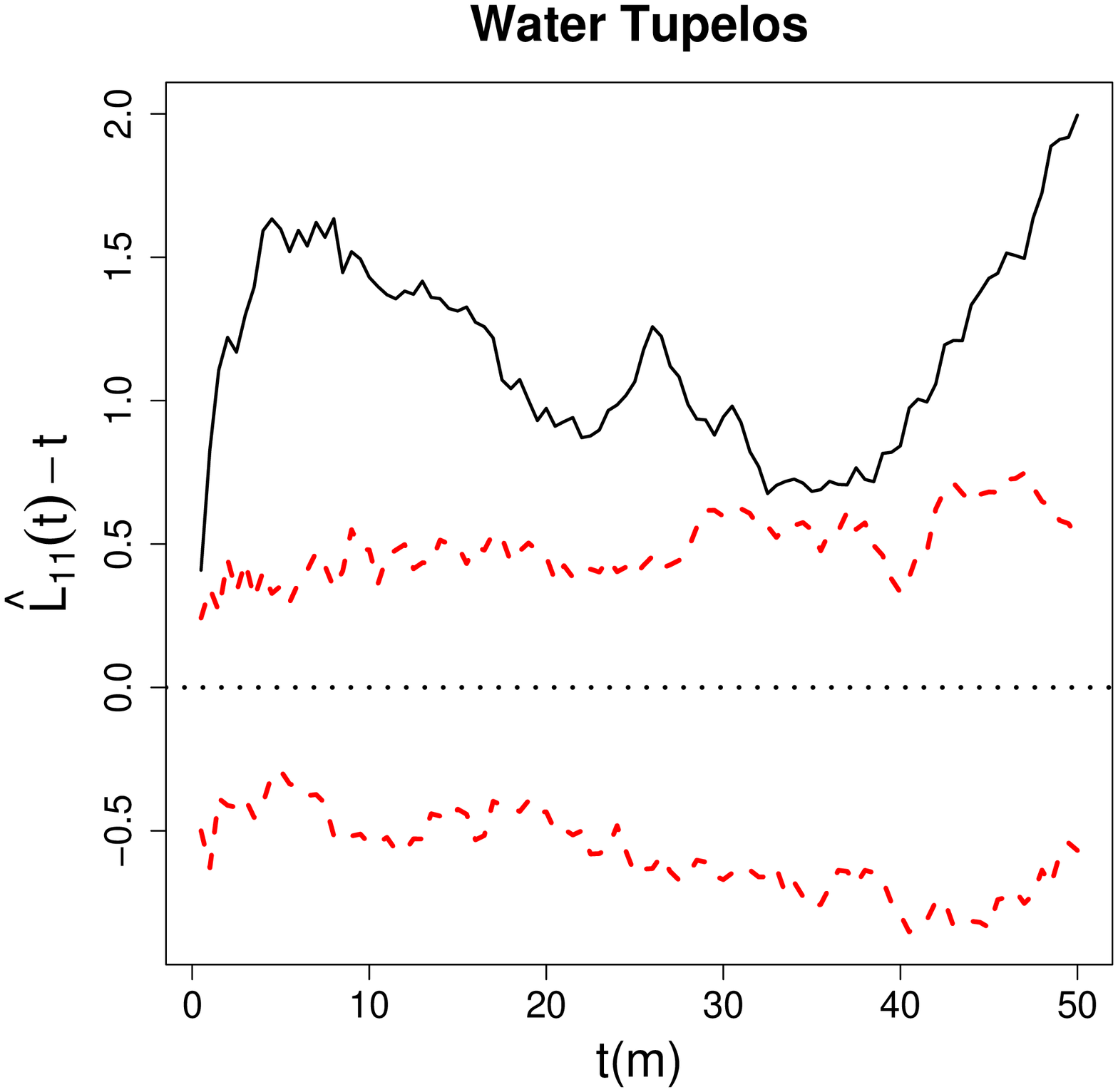} }}
\rotatebox{-90}{ \resizebox{2 in}{!}{\includegraphics{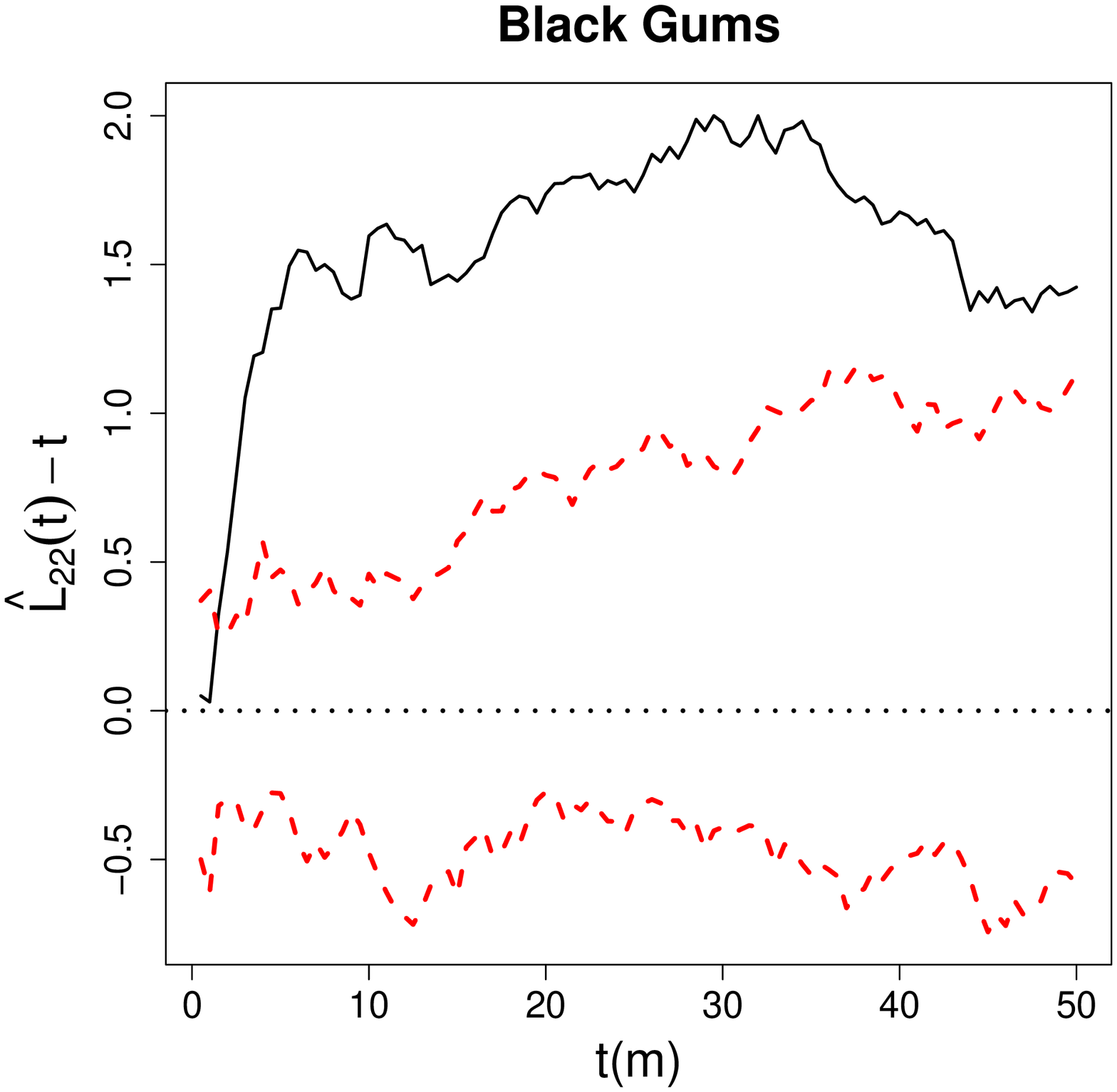} }}
\rotatebox{-90}{ \resizebox{2 in}{!}{\includegraphics{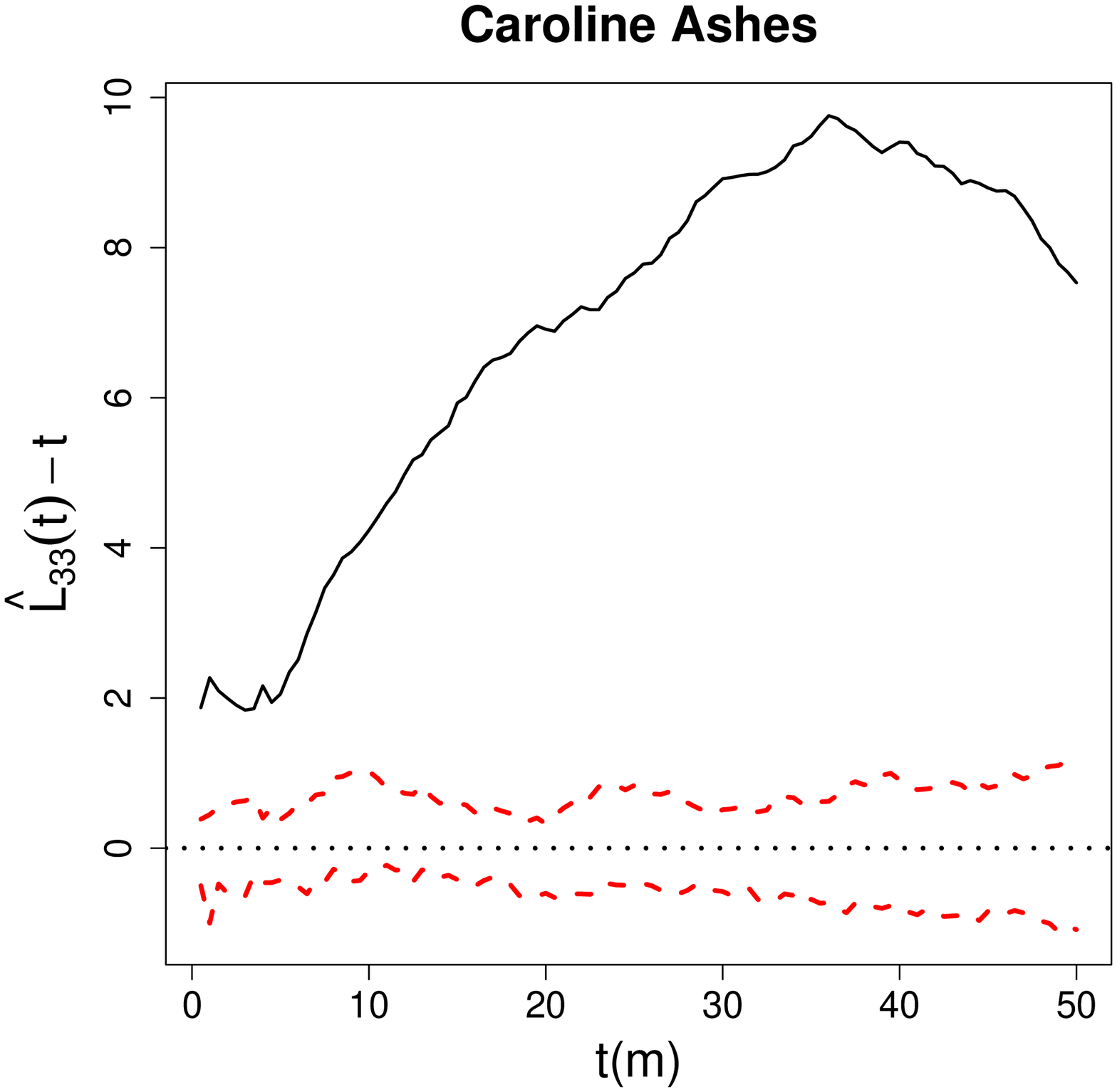} }}
\caption{
\label{fig:swamp-Liihat}
Second-order properties of swamp tree data.
Functions plotted are Ripley's univariate $L$-functions
$\widehat{L}_{ii}(t)-t$ for $i=1,2,3$,
where $i=1$ for water tupelos, $i=2$ for black gums,
and $i=3$ for Carolina ashes.
The dashed lines around 0 are the upper and lower 95 \% confidence bounds for the
$L$-functions based on Monte Carlo simulation under the CSR independence pattern.
Note also that vertical axes are not identically scaled for all plots.}
\end{figure}

\begin{figure}[t]
\centering
\rotatebox{-90}{ \resizebox{2 in}{!}{\includegraphics{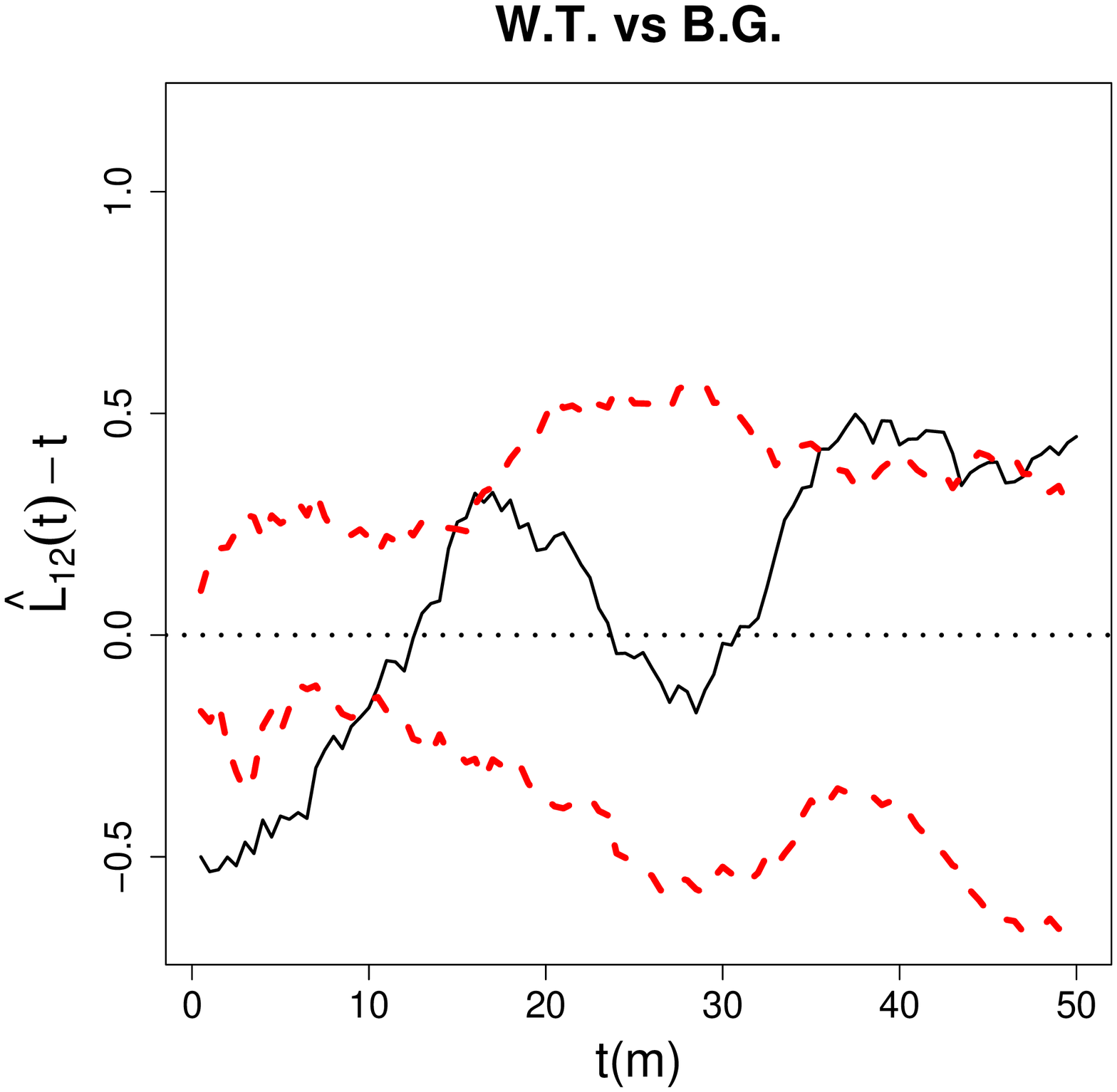} }}
\rotatebox{-90}{ \resizebox{2 in}{!}{\includegraphics{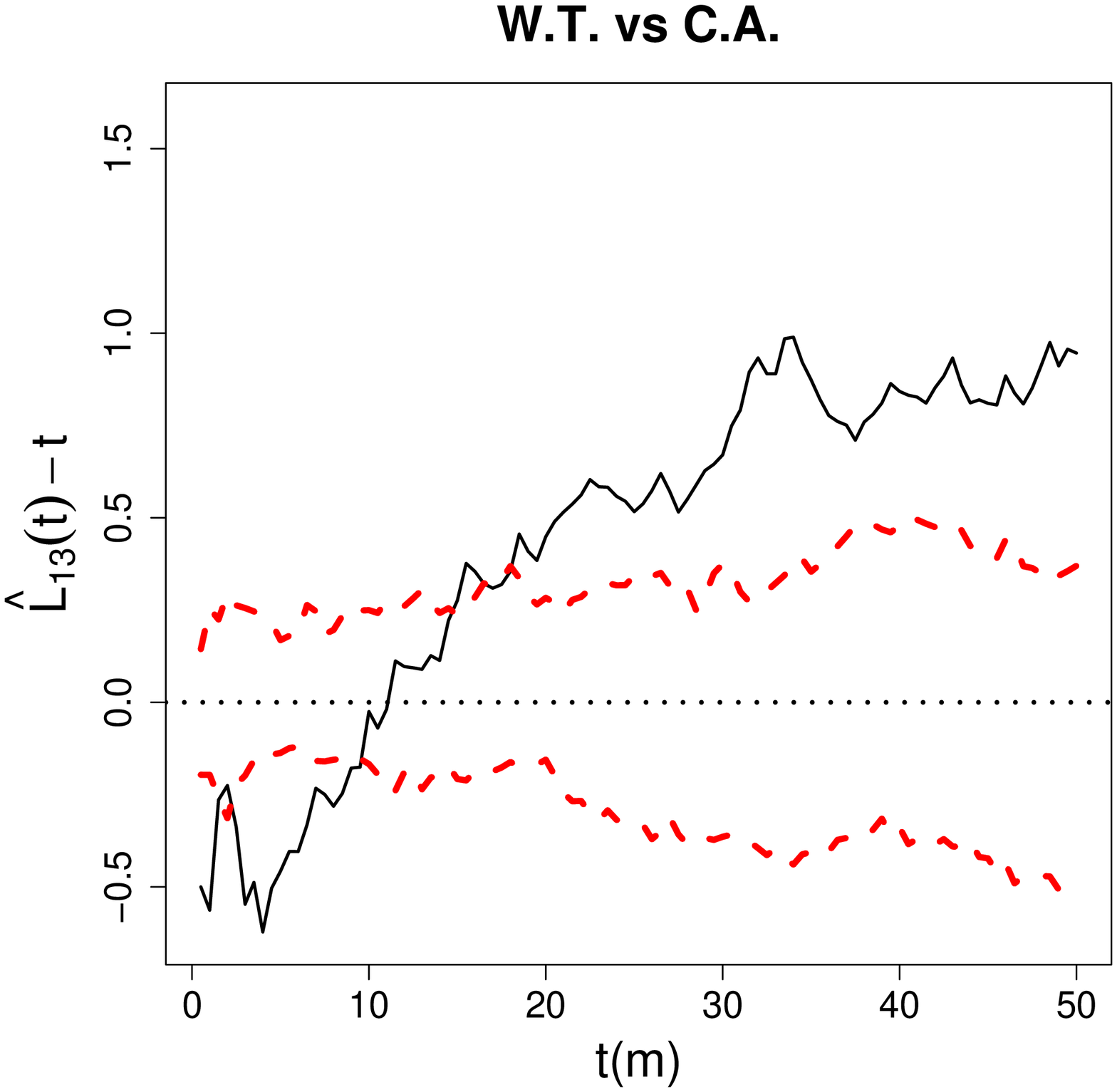} }}
\rotatebox{-90}{ \resizebox{2 in}{!}{\includegraphics{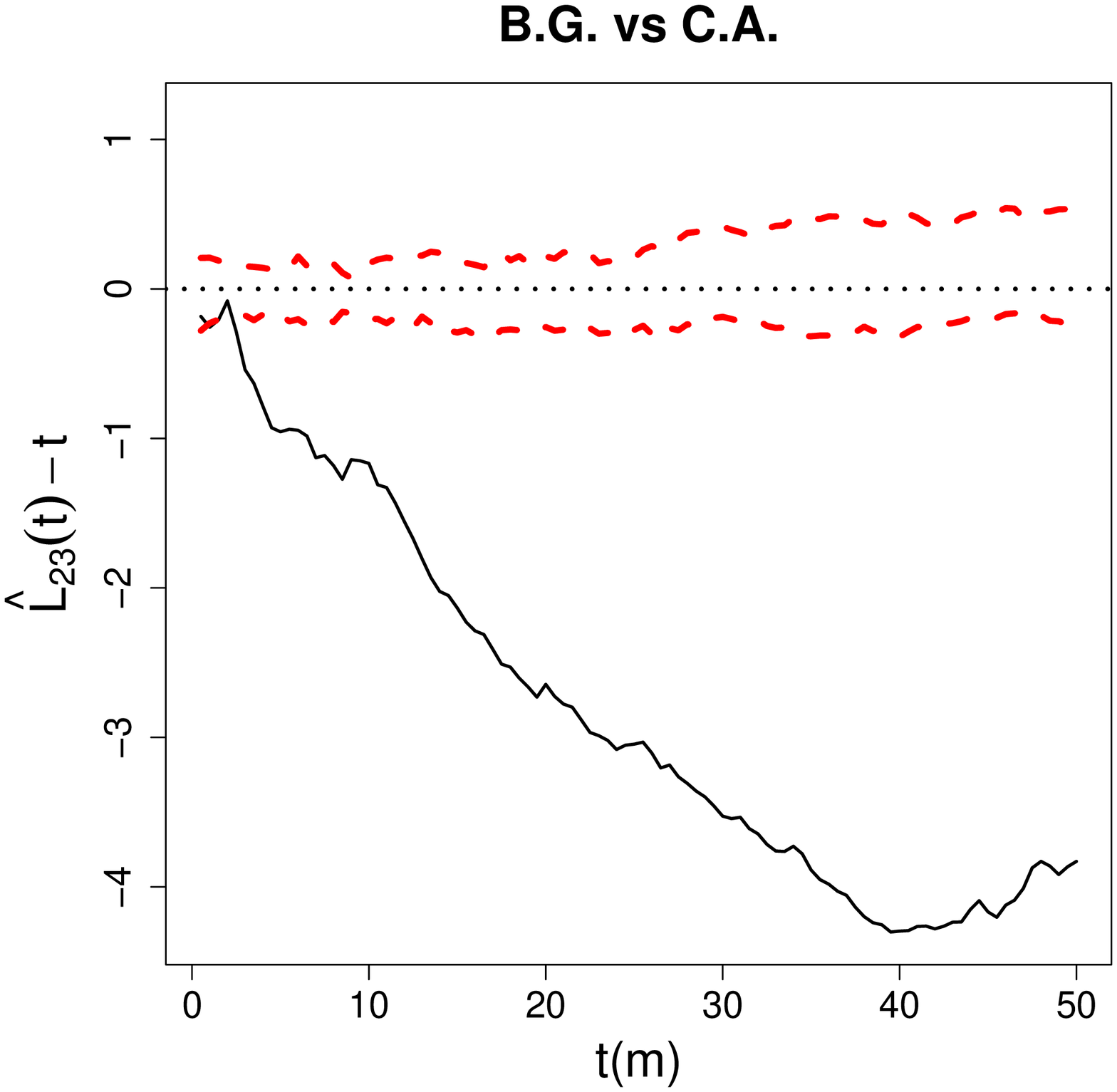} }}
\caption{
\label{fig:swamp-Lijhat}
Second-order properties of swamp tree data.
Functions plotted are Ripley's bivariate $L$-functions
$\widehat{L}_{ij}(t)-t$ for $i,j=1,2,3$ and $i \not= j$
where $i=1$ for water tupelos (W.T.), $i=2$ for black gums (B.G.),
and $i=3$ for Carolina ashes (C.A.).
The dashed lines around 0 are the upper and lower 95 \% confidence bounds for the
$L$-functions based on Monte Carlo simulations under the CSR independence pattern.
Note also that vertical axes are differently scaled.}
\end{figure}

\begin{figure}[t]
\centering
\rotatebox{-90}{ \resizebox{2 in}{!}{\includegraphics{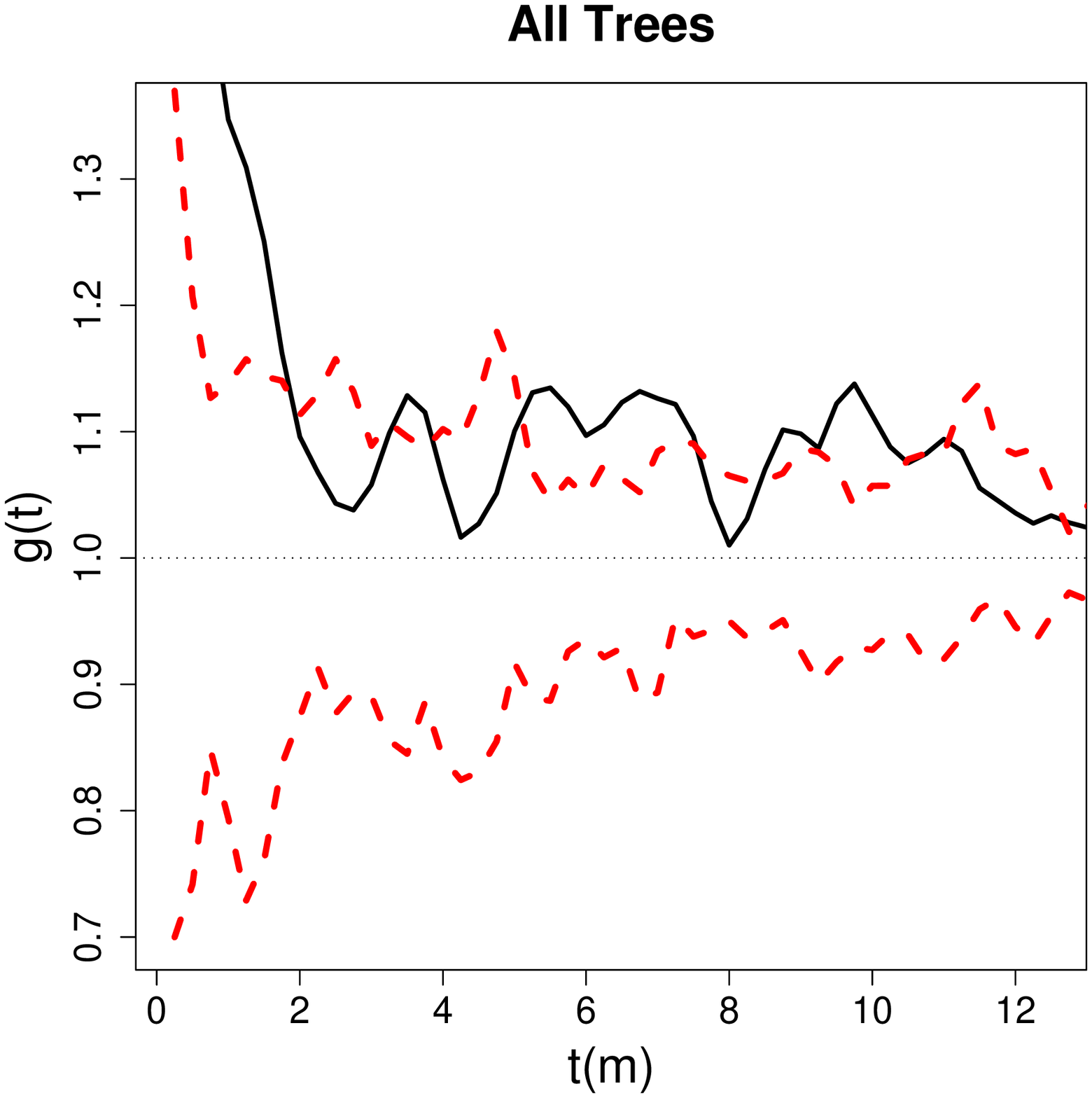} }}\\
\rotatebox{-90}{ \resizebox{2 in}{!}{\includegraphics{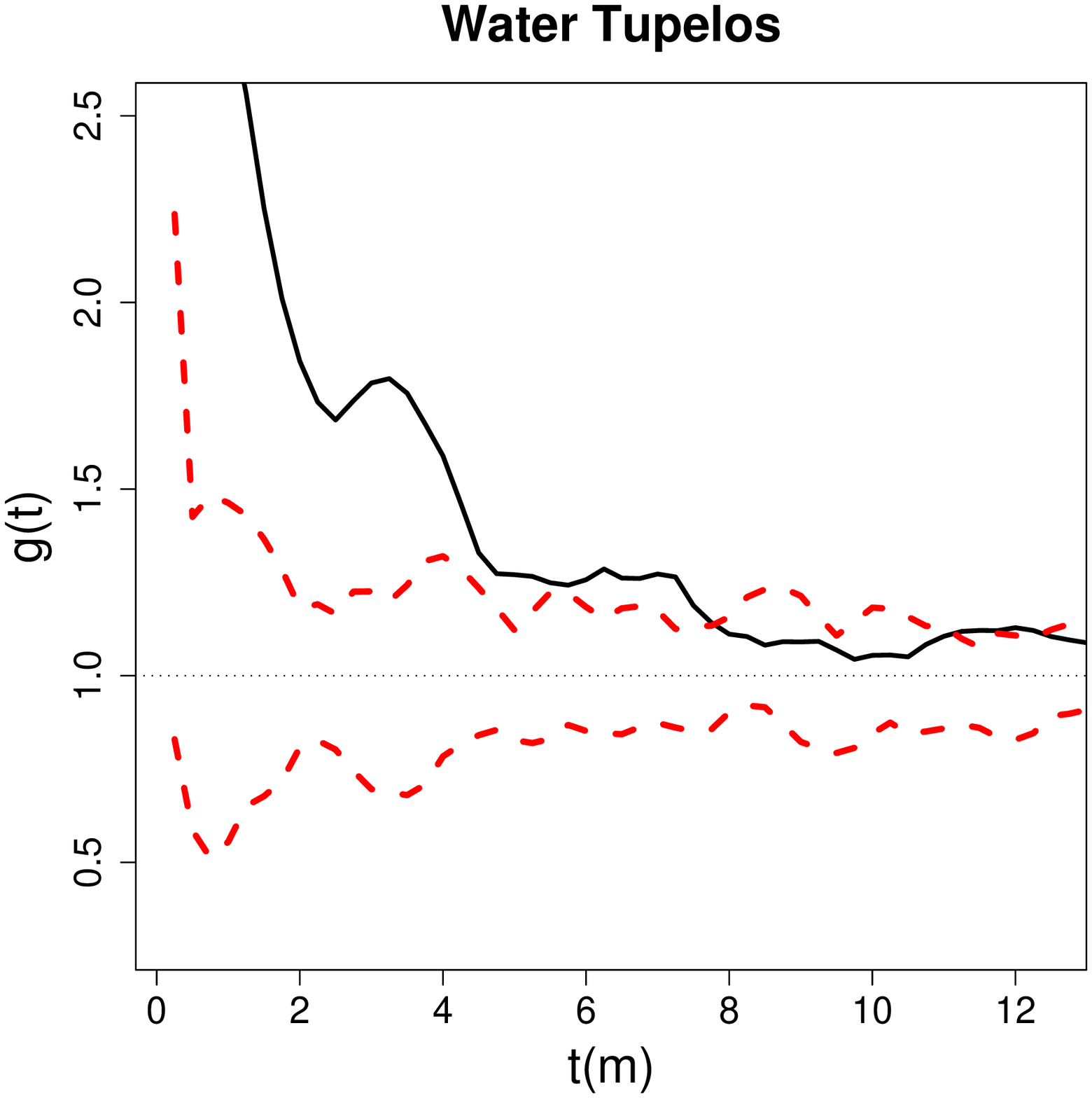} }}
\rotatebox{-90}{ \resizebox{2 in}{!}{\includegraphics{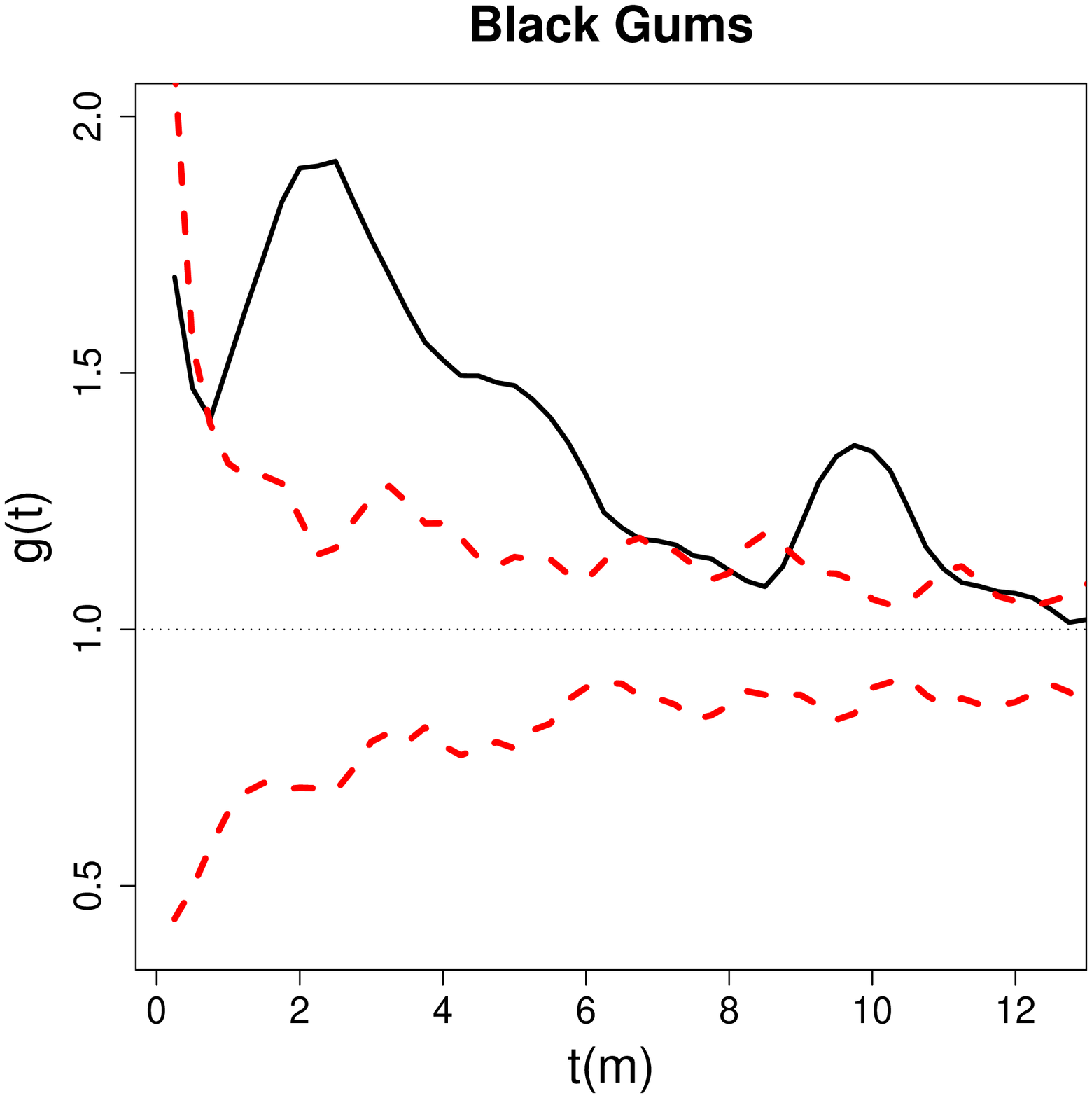} }}
\rotatebox{-90}{ \resizebox{2 in}{!}{\includegraphics{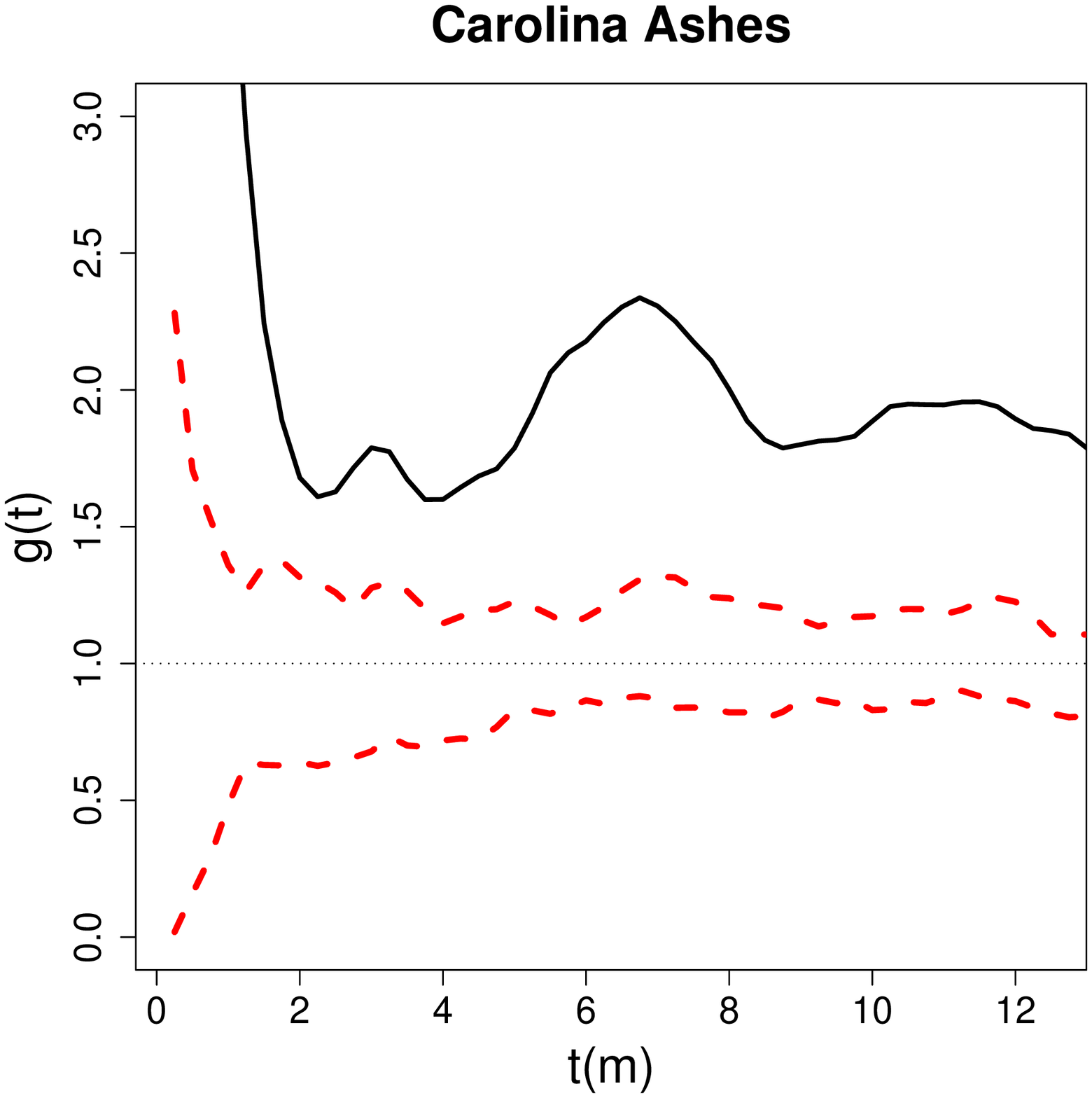} }}
\caption{
\label{fig:swamp-PCFii}
Pair correlation functions for all trees combined and for each species in the swamp tree data.
Wide dashed lines around 1 (which is the theoretical value)
are the upper and lower (pointwise) 95 \% confidence bounds for the
$L$-functions based on Monte Carlo simulation under the CSR independence pattern.
Note also that vertical axes are differently scaled.}
\end{figure}

\begin{figure}[t]
\centering
\rotatebox{-90}{ \resizebox{2 in}{!}{\includegraphics{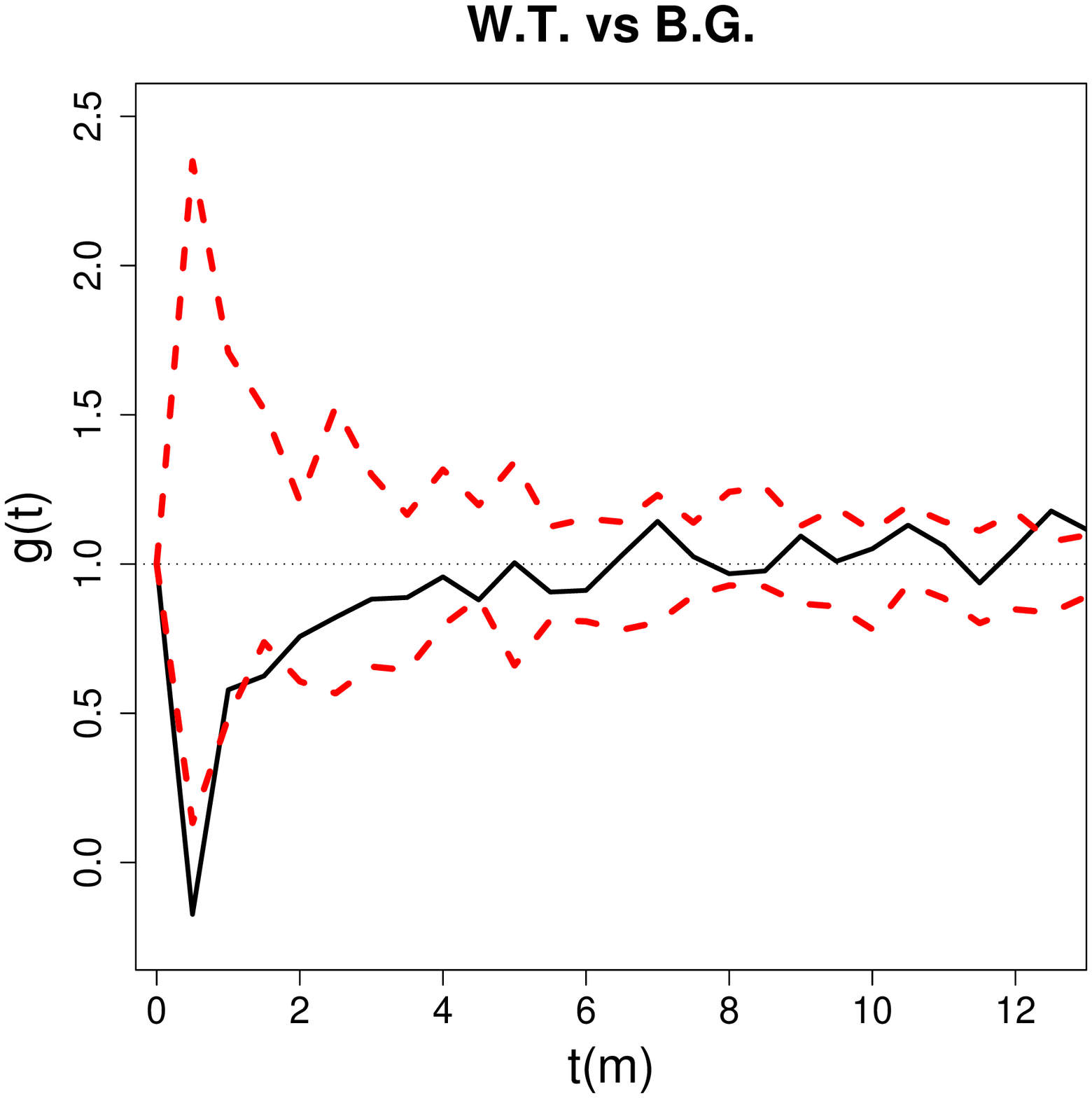} }}
\rotatebox{-90}{ \resizebox{2 in}{!}{\includegraphics{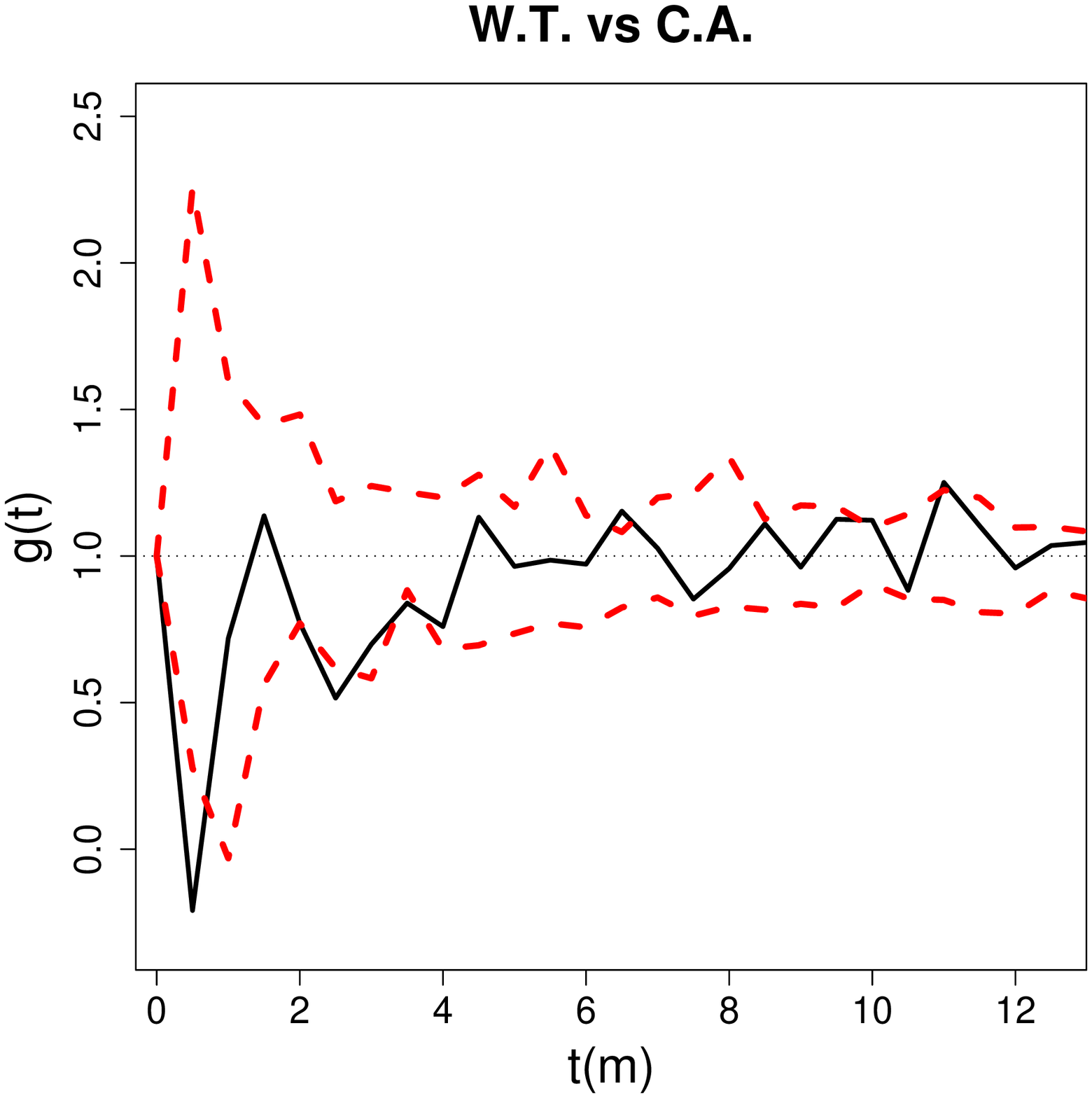} }}
\rotatebox{-90}{ \resizebox{2 in}{!}{\includegraphics{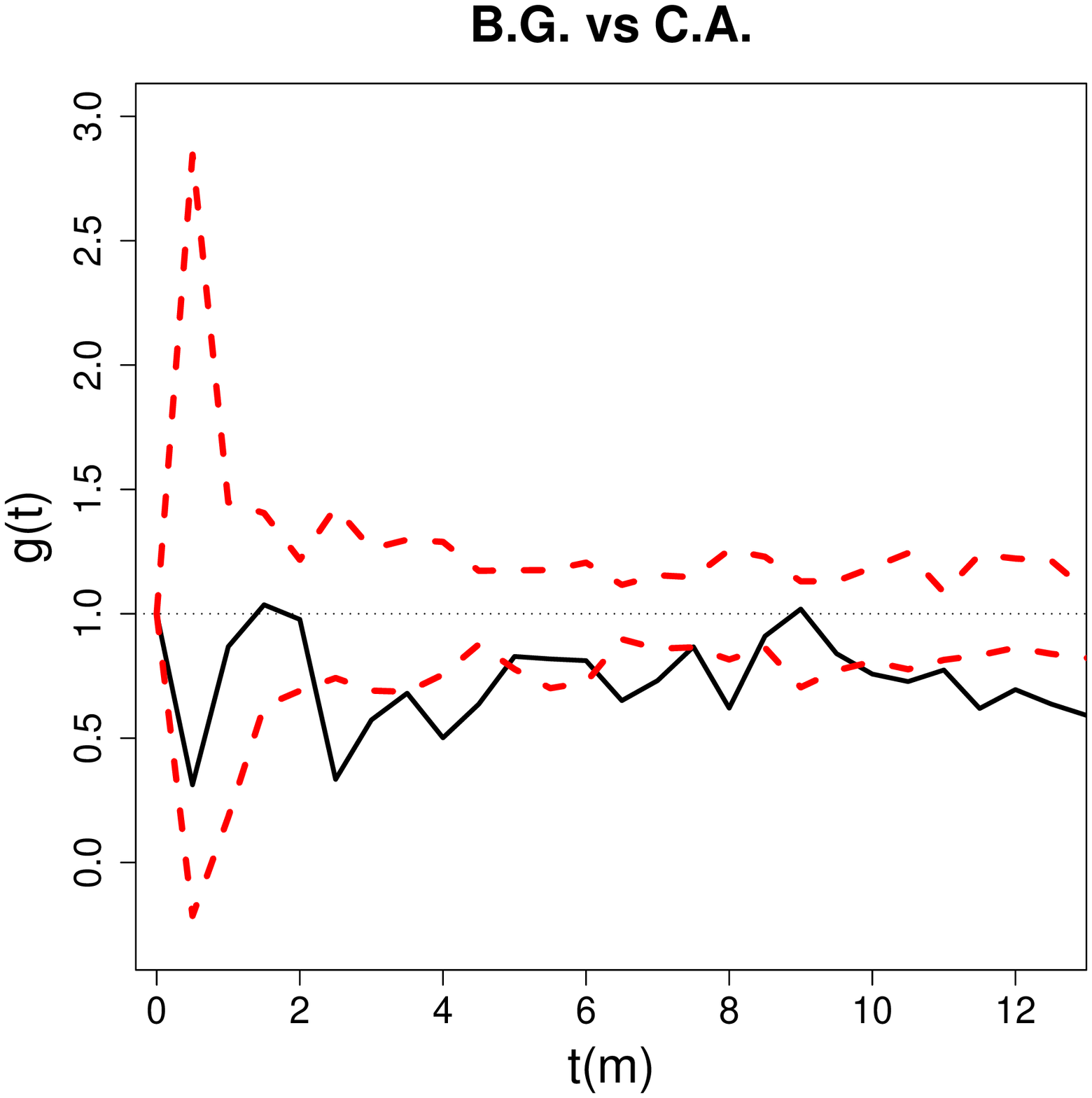} }}
\caption{
\label{fig:swamp-PCFij}
Pair correlation functions for each pair of species in the swamp tree data.
Wide dashed lines around 1 (which is the theoretical value)
are the upper and lower (pointwise) 95 \% confidence bounds for the
$L$-functions based on Monte Carlo simulations under the CSR independence pattern.
W.T. = water tupelos, B.G. = black gums and C.A. = Carolina ashes.
Note also that vertical axes are not identically scaled for all plots.}
\end{figure}

Based on the NNCT-tests and Cuzick-Edwards $k$-NN tests above, we conclude that tree species exhibit
significant deviation from the CSR independence pattern.
Considering Figure \ref{fig:SwampTrees}
and the corresponding NNCT in Table \ref{tab:NNCT-swamp},
this deviation is toward the segregation of the tree species.
However, the results of NNCT-tests pertain to small scale interaction
at about the average NN distances;
and the results of Cuzick-Edward's $k$-NN tests pertain to
interaction at about the average $k$-NN distances.
In Figure \ref{fig:swamp-Liihat},
we present the plots of $\widehat{L}_{ii}(t)-t$ functions for each species
as well as the plot of the entire data combined.
We also present the upper and lower 95 \% confidence bounds for each $\widehat{L}_{ii}(t)-t$.
Observe that 
(the $L_{00}(t)-t$ curve is above the upper confidence bound) at all scales.
Water tupelos exhibit aggregation for the range of the plotted distances;
black gums exhibit significant aggregation for distances $t>1$ m;
and Carolina ashes exhibit significant aggregation for the range of plotted distances.
Hence, segregation of the species might be due to different
levels and types of aggregation of the species in the study region.

We also calculate Ripley's bivariate $L$-function for each pair of tree species
and present them in Figure \ref{fig:swamp-Lijhat},
we present the bivariate plots of $\widehat{L}_{ij}(t)-t$ functions
together with the upper and lower 95 \% confidence bounds for each pair of species.
Due to the symmetry of $L_{ij}(t)$,
we only present the plots for 3 different pairs.
Observe that for distances up to $t \approx 10$ m,
water tupelos and black gums exhibit significant segregation
($\widehat{L}_{12}(t)-t$ is below the lower confidence bound),
for $35<t<40$ they exhibit significant association,
and for the rest of the plotted distances their interaction is not significantly
different from the CSR independence pattern;
water tupelos and Carolina ashes are significantly segregated up to about $t \approx 10$ m,
and for $t>15$ m they are significantly associated.
Black gums and Carolina ashes are significantly segregated for $t>2$ m.

Since Ripley's $K$-function is cumulative,
we also provide the pair correlation functions
for all trees and each species for the swamp tree data
in Figure \ref{fig:swamp-PCFii}.
Observe that all trees are
aggregated around distance values of 0-1,3,4-7,8-10 m;
water tupelos are aggregated for distance values of 0-7 m;
black gums are aggregated for distance values of 1-6 and 8-11 m; Carolina
ashes are aggregated for all the range of the plotted distances.
Comparing Figures \ref{fig:swamp-Liihat} and
\ref{fig:swamp-PCFii}, we see that Ripley's $L$ and pair correlation
functions detect the same patterns but with different distance
values. That is, Ripley's $L$ implies that the particular pattern is
significant for a wider range of distance values compared to $g(t)$,
since Ripley's $L$ is cumulative, so the values of $L$ at small
scales confound the values of $L$ at larger scales
(\cite{loosmore:2006}). Hence the results based on pair correlation
function $g(t)$ are more reliable.

The bivariate pair correlation functions
for the species in swamp tree data are plotted in Figure
\ref{fig:swamp-PCFij}. Observe that water tupelos and black gums are
segregated for distance values of 0-1 m; water tupelos and Carolina
ashes are segregated for values of 0-1 and 2.5 m and are associated
for values about 6 and 11 m;
black gums and Carolina ashes are segregated for 2-5, 6-8.5, and 9.5-12 meters.

Since the estimator variance and hence
the bias are considerably large for small $t$ if $g(t)>0$,
the confidence bands for smaller $t$ values are much wider compared
to those for larger $t$ values (see for example Figures \ref{fig:swamp-PCFii} and \ref{fig:swamp-PCFij}).
So pair correlation
function analysis is more reliable for larger distances and it is
safer to use $g(t)$ for distances larger than the average NN
distance in the data set.
Comparing Figure \ref{fig:swamp-Liihat}
with Figure \ref{fig:swamp-PCFii} and Figure \ref{fig:swamp-Lijhat}
with Figure \ref{fig:swamp-PCFij},
we see that Ripley's $L$ and pair
correlation functions usually detect the same large-scale pattern
but at different ranges of distance values. Ripley's $L$ suggests
that the particular pattern is significant for a wider range of
distance values compared to $g(t)$,
but at larger scales $g(t)$ is more reliable to use.

While second order analysis (using Ripley's $K$ and $L$-functions or
pair correlation function) provides information on the univariate
and bivariate patterns at all scales (i.e., for all distances),
NNCT-tests summarize the spatial interaction for the smaller scales
(for distances about the average NN distance in the data set). In
particular, for the swamp tree data average NN distance ($\pm$
standard deviation) is about 1.93 ($\pm$ 1.17) meters and notice that
Ripley's $L$-function and NNCT-tests yield similar results for
distances about 2 meters.
Further, the average $k$-NN distances $\pm$ standard deviations
for $k=2,3,4,5$ are $2.94\pm1.36$, $3.81\pm1.41$, $4.47\pm1.45$, and $5.10\pm1.49$, respectively.

\subsection{Leukemia Data}
\label{sec:leukemia-data}
\cite{cuzick:1990} considered the spatial locations of 62 cases of childhood
leukemia in the North Humberside region of the UK, between the years 1974 to 1982 (inclusive).
A sample of 143 controls are selected using the completely randomized design
from the same region.
We analyze the spatial distribution of leukemia cases and controls
in this data using a $2 \times 2$ NNCT.
We plot the locations of these points in the study region in
Figure \ref{fig:leukemia} and provide the corresponding $2 \times 2$ NNCT
together with percentages based on row and column sums in Table \ref{tab:NNCT-leukemia}.
Observe that the percentages in the diagonal cells are about the same as the
marginal (row or column) percentages of the subjects in the study,
which might be interpreted as the lack of any deviation from RL for both classes.
Figure \ref{fig:leukemia} is also supportive of this observation.

\begin{figure}[ht]
\centering
\rotatebox{-90}{ \resizebox{3.5 in}{!}{\includegraphics{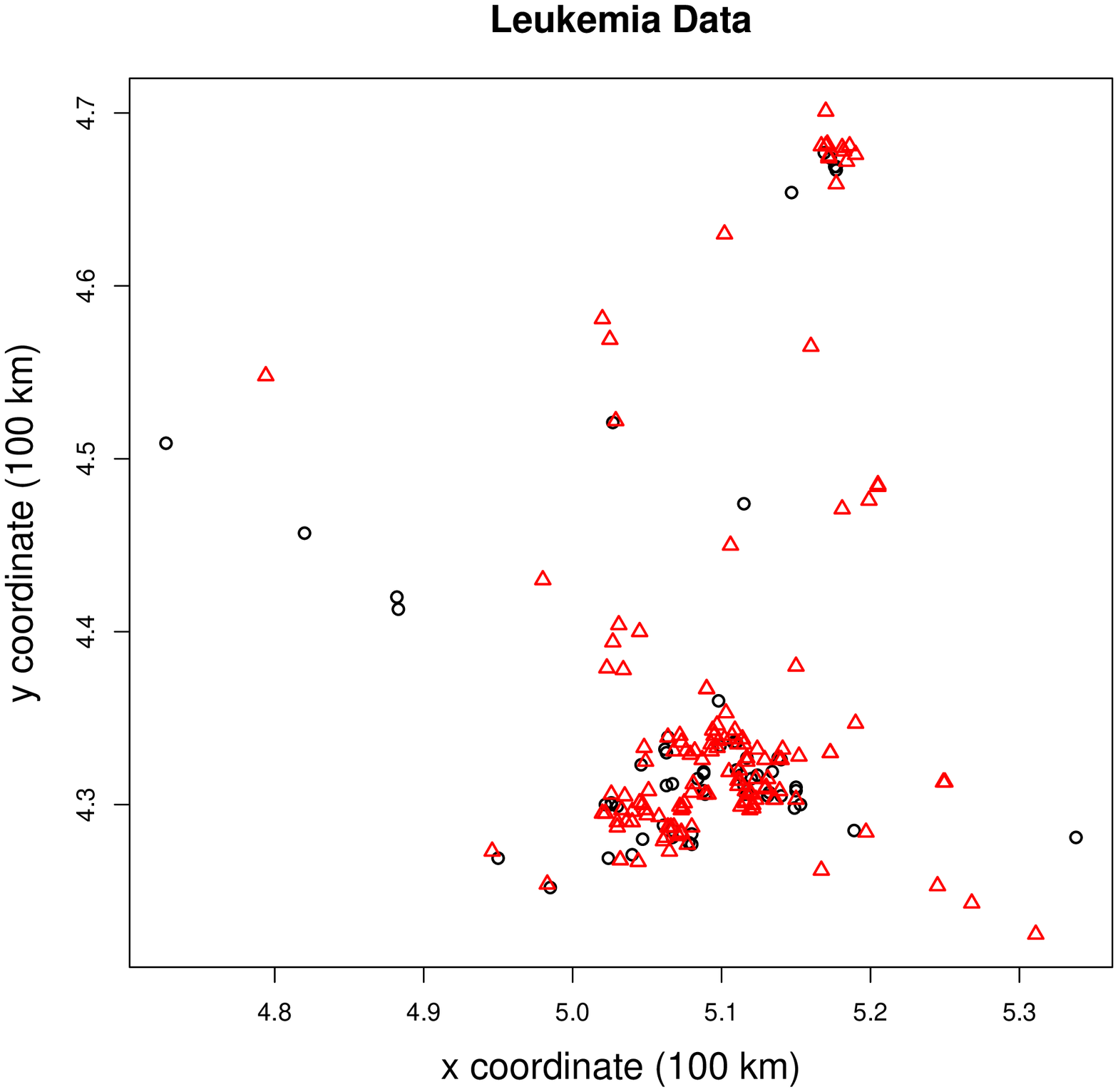} }}
 \caption{
\label{fig:leukemia}
The scatter plots of the locations of cases (circles $\circ$) and controls
(triangles $\triangle$) in North Humberside leukemia data set.}
\end{figure}

\begin{table}[ht]
\centering
\begin{tabular}{cc|cc|c}
\multicolumn{2}{c}{}& \multicolumn{2}{c}{NN}& \\
\multicolumn{2}{c}{}&    case &  control   &   sum  \\
\hline
& case &    25 (38 \%)  &   41 (62 \%)   &   66 (30 \%)  \\
\raisebox{1.5ex}[0pt]{base}
& control &    39 (26 \%)  &   113 (74 \%)   &  152 (70 \%) \\
\hline
&sum  &    64 (29 \%) &   154 (71 \%)    &  218 (100 \%)  \\
\end{tabular}
\caption{\label{tab:NNCT-leukemia}
The NNCT for the North Humberside leukemia data and
the corresponding percentages (in parentheses).}
\end{table}

\begin{table}[ht]
\centering
\begin{tabular}{|c|c|c|c|c||c|c|c|c|}
\hline
\multicolumn{9}{|c|}{Cuzick-Edward's test statistics and the associated $p$-values for Leukemia Data} \\
\hline
$T_1$ & $T_2$ & $T_3$ & $T_4$ & $T_5$ & $T^{comb}_{1-2}$ & $T^{comb}_{1-3}$ & $T^{comb}_{1-4}$ & $T^{comb}_{1-5}$ \\
\hline
25 & 53 & 78 & 95 & 116 & 2.12 & 2.46 & 2.53 & 2.58  \\
(.0647) & (.0043) & (.0014) & (.0093) & (.0099) & (.0170) & (.0068) & (.0057) & (.0048)  \\
\hline
\end{tabular}

\caption{ \label{tab:leukemia-CE-test-stat}
Test statistics and the associated $p$-values for Cuzick-Edward's
$k$-NN (i.e., $T_k$) and $T^{comb}_S$ tests for North Humberside leukemia data.}
\end{table}

It is reasonable to assume that some process affects a posteriori
the population of North Humberside region
so that some of the individuals get to be cases,
while others continue to be healthy (i.e., they are controls).
So the appropriate null hypothesis is the RL pattern.
We calculate $Q=152$ and $R=142$ for this data set.
In Tables \ref{tab:ex-NNCT-test-stat} and \ref{tab:leukemia-CE-test-stat},
we present the test statistics and the associated $p$-values.
Observe that none of the NNCT-tests yields a significant result.
On the other hand, Cuzick-Edwards $T_k$ are all significant for $k>1$,
and so are all $T^{comb}_S$ tests.
Hence, we conclude that there is no significant segregation of
cases at small scales (about NN-distances),
but cases tend to cluster significantly at larger scales.

\begin{figure}[ht]
\centering
\rotatebox{-90}{ \resizebox{3 in}{!}{\includegraphics{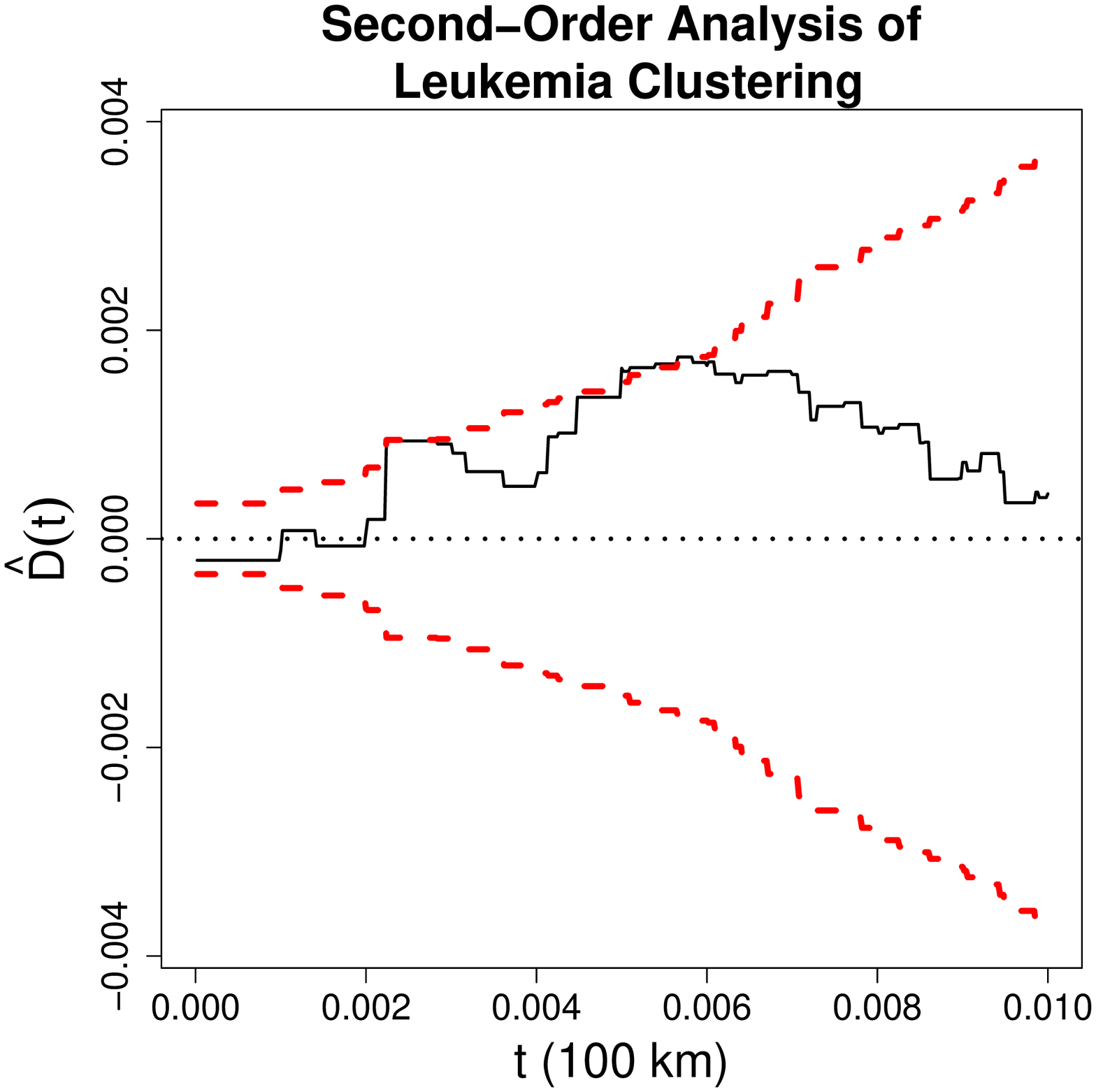} }}
 \caption{
\label{fig:leukemia-Dhat}
Second-order analysis of North Humberside childhood leukemia data:
Function plotted is Diggle's modified bivariate $K$-function
$\widehat{D}(t)=\widehat{K}_{11}(t)-\widehat{K}_{22}(t)$ with
$i=1$ for controls and $i=2$ for leukemia cases.
The dashed lines around 0 are plus and minus two standard errors
of $\widehat{D}(t)$ under RL of cases and controls.}
\end{figure}

Based on the NNCT-tests above, we conclude that the cases and controls
do not exhibit significant clustering (i.e., segregation) at small scales.
Based on Cuzick-Edward's tests, we find that the cases
are significantly segregated around $k$-NN distances for $k>1$.
However, NNCT-methods only provide information on spatial interaction
for distances about expected NN distance in the data set,
and Cuzick-Edward's tests provide information about the $k$-NN distances.
So it might be the case that the type and level of interaction
might still be different at larger or other scales (i.e., distances between the subjects' locations).
However, the locations of the subjects in this population (cases and controls together)
seem to be from an inhomogeneous Poisson process
(see also Figure \ref{fig:leukemia}).
Hence Ripley's $K$- or $L$-functions in the general form are not appropriate
to test for the spatial clustering of the cases (\cite{kulldorff:2006}).
So we use the modified version due to \cite{diggle:2003}, namely,
$D(t)=K_{11}(t)-K_{22}(t)$ where $K_{ii}(t)$ is Ripley's univariate $K$-function for class $i$.
In this setup, ``no spatial clustering" is equivalent to RL of cases and controls
on the locations in the sample, which implies $D(t)=0$,
since $K_{22}(t)$ measures the degree of spatial aggregation of the controls
(i.e., the population at risk), while $K_{11}(t)$ measures
this same spatial aggregation plus any additional clustering due to the disease.
The test statistic $D(t)$ is estimated by
$\widehat D(t)=\widehat K_{11}(t)-\widehat K_{22}(t)$,
where $\widehat K_{ii}(t)$ is as in Equation \eqref{eqn:Kiihat}.
Figure \ref{fig:leukemia-Dhat} shows the plot of $\widehat D(t)$
plus and minus two standard errors under RL.
Observe that at distances about 200 and 600 meters,
there is evidence for mild clustering of diseases
(i.e., segregation of cases from controls)
since the empirical function $\widehat D(t)$ gets close or a little
above of the upper limit.
At smaller scales, plot in Figure \ref{fig:leukemia-Dhat} is
consistent with the results of the NNCT analysis.
In particular average NN distance for leukemia data is 700 ($\pm$ 1400) m,
and NNCT analysis summarizes the pattern for about $t=1000$ m
which is depicted in Figure \ref{fig:leukemia-Dhat}.
(Further, the average $k$-NN distances $\pm$ standard deviations
for $k=2,3,4,5$ are $1342\pm2051$, $1688\pm2594$, $2152\pm3188$, and $2495\pm3810$, respectively).
This same data set was also analyzed
by (\cite{diggle:2003} pp 131-132) and similar plots and results were obtained.

\section{Discussion and Conclusions}
\label{sec:disc-conc}
In this article,
we discuss segregation or clustering tests based on nearest neighbor contingency tables (NNCTs).
Pielou's and Dixon's segregation tests are already in use in literature
(\cite{pielou:1961} and \cite{dixon:1994, dixon:EncycEnv2002, dixon:NNCTEco2002}.
Pielou's test of independence is only appropriate when the null hypothesis implies
that the NNCT is based on a random sample of (base,NN) pairs,
but not appropriate when the null case implies the NNCTs are based on data from
complete spatial randomness (CSR) independence or random labeling (RL) patterns (\cite{ceyhan:overall}).
Dixon's tests are appropriate for the null patterns of CSR independence or RL
(but they are conditional under the CSR independence pattern),
which are more realistic in practical situations.
In literature, both of Pielou's and Dixon's tests are used
for the null hypotheses of CSR independence or RL.
We propose three new tests using the correct (asymptotic)
distribution of cell counts in NNCTs and a corrected
version for Pielou's test based on empirical estimates of its mean and variance.
We also compare the NNCT-tests with Cuzick-Edward's $k$-NN and combined tests in an extensive
Monte Carlo simulation study and with Ripley's $K$ or $L$-functions,
Diggle's $D$-function and pair correlation functions
in example data sets.

For testing segregation or association against the CSR independence or RL patterns,
we recommend the disuse of Pielou's test,
as it gives more false alarms than allowed by the significance level.
As a quick fix,
one can use the Monte Carlo corrected version
for rectangular study regions for similar sample sizes.
Alternatively, one can also resort to Monte Carlo randomization for Pielou's test.
Considering the empirical significance levels, empirical power estimates,
and distributional properties,
we recommend version III of the new NNCT-tests when testing for segregation or association.
Among Cuzick-Edward's tests combined version of $T_k$ for $k=1,2,\ldots,5$
has slightly better performance than NNCT-tests for the association alternatives,
but the gain does not compensate the computational cost of this test.
Figure 4 in (\cite{dixon:1994} p 1946) shows that the acceptance regions for Pielou's
and Dixon's tests have different shapes,
so these tests are answering different questions.
But, the newly proposed NNCT-tests and Dixon's test address
the same question about the spatial interaction between the classes.
On the other hand, Cuzick-Edward's test is designed for the clustering
of the cases (or the first class in the general framework),
so NNCT-tests and Cuzick-Edward's tests also answer similar but not identical questions.


When testing against CSR independence or RL, NNCT-tests provide
information about the spatial interaction
at about the average NN distance in the data sets.
Cuzick-Edward's tests provide information about the $k$-NN distance.
Ripley's $K$ or $L$-function provides the type and level
of spatial interaction at all scales (i.e., at all distances one is interested)
when used against the CSR independence pattern.
However, due to the cumulative nature of these functions,
Stoyan's pair correlation function is preferable for large distances.
Furthermore, Diggle's $D$-function provides the level of spatial interaction
(or clustering of a class when compared to another) at all scales
when used against the RL pattern.
Among these tests, Cuzick-Edward's test is designed only for
the two class case of cases and controls.
The NNCT-tests can be used for the multivariate spatial interaction
between two or more classes.
Ripley's $L$ has univariate and bivariate versions while
Diggle's $D$ is designed for bivariate pattern analysis.
A practical concern about these tests is the lack
of code in some statistical software
in a way that others could use.
The methods outlined here have been implemented in $R$ version 2.6.2,
and the relevant code is available from the author upon request.

In this article, we have only considered spatial patterns of two
classes in a study region.
Dixon has extended his tests
into multi-class situation with three or more classes (species) (\cite{dixon:NNCTEco2002}).
On the other hand, Pielou's test is defined
and has only been used for the two-class spatial patterns. Its
inappropriateness discourages the immediate extension to
multi-class patterns.
However, the newly introduced versions can easily be extended to the multi-class case.

\section*{Acknowledgments}
I would like to thank Prof Bo Henry Lindqvist,
an associate editor, and two anonymous referees,
whose constructive remarks and suggestions greatly improved
the presentation and flow of this article.
Most of the Monte Carlo simulations presented in this article
were executed on the Hattusas cluster of
Ko\c{c} University High Performance Computing Laboratory.


\begin{thebibliography}{}

\bibitem[Armstrong and Irvine, 1989]{armstrong:1989}
Armstrong, J.~E. and Irvine, A.~K. (1989).
\newblock Flowering, sex ratios, pollen-ovule ratios, fruit set, and
  reproductive effort of a dioecious tree, \emph{{M}yristica {I}nsipida}
  (\emph{Myristicacea}), in two different rain forest communities.
\newblock {\em American Journal of Botany}, 76:75--85.

\bibitem[Baddeley et~al., 2000]{baddeley:2000b}
Baddeley, A., M{\o}ller, J., and Waagepetersen, R. (2000).
\newblock Non- and semi-parametric estimation of interaction in inhomogeneous
  point patterns.
\newblock {\em Statistica Neerlandica}, 54(3):329–--350.

\bibitem[Barot et~al., 1999]{barot:1999}
Barot, S., Gignoux, J., and Menaut, J.~C. (1999).
\newblock Demography of a savanna palm tree: predictions from comprehensive
  spatial pattern analyses.
\newblock {\em Ecology}, 80:1987--2005.

\bibitem[Ceyhan, 2006]{ceyhan:overall}
Ceyhan, E. (2006).
\newblock On the use of nearest neighbor contingency tables for testing spatial
  segregation.
\newblock {\em Accepted for publication in Environmental and Ecological Statistics.}
\newblock Also available as Technical Report \# KU-EC-08-4, Ko\c{c} University, Istanbul, Turkey or
online as arXiv:0807.4236 [stat.ME].

\bibitem[Ceyhan, 2007]{ceyhan:segreg-edge-correctAS2007}
Ceyhan, E. (2007).
\newblock Edge correction for segregation tests based on nearest neighbor
  contingency tables.
\newblock In {\em Proceedings of the Applied Statistics 2007 International
  Conference, Ribno (Bled), Slovenia}.

\bibitem[Ceyhan, 2008a]{ceyhan:cell}
Ceyhan, E. (2008a).
\newblock Overall and pairwise segregation tests based on nearest neighbor
  contingency tables.
\newblock {\em Accepted for publication in Computational Statistics \& Data Analysis.}
\newblock Available as Technical Report \# KU-EC-08-1, Ko\c{c} University, Istanbul, Turkey or
online as arXiv:0805.1629v2 [stat.ME].

\bibitem[Ceyhan, 2008b]{ceyhanarXivQRAdjust:2008}
Ceyhan, E. (2008b).
\newblock {QR}-adjustment for clustering tests based on nearest neighbor
  contingency tables.
\newblock Also available as Technical Report \# KU-EC-08-5, Ko\c{c} University, Istanbul, Turkey or
online as arXiv:0807.4231v1 [stat.ME].

\bibitem[Coomes et~al., 1999]{coomes:1999}
Coomes, D.~A., Rees, M., and Turnbull, L. (1999).
\newblock Identifying aggregation and association in fully mapped spatial data.
\newblock {\em Ecology}, 80(2):554--565.

\bibitem[Cuzick and Edwards, 1990]{cuzick:1990}
Cuzick, J. and Edwards, R. (1990).
\newblock Spatial clustering for inhomogeneous populations (with discussion).
\newblock {\em Journal of the Royal Statistical Society, Series B}, 52:73--104.

\bibitem[Diggle, 2003]{diggle:2003}
Diggle, P.~J. (2003).
\newblock {\em Statistical Analysis of Spatial Point Patterns}.
\newblock Hodder Arnold Publishers, London.

\bibitem[Dixon, 1994]{dixon:1994}
Dixon, P.~M. (1994).
\newblock Testing spatial segregation using a nearest-neighbor contingency
  table.
\newblock {\em Ecology}, 75(7):1940--1948.

\bibitem[Dixon, 2002a]{dixon:NNCTEco2002}
Dixon, P.~M. (2002a).
\newblock Nearest-neighbor contingency table analysis of spatial segregation
  for several species.
\newblock {\em Ecoscience}, 9(2):142--151.

\bibitem[Dixon, 2002b]{dixon:EncycEnv2002}
Dixon, P.~M. (2002b).
\newblock Nearest neighbor methods.
\newblock {\em Encyclopedia of Environmetrics, edited by Abdel H. El-Shaarawi
  and Walter W. Piegorsch, John Wiley \& Sons Ltd., NY}, 3:1370--1383.

\bibitem[Good and Whipple, 1982]{good:1982}
Good, B.~J. and Whipple, S.~A. (1982).
\newblock Tree spatial patterns: {S}outh {C}arolina bottomland and swamp
  forests.
\newblock {\em Bulletin of the Torrey Botanical Club}, 109:529--536.

\bibitem[Goreaud and P\'{e}lissier, 2003]{goreaud:2003}
Goreaud, F. and P\'{e}lissier, R. (2003).
\newblock Avoiding misinterpretation of biotic interactions with the intertype
  ${K}_{12}$-function: population independence vs. random labelling hypotheses.
\newblock {\em Journal of Vegetation Science}, 14(5):681–--692.

\bibitem[Haase, 1995]{haase:1995}
Haase, P. (1995).
\newblock Spatial pattern analysis in ecology based on {R}ipley's
  {$K$}-function: {I}ntroduction and methods of edge correction.
\newblock {\em The Journal of Vegetation Science}, 6:575--582.

\bibitem[Hamill and Wright, 1986]{hamill:1986}
Hamill, D.~M. and Wright, S.~J. (1986).
\newblock Testing the dispersion of juveniles relative to adults: A new
  analytical method.
\newblock {\em Ecology}, 67(2):952--957.

\bibitem[Herler and Patzner, 2005]{herler:2005}
Herler, J. and Patzner, R.~A. (2005).
\newblock Spatial segregation of two common {G}obius species ({T}eleostei:
  {G}obiidae) in the {N}orthern {A}driatic {S}ea.
\newblock {\em Marine Ecology}, 26(2):121--129.

\bibitem[Herrera, 1988]{herrera:1988}
Herrera, C.~M. (1988).
\newblock Plant size, spacing patterns, and host-plant selection in
  \emph{{O}syris quadripartita}, a hemiparasitic dioecious shrub.
\newblock {\em Journal of Ecology}, 76:995--1006.

\bibitem[Kulldorff, 1997]{kulldorff:1997}
Kulldorff, M. (1997).
\newblock A spatial scan statistic.
\newblock {\em Communications in Statistics - Theory and Methods},
  26:1481--1496.

\bibitem[Kulldorff, 2006]{kulldorff:2006}
Kulldorff, M. (2006).
\newblock Tests for spatial randomness adjusted for an inhomogeneity: A general
  framework.
\newblock {\em Journal of the American Statistical Association},
  101(475):1289--1305.

\bibitem[Loosmore and Ford, 2006]{loosmore:2006}
Loosmore, N. and Ford, E. (2006).
\newblock Statistical inference using the $g$ or $k$ point pattern spatial
  statistics.
\newblock {\em Ecology}, 87:1925--1931.

\bibitem[Meagher and Burdick, 1980]{meagher:1980}
Meagher, T.~R. and Burdick, D.~S. (1980).
\newblock The use of nearest neighbor frequency analysis in studies of
  association.
\newblock {\em Ecology}, 61(5):1253--1255.

\bibitem[Moran, 1948]{moran:1948}
Moran, P. A.~P. (1948).
\newblock The interpretation of statistical maps.
\newblock {\em Journal of the Royal Statistical Society, Series B},
  10:243--251.

\bibitem[Nanami et~al., 1999]{nanami:1999}
Nanami, S.~H., Kawaguchi, H., and Yamakura, T. (1999).
\newblock Dioecy-induced spatial patterns of two codominant tree species,
  \emph{{P}odocarpus nagi} and \emph{{N}eolitsea aciculata}.
\newblock {\em Journal of Ecology}, 87(4):678--687.

\bibitem[Orton, 1982]{orton:1982}
Orton, C.~R. (1982).
\newblock Stochastic process and archeological mechanism in spatial analysis.
\newblock {\em Journal of Archeological Science}, 9:1--23.

\bibitem[Perry et~al., 2006]{perry:2006}
Perry, G., Miller, B., and Enright, N. (2006).
\newblock A comparison of methods for the statistical analysis of spatial point
  patterns in plant ecology.
\newblock {\em Plant Ecology}, 187(1):59–--82.

\bibitem[Pielou, 1961]{pielou:1961}
Pielou, E.~C. (1961).
\newblock Segregation and symmetry in two-species populations as studied by
  nearest-neighbor relationships.
\newblock {\em Journal of Ecology}, 49(2):255--269.

\bibitem[Ripley, 2004]{ripley:2004}
Ripley, B.~D. (2004).
\newblock {\em Spatial Statistics}.
\newblock Wiley-{I}nterscience, New York.

\bibitem[Searle, 2006]{searle:2006}
Searle, S.~R. (2006).
\newblock {\em Matrix Algebra Useful for Statistics}.
\newblock Wiley-{I}ntersciences.

\bibitem[Song and Kulldorff, 2003]{song:2003}
Song, C. and Kulldorff, M. (2003).
\newblock Power evaluation of disease clustering tests.
\newblock {\em International Journal of Health Geographics}, 2(9).

\bibitem[Stoyan and Stoyan, 1994]{stoyan:1994}
Stoyan, D. and Stoyan, H. (1994).
\newblock {\em Fractals, random shapes and point fields: methods of geometrical
  statistics.}
\newblock John Wiley and Sons, New York.

\bibitem[Stoyan and Stoyan, 1996]{stoyan:1996}
Stoyan, D. and Stoyan, H. (1996).
\newblock Estimating pair correlation functions of planar cluster processes.
\newblock {\em Biometrical Journal}, 38(3):259--271.

\bibitem[van Lieshout and Baddeley, 1996]{lieshout:1996}
van Lieshout, M.~N.~M. and Baddeley, A.~J. (1996).
\newblock A nonparametric measure of spatial interaction in point patterns.
\newblock {\em Statistica Neerlandica}, 50:344--361.

\bibitem[van Lieshout and Baddeley, 1999]{lieshout:1999}
van Lieshout, M.~N.~M. and Baddeley, A.~J. (1999).
\newblock Indices of dependence between types in multivariate point patterns.
\newblock {\em Scandinavian Journal of Statistics}, 26:511--532.

\bibitem[Waller and Gotway, 2004]{waller:2004}
Waller, L.~A. and Gotway, C.~A. (2004).
\newblock {\em Applied Spatial Statistics for Public Health Data}.
\newblock Wiley-Interscience, NJ.

\bibitem[Whipple, 1980]{whipple:1980}
Whipple, S.~A. (1980).
\newblock Population dispersion patterns of trees in a {S}outhern {L}ouisiana
  hardwood forest.
\newblock {\em Bulletin of the Torrey Botanical Club}, 107:71--76.

\bibitem[Wiegand et~al., 2007]{wiegand:2007}
Wiegand, T., Gunatilleke, S., and Gunatilleke, N. (2007).
\newblock Species associations in a heterogeneous {S}ri {L}ankan dipterocarp
  forest.
\newblock {\em The {A}merican {N}aturalist}, 170(4):77--95.

\bibitem[Yamada and Rogersen, 2003]{yamada:2003}
Yamada, I. and Rogersen, P.~A. (2003).
\newblock An empirical comparison of edge effect correction methods applied to
  {$K$}-function analysis.
\newblock {\em Geographical Analysis}, 35(2):97--109.

\end{thebibliography}

\section*{Appendix: Details of Empirical Correction of Pielou's Test of Segregation}
In this section, we provide the details concerning
the estimation of the mean and variance of
Pielou's test of segregation.
First, we plot the kernel density estimates of Pielou's test statistic
for various sample size combinations, and the
density of the corresponding asymptotic distribution.
Then, we report the means and variances of the test statistic
for each sample size combination, and suggest a transformation for the
test statistic based on these means and variances.

In Figure \ref{fig:Piel-chi-scores},
we plot the kernel density estimates for Pielou's test statistic obtained for
each sample size combination and the
density plot of the $\chi^2_1$-distribution.
Note that the discrepancy between
the density plot of $\chi^2_1$-distribution and the kernel density
estimates around 0 is because of the kernel
smoothing in density estimation.
Otherwise, for larger ---than 0--- values,
kernel density estimates follow the trend of a $\chi^2$ distribution,
but perhaps requires an adjustment for location and scale.
Notice also that, the kernel density curves are smaller for balanced (i.e., similar) sample sizes
and larger for unbalanced (i.e., very different) sample sizes compared to the pdf
of $\chi^2_1$-distribution.

\begin{figure}[h]
\centering
 \rotatebox{-90}{ \resizebox{3.1 in}{!}{\includegraphics{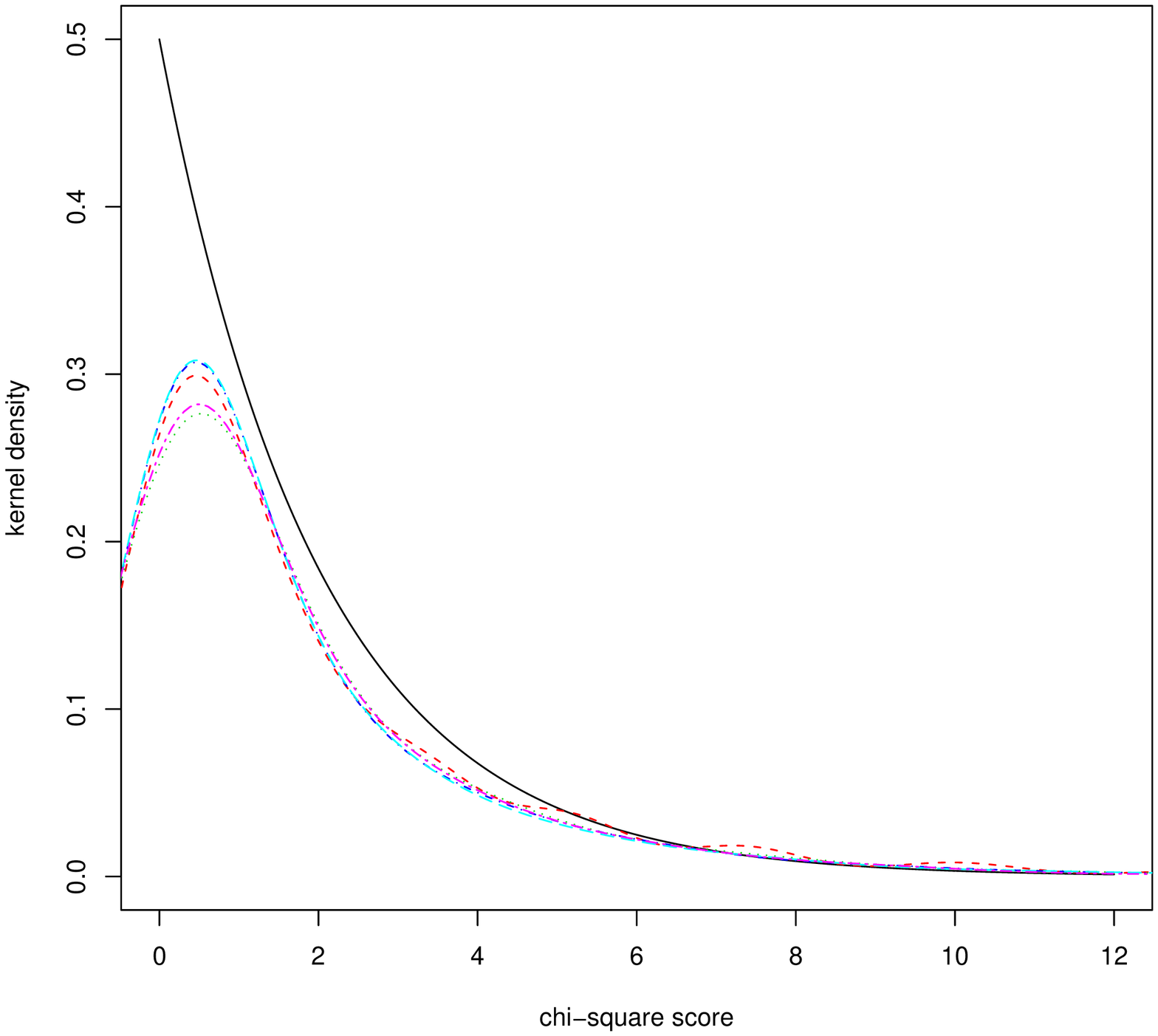} }}
 \rotatebox{-90}{ \resizebox{3.1 in}{!}{\includegraphics{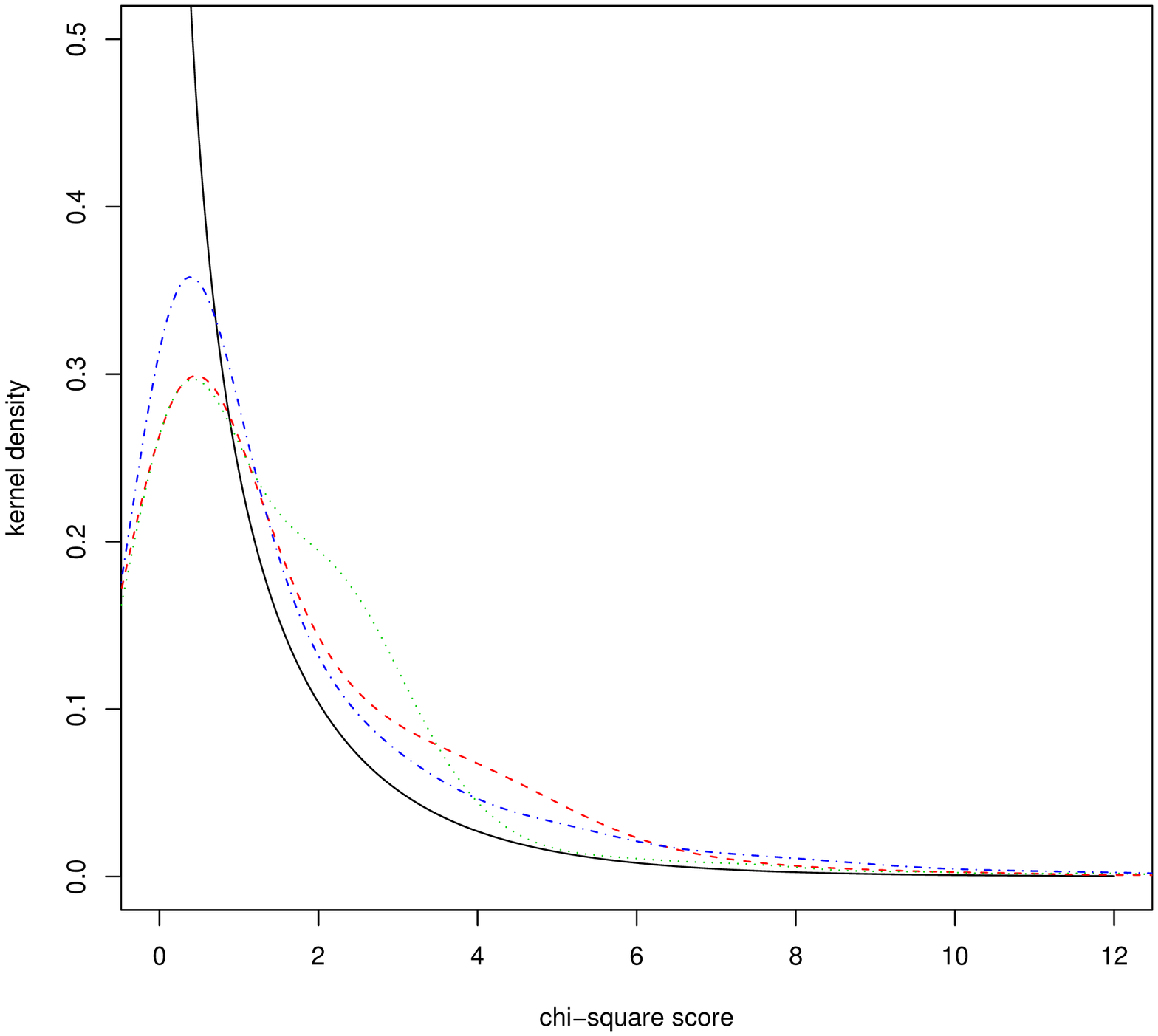} }}
 \caption{
\label{fig:Piel-chi-scores}
The density plots of of the $\chi^2_1$-distribution (solid line) and
the kernel density estimates of the Pielou's test scores, $\X^2_P$,
for balanced (left) and unbalanced (right) sample size combinations.}
\end{figure}

\begin{table}[]
\centering
\begin{tabular}{|c||c|c||c|c|}
\hline
\multicolumn{5}{|c|}{Empirical Means and Variances} \\
\multicolumn{5}{|c|}{of the Test Statistics} \\
\hline
sizes &\multicolumn{2}{|c||}{Means}&\multicolumn{2}{|c|}{Variances} \\
\hline
$(n_1,n_2)$ & $\mathbf M[\X^2_P]$ & $\mathbf M[C_D]$ &
$\mathbf V[\X^2_P]$ & $\mathbf V[C_D]$ \\
\hline
 (10,10) &  1.793 & 2.021 & 5.698 & 3.420\\
 \hline
 (10,30) &  1.647 & 2.020 & 4.233 & 3.660\\
 \hline
 (10,50) &  1.575 & 1.997 & 4.481 & 3.857\\
 \hline
 (30,30) &  1.654 & 2.007 & 5.201 & 3.774\\
 \hline
 (30,50) &  1.653 & 2.009 & 5.237 & 3.787\\
 \hline
 (50,50) &  1.647 & 2.016 & 5.328 & 3.905\\
 \hline
 (100,100) &  1.646 & 2.010 & 5.304 & 3.837\\
 \hline
 (200,200) &  1.628 & 2.005 & 5.409 & 4.053\\
\hline
\end{tabular}
\caption{
\label{tab:means-variances}
The empirical means and variances for Pielou's and Dixon's segregation tests.}
\end{table}

Let $\mathbf M[\X^2_P]$ be the sample mean and
$\mathbf V[\X^2_P]$ be the sample variance of the calculated $\X^2_P$ values.
We present the empirical means and variances of
Pielou's and Dixon's test statistics for each sample size
combination in Table \ref{tab:means-variances}
which suggests that $\mathbf M[\X^2_P] \approx 1.63$ and
$\mathbf V[\X^2_P]\approx5.40$.
Since, the critical values based on $\chi^2_1$-distribution is used for Pielou's test,
it is desirable to have the corrected scores to be approximately
distributed as $\chi^2_1$.
We transform the $\X^2_P$ scores by adjusting for
location and scaling as
$\displaystyle \X^2_{P,mc}:= \frac{\X^2_P-\gamma_P}{\delta_P}$ so that
$\displaystyle \E\left[\frac{\X^2_P-\gamma_P}{\delta_P} \right]\approx 1$ and
$\displaystyle \Var\left[\frac{\X^2_P-\gamma_P}{\delta_P} \right] \approx 2$ would hold.
Such a transformation will convert the $\X^2_P$ values into a variable
approximately distributed as $\chi^2_1$.
Using the sample estimates $\mathbf M[\X^2_P]$ and $\mathbf V[\X^2_P]$ for $\E[\X^2_P]$ and
$\Var[\X^2_P]$, and solving for $\gamma_P$ and $\delta_P$ by simple
algebra yields $\delta_P=\sqrt{\mathbf V[\X^2_P]/2}=1.643$ and
$\gamma_P=\mathbf M[\X^2_P]-\delta_P=-0.013$.

\end{document}